\documentclass[float,superscriptaddress,nofootinbib,twocolumn,amsmath,amssymb,aps,prc]{revtex4-1}
\usepackage{graphicx}
\usepackage{bm}
\usepackage{color}
\usepackage{subfigure}
\usepackage{times}
\usepackage[colorlinks,
            linkcolor=red,
            anchorcolor=blue,
            citecolor=blue
            ]{hyperref}

\begin{document}

\title{A hydrodynamic study of hyperon spin polarization in relativistic heavy ion collisions}

\author{Baochi Fu}
\affiliation{Department of Physics and State Key Laboratory of Nuclear Physics and Technology, Peking University, Beijing 100871, China}
\affiliation{Collaborative Innovation Center of Quantum Matter, Beijing 100871, China}
\author{Kai Xu}
\affiliation{Department of Physics and State Key Laboratory of Nuclear Physics and Technology, Peking University, Beijing 100871, China}
\affiliation{Collaborative Innovation Center of Quantum Matter, Beijing 100871, China}
\author{Xu-Guang Huang}
\email{huangxuguang@fudan.edu.cn}
\affiliation{Department of Physics and Center for Field Theory and Particle Physics, Fudan University, Shanghai, 200433, China}
\affiliation{Key Laboratory of Nuclear Physics and Ion-beam Application (MOE), Fudan University, Shanghai 200433, China}
\author{Huichao Song}
\email{huichaosong@pku.edu.cn}
\affiliation{Department of Physics and State Key Laboratory of Nuclear Physics and Technology, Peking University, Beijing 100871, China}
\affiliation{Collaborative Innovation Center of Quantum Matter, Beijing 100871, China}
\affiliation{Center for High Energy Physics, Peking University, Beijing 100871, China}

\date{\today}

\begin{abstract}
    We perform a systematic study of the spin polarization of hyperons in heavy-ion collisions using the MUSIC hydrodynamic model with A Multi-Phase Transport (AMPT) pre-equilibrium dynamics. Our model calculations nicely describe the measured  collision-energy, centrality, rapidity, and $p_T$ dependence of $\Lambda$ polarization.
    We also study and predict the global spin polarization of $\Xi^-$ and $\Omega^-$ as a function of collision energy, which provides a baseline for the studies of the magnetic moment, spin, and mass dependence of the spin polarization.
    For the local spin polarization, we calculate the radial and azimuthal components of the transverse $\Lambda$ polarization and find specific modulating behavior which could reflect the circular vortical structure.
    However, our model fails to describe the azimuthal-angle dependence of the longitudinal and transverse  $\Lambda$ polarization, which indicates that the hydrodynamic framework with the spin Cooper-Frye formula under the assumption of thermal equilibrium of spin degree of freedom needs to be improved.
\end{abstract}

\maketitle

\section{\label{sec:Intro}Introduction}
In peripheral high-energy heavy-ion collisions, the colliding system contains a large amount of orbital angular momentum perpendicular to the reaction plane, a portion of which is carried by the produced quark-gluon plasma (QGP) in the form of fluid vorticity.
Such orbital angular momentum of QGP can be transferred into the spin of the constituent particles via the spin-vorticity coupling. As a results, the final emitted hyperons are globally spin-polarized along the direction of initial angular momentum~\cite{Liang:2004ph,Voloshin:2004ha}.


In experiments, the spin polarization of $\Lambda$ and $\bar{\Lambda}$ hyperons (``$\Lambda$ polarization") can be determined by measuring the angular distribution of their weak-decay products. In such a way, the global $\Lambda$ polarization (i.e., the spin polarization integrated over kinematics) has been successfully observed by STAR Collaboration in 7.7- 200 A GeV Au + Au collisions at Relativistic Heavy Ion Collider(RHIC)~\cite{STAR:2007, STAR:2017ckg,STAR:2018ivw,Niida:2018hfw}, while was also reported to be consistent with zero in 2.76 A TeV and 5.02 A TeV Pb + Pb collisions at the Large Hadron Collider (LHC)~\cite{ALICE:2019}. Besides the global polarization, the differential $\Lambda$ polarization has also been measured (dubbed local $\Lambda$ polarization)~\cite{Niida:2018hfw,STAR:2019srw,ALICE:2019}. In particular, it is found that the longitudinal component of $\Lambda$ polarization shows a quadrupole pattern in the transverse plane\cite{STAR:2019srw}.
Different from the global behavior, such quadrupole structure of the longitudinal $\Lambda$ polarization survives at very high collision energies indicating that the spin polarization might be influenced by not only the initial orbital angular momentum
but also other contributions like, e.g., the inhomogeneous expansion of the fireball.

In theory, the global spin polarization in heavy-ion collisions was first proposed by Liang and Wang \cite{Liang:2004ph} and was noted by Voloshin from the experimental side \cite{Voloshin:2004ha}.
Early works focused on calculations of spin polarization of quarks and anti-quarks and the polarization of final hadrons was estimated through recombination or fragmentation mechanism \cite{Liang:2004ph,Gao:2007bc,Betz:2007kg,Huang:2011ru}. Later on, an explicit formula, called spin Cooper-Frye formula, that connects the spin polarization of hadrons to the thermal vorticity at freeze-out hypersurface was established for the thermal equilibrium system~\cite{Becattini:2013fla} (see also Refs.~\cite{Fang:2016vpj, Liu:2020flb} for derivations in kinetic theory). It can be used to calculate $\Lambda$ polarization with  hydrodynamics or transport model simulations and roughly fit the global $\Lambda$ polarization~\cite{Becattini:2013vja,Becattini:2015ska,Becattini:2016gvu,Karpenko:2016jyx,Xie:2016fjj,Xie:2017upb,Li:2017slc,Ivanov:2019ern,Shi:2017wpk,Wei:2018zfb, Ivanov:2019ern, Ivanov:2020wak}. However, such spin Cooper-Frye formula fails to reproduce the azimuthal-angle dependence of $\Lambda$ polarization and results in a sign difference compared with the experimental data~\cite{Becattini:2017gcx, Florkowski:2019voj,Wu:2019eyi, Xie:2019jun, Xia:2019fjf,Becattini:2019ntv,Liu:2019krs}. (For reviews, please refer to Refs.~\cite{Huang:2020xyr,Liu:2020ymh,Becattini:2020ngo,Gao:2020vbh,Gao:2020lxh,Huang:2020dtn}.) Besides the $\Lambda$ polarization, the spin polarization of quarks can also lead to other potential observables, such as  the global and local spin alignment of vector mesons~\cite{Liang:2004xn,Sheng:2019kmk,Sheng:2020ghv,Xia:2020tyd}, enhancement of spinful hadrons' yields~\cite{Taya:2020sej}, baryonic spin hall effect~\cite{Liu:2020dxg}.

In this work, we perform a systematic study on the $\Lambda$ polarization in 7.7 A GeV- 200 A GeV Au + Au collisions, using the MUSIC hydrodynamic model with A Multi-Phase Transport (AMPT) pre-equilibrium initial condition.
Such hydrodynamic models and hybrid approaches have been widely used in relativistic heavy-ion collision at RHIC and the LHC energies, which successfully describe various soft hadronic observables such as the particle yields, transverse momentum spectra, flows anisotropies~\cite{Schenke:2010rr,Schenke:2011bn, Gale:2012rq,Schenke:2014zha,Karpenko:2015xea,Song:2017wtw,Zhao:2017yhj,Zhu:2016puf, Zhao:2018lyf, Denicol:2018wdp,Schenke:2020mbo,Schenke:2019pmk}.
In hydrodynamic calculations, the dynamic variables are local temperature, flow velocity, and conserved charge densities. Using the spin Cooper-Frye formula, the spin polarization vector can be calculated with these hydrodynamics variables on the freeze-out surface. In~\cite{Karpenko:2016jyx,Xie:2017upb,Ivanov:2019ern}, the energy dependence of global $\Lambda$ polarization has been reproduced with hydrodynamic calculations.  However, for the local $\Lambda$ polarization, hydrodynamics fails to reproduce the sign of the azimuthal-angle dependence of both transverse and longitudinal polarization \cite{Xie:2017upb, Karpenko:2016jyx, Becattini:2017gcx, Florkowski:2019voj,Wu:2019eyi}, which is a pending puzzle.
In this work, we will further study the global and local spin polarization within the AMPT+MUSIC hydrodynamic framework. For the global polarization, besides demonstrating that AMPT+MUSIC model can nicely describe the measured  collision-energy, centrality, rapidity, and $p_T$ dependence of $\Lambda$ polarization, we will predict the spin polarization of $\Xi^-$ and $\Omega^-$,  which provides a baseline for the studies of the magnetic moment, spin, and mass dependence of the spin polarization. For the local polarization, we focus on investigating how the initial condition and hydrodynamic evolution influence the local polarization and how different vorticity terms influence the local polarization of $P_z$.  We also calculate the radial and azimuthal components of the transverse $\Lambda$ polarization and find specific modulating behavior which could reflect the circular vortical structure.

This paper is organized as follows. Section~\ref{sec:Model} gives a brief introduction to AMPT+MUSIC hybrid model and the spin polarization formula used in this work.
Section~\ref{sec:num} presents and discusses the results of global and local spin polarization.
Section~\ref{sec:Summary}  concludes and summarizes the paper.

\section{\label{sec:Model}Model setup}
In this work, we implement AMPT+MUSIC hybrid model to study the global and local $\Lambda$-polarization in 7.7- 200 A GeV Au + Au collisions at RHIC. AMPT is A Multi-Phase Transport Model developed in~\cite{Lin:2004en}, which is implemented here to simulate the pre-equilibrium dynamics and generate the initial profiles for the succeeding MUSIC hydrodynamic evolution. MUSIC is a 3+1 dimensional relativistic viscous hydrodynamic model~\cite{Schenke:2010nt,Schenke:2011bn,Gale:2012rq} to describe the evolution of the QGP fireball. The MUSIC hydrodynamics can also  be followed by a UrQMD afterburner~\cite{Bass:1998ca, Bleicher:1999xi} to describe the scattering and evolution of dilute hadronic matter.

Compared with early hybrid model simulations with AMPT initial energy density profiles~\cite{Xu:2016hmp,Zhao:2017rgg,Zhao:2020pty}, we generate the whole energy-momentum tensor $T^{\mu\nu}$ from AMPT for the hydrodynamic evolution since initial flow is essential for the development of fluid vorticity for the study of spin polarization.

\subsection{\label{subsec:AMPT}AMPT Initial condition}
In this paper, we use AMPT model to generate the initial energy-momentum tensor $T^{\mu\nu}$
and net baryon current $N^\mu$ for the succeeding MUSIC hydrodynamic evolution.
In the string melting version of AMPT~\cite{Lin:2004en}, the sub-program HIJING \cite{Wang:1991hta,Gyulassy:1994ew} samples initial partons from mini-jets and excitation strings.
Then, the phase-space information of partons is imported to sub-program ZPC for the subsequent partonic  evolution. Here we do not evolve this partonic stage till hadronization but take the information of partons at a certain switching hypersurface with a constant proper time $\tau_0$ (the initial time of hydrodynamics).

On the switching hypersurface, the energy-momentum tensor $T^{\mu\nu}$ and net baryon density $n$ are constructed from Gaussian smearing in Milne coordinate as~\cite{Pang:2012he, Pang:2016igs,Xu:2016hmp}:
\begin{subequations}
\begin{eqnarray}
&&T^{\mu \nu}(\tau_0, x, y, \eta_s) = \sum_{i} {\frac{p^\mu_i p^\nu_i}{p^\tau_i}}\Phi_G(\tau_0,{\bm x}; {\bm x}_i),\\
&&n(\tau_0, x, y, \eta_s) =  \sum_{i} Q_i \Phi_G(\tau_0,{\bm x}; {\bm x}_i),
\end{eqnarray}%
\end{subequations}
where the smearing function $\Phi_G$ is given by
\begin{eqnarray}
&&\Phi_G(\tau_0,{\bm x}; {\bm x}_i) = {\frac{K}{\tau_0 \sqrt{2 \pi \sigma^2_{\eta_s}}}}{\frac{1}{2 \pi \sigma^2_r}}\nonumber\\
&&\;\;\;\;\;\;\times
\exp\left[- {\frac{( x - x_i)^2 + (y - y_i)^2}{2 \sigma^2_r}} - {\frac{(\eta_s - \eta_{is})^2}{2 \sigma^2_{\eta_s}}}\right].\;\;\;
\end{eqnarray}%
Here, $p^\mu_i$ denotes the 4-momentum vector of the $i$-th parton with
$p^\tau_i = m_{iT}\cosh(Y_i - \eta_{is})$, $p^x_i = p_{ix}$, $p^y_i = p_{iy}$, $p^\eta_i = m_{iT}\sinh(Y_i - \eta_{is})/\tau_0$, $m_{iT}$ the transverse mass, $Y_i$ the rapidity, and $\eta_{is}$ the spacetime rapidity of the $i$-th parton. $Q_i$ is the baryon charge of $i$-th parton. The Gaussian smearing factors $\sigma_r, \sigma_{\eta_s}$, and scale factor $K$ are tuned
to fit the spectra and flow of all charged hadrons at the most central collisions and are kept fixed for other centrality classes (see also Refs.~\cite{Pang:2012he,Xu:2016hmp}).
In each centrality bin, we average 1,000 AMPT events to get smooth initial $T^{\mu \nu}$ and $n$ profiles for the succeeding hydrodynamic evolution.

\subsection{MUSIC \ Hydrodynamics}
MUSIC is a 3+1 dimensional viscous hydrodynamic program which solves the conservation equations for energy momentum tensor $T^{\mu\nu}$  and charge current $N^{\mu}$~\cite{Schenke:2010nt, Schenke:2010rr, Schenke:2011bn}:
\begin{subequations}
\label{hydro0}
\begin{eqnarray}
\label{hydro1}
&&\partial_\mu T^{\mu\nu}(x) = 0,  \\
&&\partial_\mu N^{\mu} (x) = 0,
\end{eqnarray}%
\end{subequations}
and the 2nd order Israel-Stewart equations for shear stress tensor $\pi^{\mu\nu}$ and bulk pressure $\Pi$~\cite{Denicol:2012cn, Denicol:2014vaa}:
\begin{subequations}
\label{eq:IS}
\begin{eqnarray}
&&\;\;\;\;\;\;\;\; \tau_\Pi \dot\Pi + \Pi = - \zeta \theta - \delta_{\Pi\Pi}\Pi\theta + \lambda_{\Pi_\pi} \pi^{\mu\nu} \sigma_{\mu\nu}, \label{IS1}\\
&&\tau_\pi \dot \pi^{\langle \mu\nu \rangle} + \pi^{\mu \nu} = 2\eta \sigma^{\mu\nu} - \delta_{\pi\pi} \pi^{\mu\nu} \theta + \phi_7 \pi_\alpha ^{\langle \mu} \pi^{\nu \rangle \alpha} \nonumber\\
&&\;\;\;\;\;\;\;\;\;\;\;\;\;\;\;\;\;\;\;\;\;\;\;\;\;\; - \tau_{\pi \pi} \pi_\alpha ^{\langle \mu} \sigma^{\nu \rangle \alpha} + \lambda_{\pi_\Pi} \Pi \sigma^{\mu \nu}.        \label{IS2}
\end{eqnarray}%
\end{subequations}
Here, $T^{\mu\nu}$ and $N^\mu$ are decomposed as $T^{\mu\nu} = \epsilon u^\mu u ^\nu - (p + \Pi)\Delta^{\mu\nu} + \pi^{\mu\nu}$, $N^\mu= n u^\mu$, where $u^\mu$ is the flow 4-velocity, $\epsilon$ is the local energy density, $p$ is pressure, $n$ is the net baryon density, and $\Delta^{\mu\nu} = g^{\mu\nu} - u^\mu u^\nu$ is the transverse projector to the flow velocity $u^\mu$. In Eqs.~(\ref{eq:IS}),  $\eta$ and $\zeta$ are shear and bulk viscosities, $A^{\langle \cdot \cdot \rangle}$ is the symmetrized and traceless projection, $\theta = \nabla_\mu  u^\mu$ and $\sigma^{\mu\nu} = \frac{1}{2}\big[ \nabla^\mu u^\nu + \nabla^\nu u^\mu - \frac{2}{3} \Delta^{\mu\nu} (\nabla_\alpha u^\alpha) \big]$ with $\nabla_\mu=\Delta_{\mu\nu}\partial^\nu$.
The transport coefficients $\tau_\Pi$, $\delta_{\Pi\Pi}$, $\lambda_{\Pi_\pi}$, $\tau_\pi$, $\delta_{\pi\pi}$, $\phi_7$, $\tau_{\pi\pi}$ and $\lambda_{\pi\Pi}$ are fixed by the Boltzmann equations~\cite{Denicol:2014vaa, Ryu:2015vwa}.
Note that we have neglected the baryon diffusion effect in this study so that $N^\mu$ only contains the ideal part.  To close the system, we input a crossover-type equation of state (EOS) as used in~\cite{Denicol:2018wdp}.

At the freeze-out hypersurface defined by a given constant energy density $E_{\rm sw}$, various hadron resonances are generated by the Monte-Carlo event generator iSS~\cite{Shen:2014vra} according to the differential Cooper-Frye formula:
\cite{Cooper:1974mv}:
\begin{equation}
\label{eq:cooper-frye0}
E \frac{d N_i}{d^3 p}(x) = \frac{g_i}{(2\pi)^3} p \cdot d^3 \Sigma (x) f_i(x,p),
\end{equation}%
where $f_i= f_{{\rm eq},i}(x,p) + \delta f_i (x,p)$ is the distribution function of the particle species $i$ with the equilibrium and off equilibrium parts taken the form $f_{{\rm eq},i}(x,p) = \frac{1}{e^{[p \cdot u(x) \mp \mu_B ]/T} + 1}$ and $\delta f_i(x,p) = f_{{\rm eq},i}(x,p)(1 - f_{{\rm eq},i}(x,p)) \frac{p^\mu p^\nu \pi^{\mu\nu}}{2 T^2 (\epsilon + P)}$ \cite{doi:10.1002, Shen:2014vra}.

With these emitted hadrons, the hydrodynamic evolution can be followed by the UrQMD~\cite{Bass:1998ca, Bleicher:1999xi} to describe the scatterings and decays of dilute hadronic matter. After all the
interactions cease, the information of these hadrons are taken to obtain various soft hadron observables.

Table \ref{tab:parameter} lists the  parameter setups in our calculations. In more details, the specific shear viscosity $\eta/s$ is set to be a constant and the specific bulk viscosity, $\zeta/s$ is taken a form with temperature dependence, which reach a peak near the phase transition as in Ref.~\cite{Ryu:2015vwa}. Following Ref. \cite{Karpenko:2015xea}, the initial time $\tau_0$ of hydrodynamic evolution is set to be 0.4 fm/c for 200 A GeV Au+Au collisions and gradually increase to 2.0 fm/c  for 7.7 A GeV Au+Au collisions. The switching energy density $E_{\rm sw}$ changes from 0.65 GeV/fm$^3$ to 0.35 GeV/fm$^3$, which roughly matches the chemical freeze-out temperature obtained from the statistical model from high to low collision energies.
\begin{table}[h]
\caption{\label{tab:parameter}%
Parameter setups
}
\begin{ruledtabular}
\begin{tabular}{llllll}
\textrm{$\sqrt{s_{NN}}$ (GeV)}&
\textrm{$\tau_0$ (fm)}&
\textrm{$\sigma_{\eta_s}$} &
\textrm{$\sigma_r$ (fm)} &
\textrm{$\eta /s$} &
\textrm{$E_{\rm sw}$ (GeV/fm$^3$)}\\
\colrule
200 & 0.4 & 0.6 & 0.6 & 0.08 & \hspace{2em}0.65 \\
62.4 & 0.6 & 0.6 & 0.6 & 0.08 & \hspace{2em}0.6 \\
39 & 0.8 & 0.4 & 0.6 & 0.08 & \hspace{2em}0.55 \\
27 & 1.0 & 0.4 & 0.4 & 0.10 & \hspace{2em}0.5 \\
19.6 & 1.2 & 0.2 & 0.4 & 0.12 & \hspace{2em}0.45 \\
7.7 & 2.0 & 0.2 & 0.3 & 0.16 & \hspace{2em}0.35 \\
\end{tabular}
\end{ruledtabular}
\end{table}

\subsection{\label{subsec:spin}Spin polarization}
For a many-body system with fermions, the presence of fluid vorticity would polarize the spin of fermions due to the quantum mechanical spin-orbital coupling. At thermal equilibrium, the mean spin vector is determined  by the local thermal vorticity, which at the leading order, is given by \cite{Becattini:2013fla,Fang:2016vpj,Liu:2020flb}:
\begin{equation} \label{eq:defP}
S^\mu (x,p) = - \frac{1}{2m} \frac{S(S + 1)}{3}[1 - f(x,p)] \epsilon ^{\mu\nu\rho\sigma} p_\sigma \varpi_{\nu\rho},
\end{equation}
where $S$ is the spin quantum number, $m$ is the mass of fermion, $f(x,p)$ is the equibrium Fermi-Dirac distribution function (In this spin polarization formula, the non-equilibrium corrections $\delta f$ is neglected) and $\varpi_{\mu\nu}$ is the thermal vorticity defined as:
\begin{equation}
\varpi_{\mu\nu} = - \frac{1}{2} (\partial_\mu \beta_\nu - \partial_\nu \beta_\mu),\;\; \text{with  }\;\; \beta_\mu = u_\mu / T.
\end{equation}
The spin polarization vector $P^\mu$ is defined as:
\begin{equation}
P^\mu(x,p) = \frac{1}{S} S^\mu(x,p).
\end{equation}
With an average of $P^\mu(x,p)$ over the freeze-out hypersurface, the spin polarization in momentum space $P^\mu (p)$ is given by:
\begin{equation}
\label{eq:cooper}
P^\mu (p) = \frac{\int d\Sigma_\nu p^\nu f(x,p) P^\mu(x,p)}{\int d\Sigma_\nu p^\nu f(x,p)}.
\end{equation}
This is the spin Cooper-Frye formula which links the spin polarization of final hadrons in momentum space to the thermal vorticity on the freeze-out hypersurface~\cite{Becattini:2013fla,Fang:2016vpj,Liu:2020flb}.

The global spin polarization is obtained by integrating the numerator and denominator of Eq.~(\ref{eq:cooper}) over the momentum:
\begin{equation} \label{eq:p_integration}
P^\mu = \langle P^\mu (p)\rangle=\frac{\int \frac{d^3 p}{E} \int d\Sigma_\nu p^\nu f(x,p) P^\mu(x,p)}{\int \frac{d^3 p}{E} \int d\Sigma_\nu p^\nu f(x,p)}.
\end{equation}
Note that the spin polarization measured in experiments is defined in the rest frame of the decay particle. To compare with experiment data, the expression of $P^\mu(x,p)$ or $S^\mu(x,p)$ is Lorentz transformed to particle's rest frame as follows:
\begin{equation}
\label{spinboost}
{\bm{S}}^* =
 \bm{S} - \frac{\bm{p} \cdot \bm{S}}{E(E + m)} \bm{p}.
\end{equation}
Since the spin degree of freedom is not included in the dynamics of the hadronic transport model UrQMD~\cite{Bass:1998ca, Bleicher:1999xi, Karpenko:2016jyx}. In this work, we calculate the spin polarization on the freeze-out hypersurface $E_{\rm sw}$ according to Eqs.~(\ref{eq:defP}-\ref{spinboost}) without the UrQMD afterburner and the following spin polarization results are all for primary particles. For other soft hadron observables, such as the spectra and $v_2$ of identified hadrons shown in the Appendix \ref{app:soft}, the UrQMD afterburner is implemented.

\begin{figure*}[htb]  
\subfigure{
\begin{minipage}[b]{0.49\linewidth}
\label{fig:global_a}
\centering
\center{\includegraphics[trim={2cm 0.5cm 2cm 2cm},clip, width=0.95\columnwidth]  {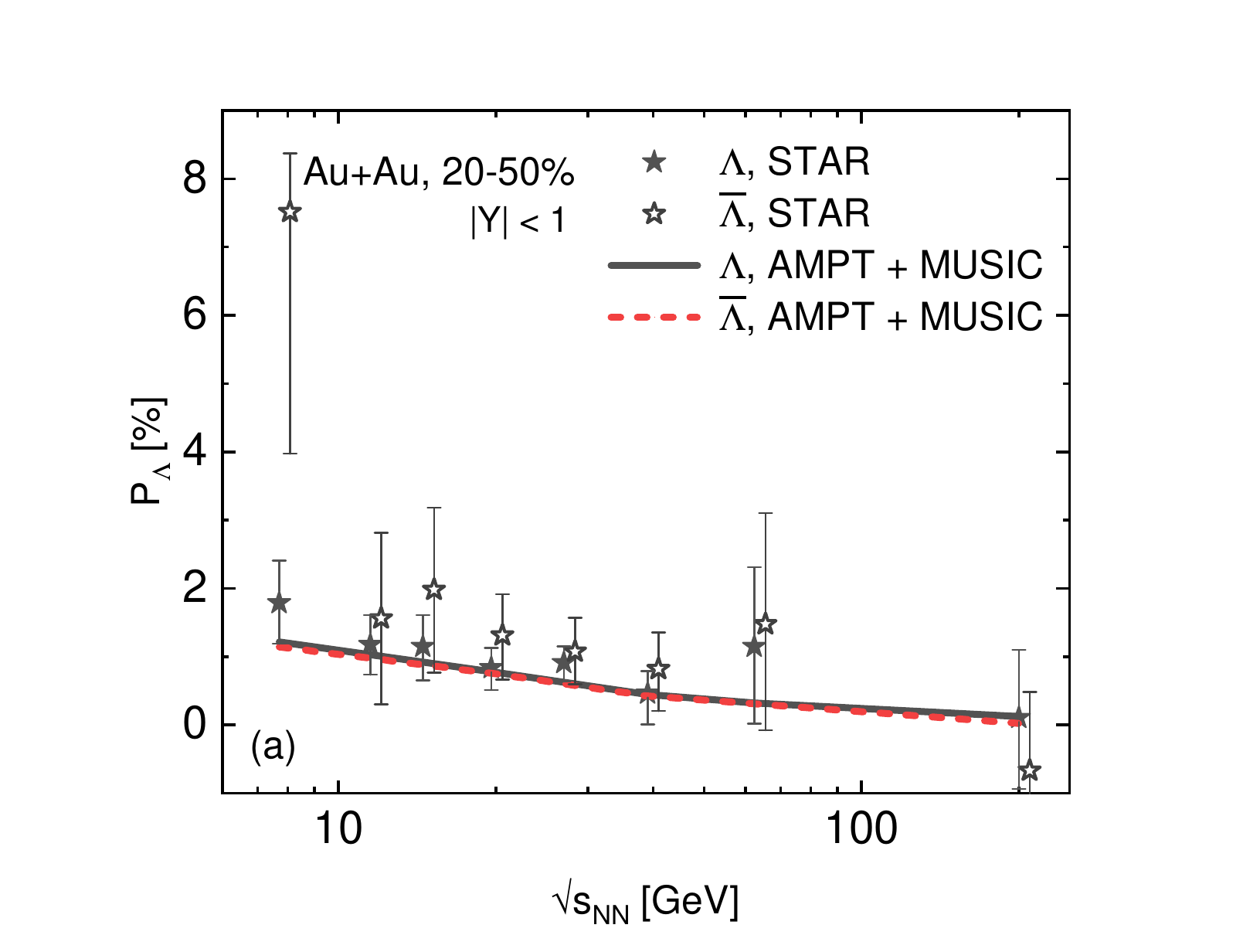}}
\end{minipage}%
}%
\subfigure{
\begin{minipage}[b]{0.49\linewidth}
\label{fig:global_b}
\center{\includegraphics[trim={2cm 0.5cm 2cm 2cm},clip, width=0.95\columnwidth]  {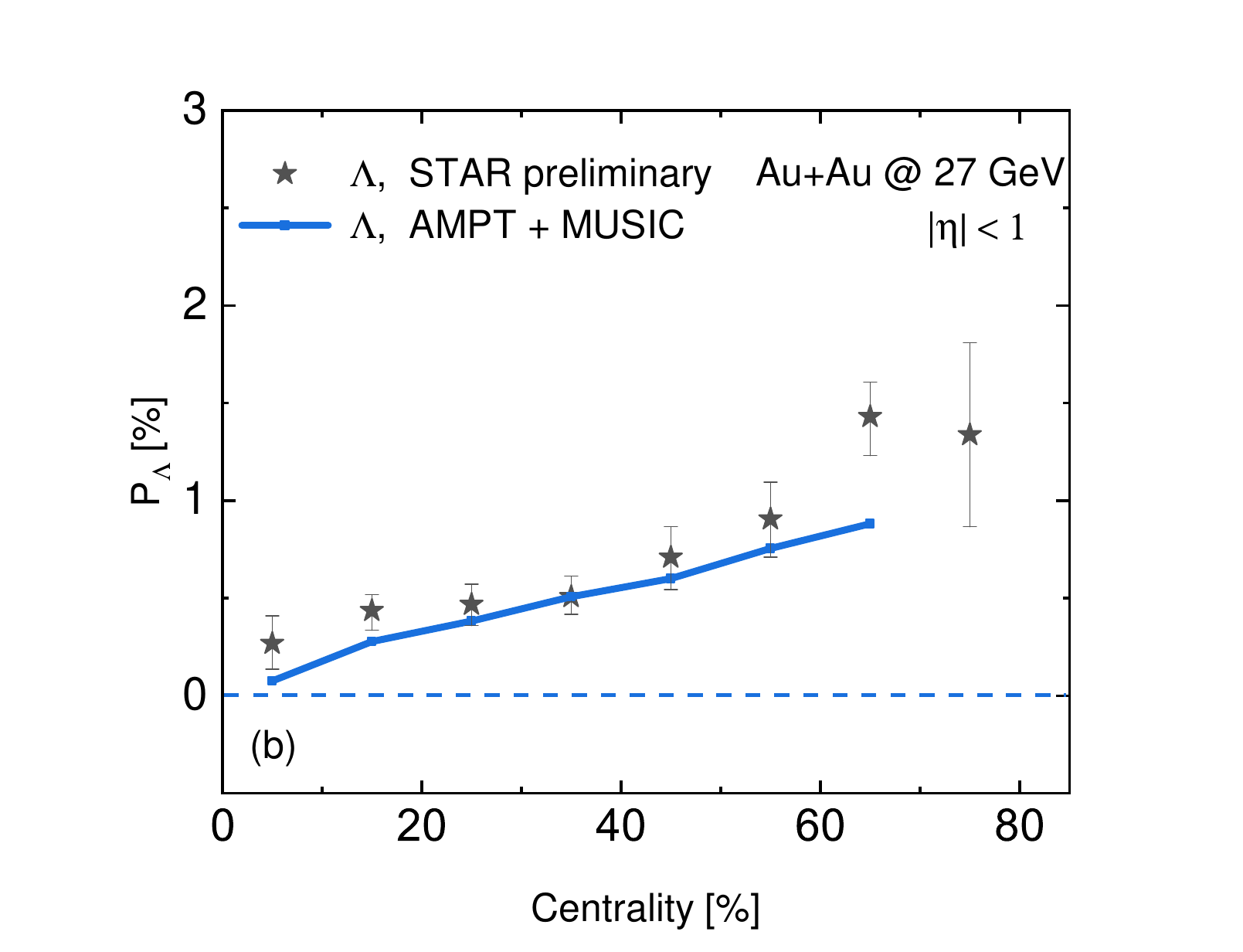}}
\end{minipage}}
\caption{\label{fig:global}
(a) Global $\Lambda$ polarization along the $-y$ direction as a function of collision energy.
(b) Centrality dependence of global $\Lambda$ polarization in 27 A GeV Au + Au collisions. The kinematic cut for the left and right panels are: $0.4 < p_T < 3$ GeV, $|Y| < 1$ and  $0.4 < p_T < 3.0$ GeV, $|\eta| < 1$, respectively. The calculated spin polarization is for primary $\Lambda$ and $\bar{\Lambda}$. The STAR data are taken from~\cite{STAR:2017ckg} (As the hyperon decay parameter $\alpha_\Lambda$ is updated, here and in the following figures, the published STAR results are scaled by a factor 0.87~\cite{Adams:AUM}.}
\end{figure*}

\section{Results and Discussions}\label{sec:num}
In this section,  we implement AMPT+MUSIC hydrodynamic model to simulate the QGP fireball evolution and then use the spin Cooper-Frye formula described in Sec. \ref{subsec:spin}  to calculate and
study the global and local spin polarization in 7.7 - 200 A GeV  Au + Au collisions at RHIC. Here we also emphasis that with the parameter listed in table \ref{tab:parameter} and a UrQMD hadronic afterburner, this hybrid model can also nicely describe many soft hadron observables, such as the spectra and $v_2$ of identified hadrons. Please refer to the Appendix \ref{app:soft} for details.

\subsection{\label{sec:Results}Global spin polarization}
Fig.\ref {fig:global_a} shows the global spin polarization of primary $\Lambda$ (without feed-down decay contributions) at mid-rapidity ($Y<1$) along $-y$ direction. The calculated global $\Lambda$ polarization from AMPT+MUSIC model decreases with the increase of collision energy and fits the experiment data within the error bars.
The decreasing feature can be understood by the fact that, at mid-rapidity, the system behaves more likely to a boost-invariant fluid with smaller vorticity, although the total angular momentum is larger, at higher energy~\cite{Deng:2016gyh,Jiang:2016woz,Deng:2020ygd}.  Figure \ref{fig:global_a} also shows that $P_{\bar\Lambda}$ is very close to $P_\Lambda$, since  the difference
between $\Lambda$ and $\bar{\Lambda}$ in our calculation only comes from the finite baryon chemical potential $\mu_B$ in Eqs. (\ref{eq:cooper-frye0}) and (\ref{eq:p_integration}). Although finite $\mu_B$ makes $P_{\bar\Lambda}$ larger than $P_\Lambda$, it also leads to more $\Lambda$ production than $\bar\Lambda$ at earlier time, where the thermal vorticity effect is more significant ~\cite{Note1}. These two effects cancel each other and makes $P_\Lambda$ and $P_{\bar{\Lambda}}$ almost identical. For this reason, in the following calculations, we will only show results for $\Lambda$.

Fig.~\ref{fig:global_b} shows  the global $\Lambda$ polarization as a function of centrality in 27 A GeV Au + Au collisions. Our calculation gives the same trend as the experimental measurements and almost fits the experimental data at different centralities. With the collision energy fixed, the global $\Lambda$ polarization is positively correlated to the total angular momentum, which increases with the centrality.

\begin{figure}[htb] 
\center{\includegraphics[trim={2cm 0.5cm 2cm 2cm},clip, width=0.95\columnwidth]  {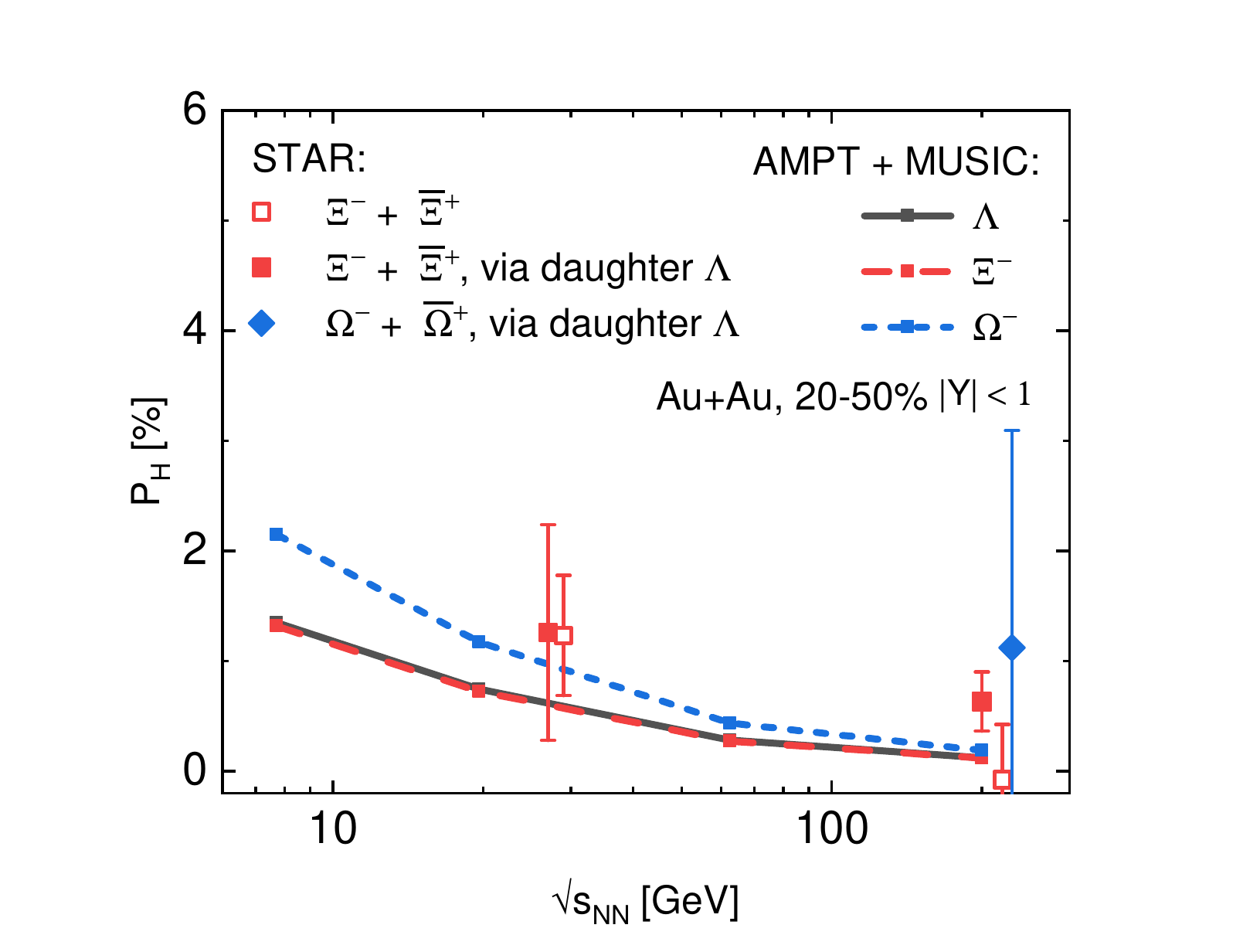}}
\caption{\label{fig:Xi}
Global spin polarization of $\Lambda$, $\Xi^-$, and $\Omega^-$ in Au+Au collisions at various collision energies.  The kinematic setup is the same as Fig. \ref{fig:global}. The STAR data are taken from Ref. \cite{Adams:AUM}. For the  measurements in 200 A GeV  and 27 A  GeV Au+Au collisions, the centrality is $20$-$80\%$ and $20$-$50\%$, respectively. For the AMPT+MUSIC calculations, the centrality is $20$-$50\%$.
}
\end{figure}

\begin{figure*}[htb] 
\subfigure{
\begin{minipage}[b]{0.49\linewidth}
\centering
\center{\includegraphics[trim={2cm 0cm 2cm 2cm},clip, width=0.95\columnwidth]  {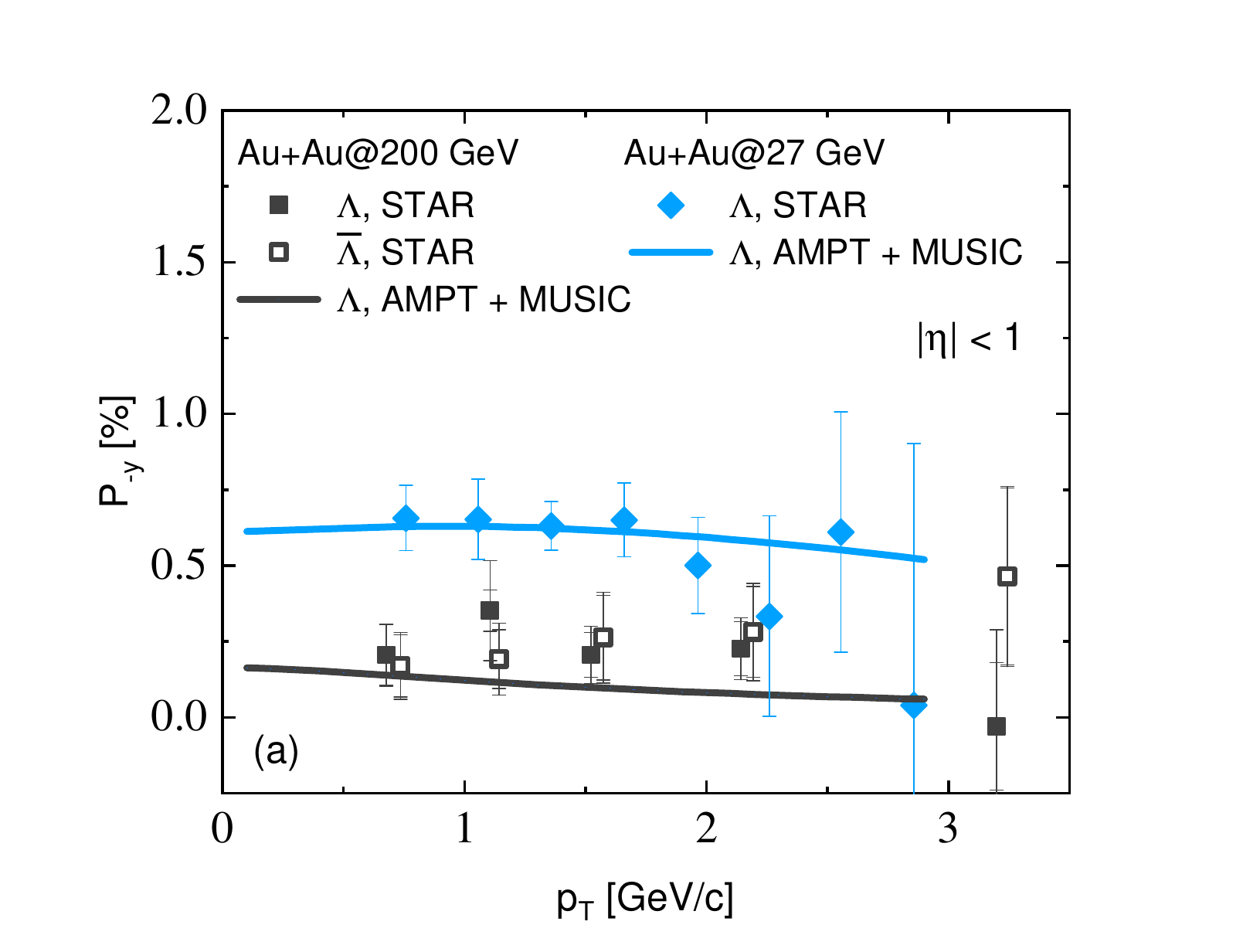}}
\end{minipage}%
}%
\subfigure{
\begin{minipage}[b]{0.49\linewidth}
\center{\includegraphics[trim={2cm 0cm 2cm 2cm},clip, width=0.95\columnwidth]  {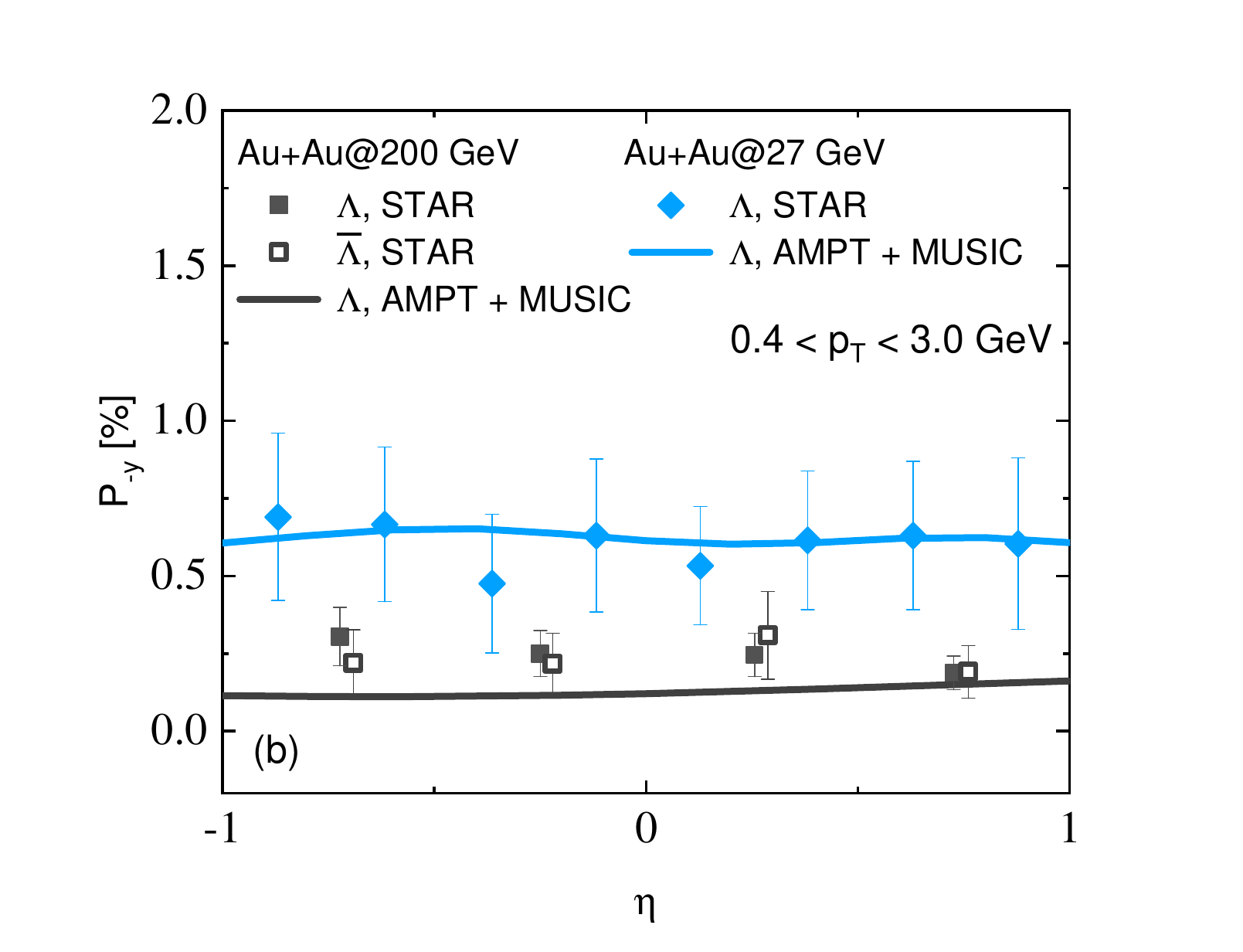}}
\end{minipage}}
\caption{\label{fig:pt_and_y}
(a) Transverse momentum and (b) pseudo-rapidity dependence of $\Lambda$ polarization along $-y$ direction in 200 A GeV Au + Au collisions at 20-60\% centrality and in 27 A GeV Au + Au collisions at 15-75\% centrality. The STAR data are taken from~\cite{STAR:2018ivw,Adams:qm}.}
\end{figure*}%

In Fig. \ref{fig:Xi}, we study the global spin polarization of $\Xi^-(1322)$ and $\Omega^-(1672)$ with the purpose of providing a baseline for the studies of the magnetic moment, spin, and mass dependence of the spin polarization. The related spin ratios for these three baryons are $S_{\Omega^-}: S_{\Xi^-}: S_{\Lambda}=3:1:1$ and the magnetic moment ratios are $|M_{\Omega^-}|: |M_{\Xi^-}|: |M_{\Lambda}|\approx 3:1:1$, while the mass ratios are $m_{\Omega^-}: m_{\Xi^-}: m_{\Lambda} = 1.5: 1.2: 1$. Therefore, by applying the spin Cooper-Frye formula Eq. (\ref{eq:p_integration}), we expect a global-polarization ordering: $P_{\Omega^-}>P_{\Xi^-}\simeq P_{\Lambda}$, as demonstrated in Fig. \ref{fig:Xi}.
In our hydrodynamic calculations, the difference between the spin polarization of $\Xi^-$ and $\Lambda$ only comes from the mass difference in
Eq. (\ref{eq:p_integration}), which is too small to discern. Note that the spin polarizations of $\Omega^-$ and $\Xi^0$ were calculated in AMPT model in Ref. \cite{Wei:2018zfb}, in which a visible difference between $\Xi^0$ and $\Lambda$ is seen. This is probably due to that, in AMPT calculation of Ref. \cite{Wei:2018zfb}, the spin polarization depends also on the momentum distribution of the hyperons which are hadronized all the time rather than on specific particlization hypersurface. Nevertheless, in both the hydrodynamic and AMPT calculations, the global spin polarizations of $\Xi$ and $\Lambda$ are close to each other. For the spin-3/2 particle $\Omega^-$, the spin polarization is clearly distinct from spin-1/2 $\Xi^-$ and $\Lambda$. We emphasize that if a strong magnetic field is present (which is not taken into account in the current simulation), the spin Cooper-Frye formula is modified with the replacement $\varpi_{\rho\sigma}\rightarrow \varpi_{\rho\sigma} +M_H F_{\rho\sigma}/(ST)$ with $M_H$ the magnetic moment of hyperon $H$, $S$ the spin quantum number of $H$, and $F_{\rho\sigma}$ the electromagnetic field~\cite{Becattini:2016gvu}. Since the ratios of the magnetic moment over spin for $\Omega^-$, $\Lambda$, and $\Xi^-$ are roughly the same, the magnetic-field induced polarization would be also similar. Besides, the magnetic field can also induce splitting of the spin polarization between hyperons and anti-hyperons~\cite{Guo:2019joy}. (See e.g. Ref. \cite{Ambrus:2020oiw} for another possible explanation of hyperon and anti-hyperon spin polarization splitting.) Currently, the preliminary experimental data shown in Fig.~\ref{fig:Xi} have large error bars.  Even precise measurements in the future, when compared with our current results, may provide a novel access to the detection of the magnetic fields at freeze-out in heavy-ion collisions.

\begin{figure*}[htbp] 
  \centering
  \subfigure{\label{fig:circle_a}
  \hspace*{-0.5cm}
  \begin{minipage}[b]{0.40\linewidth}
  \centering
  \hspace*{-0.1cm}
  \includegraphics[trim={0cm 0cm 0cm 2cm},clip, width=1.05\columnwidth]{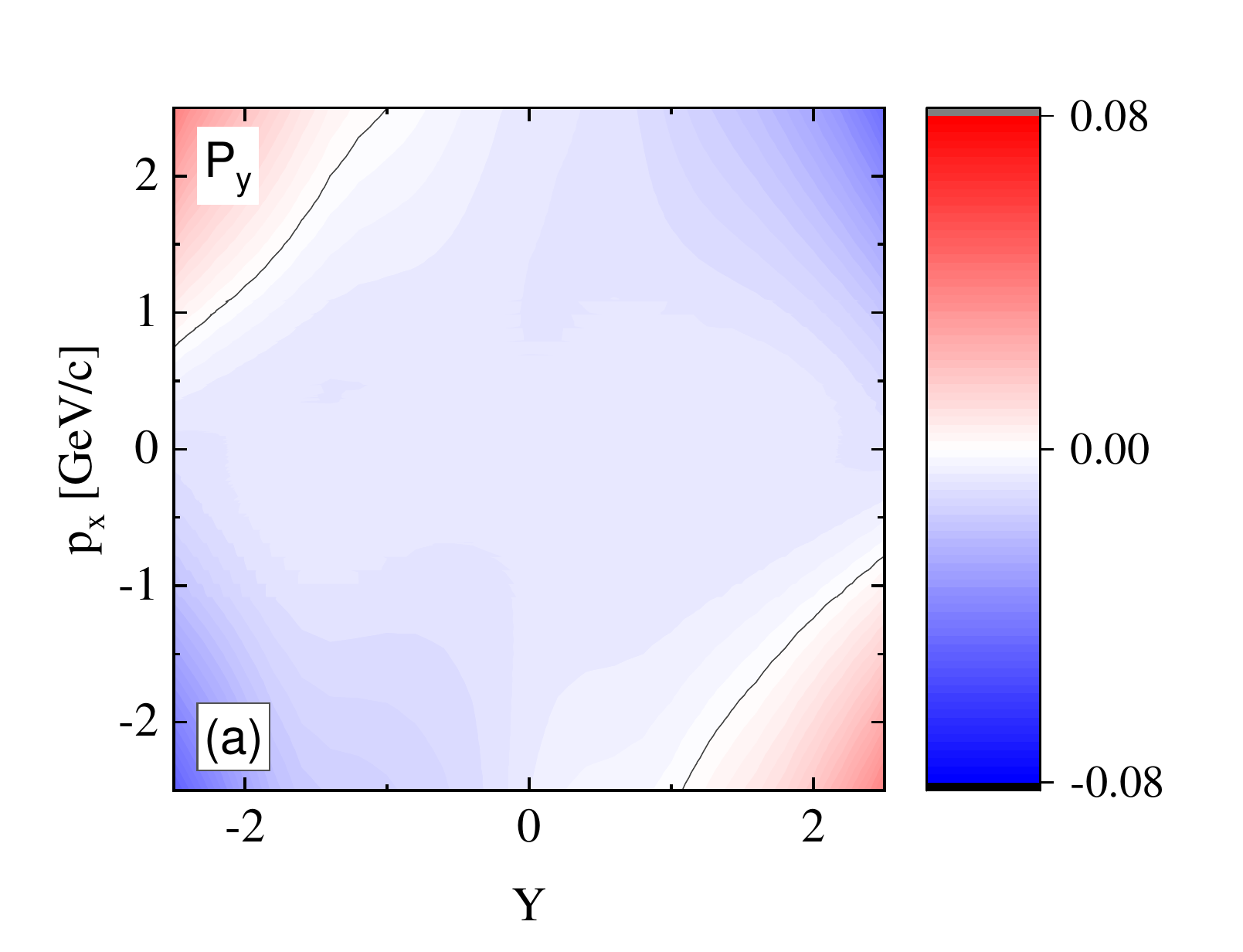}\vspace{10pt}
  \includegraphics[trim={2cm 0cm 2cm 2cm},clip, width=1.0\columnwidth]{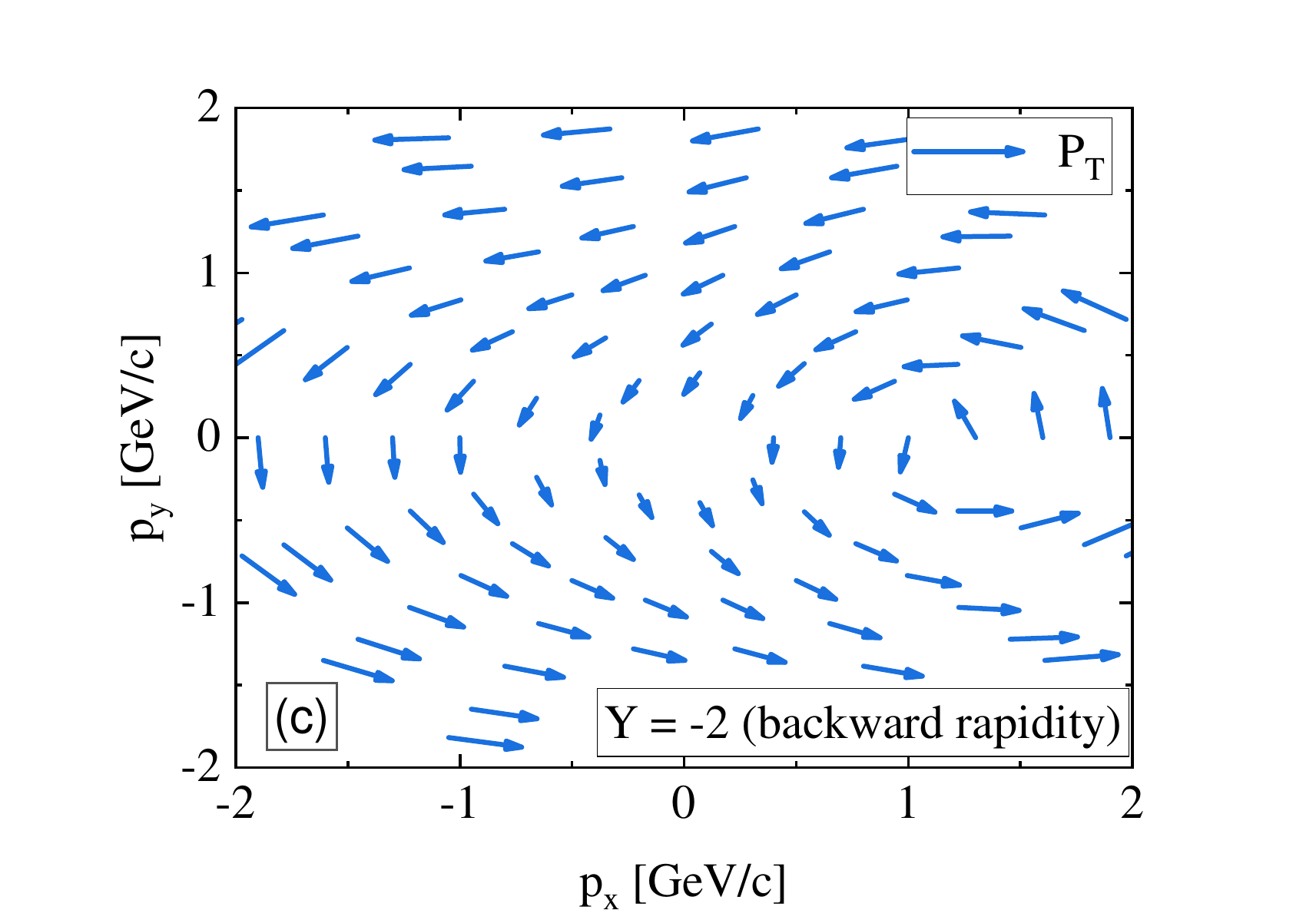}\vspace{4pt}
  \end{minipage}%
  }%
  \subfigure{\label{fig:circle_b}
  \hspace*{0cm}
  \begin{minipage}[b]{0.40\linewidth}
  \centering
  \includegraphics[trim={2cm 0cm 2cm 2cm},clip, width=1.0\columnwidth]{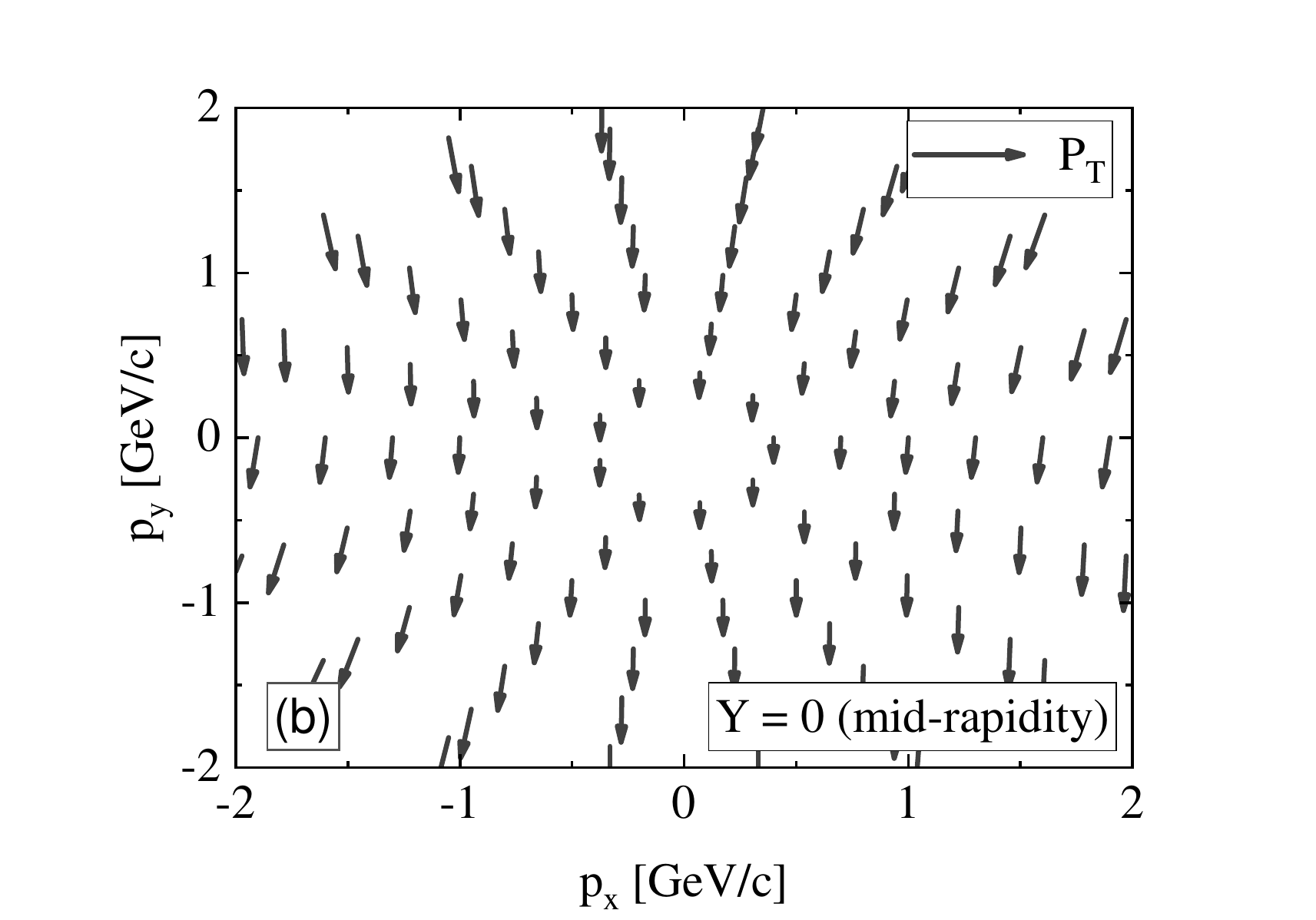}\vspace{10pt}
  \includegraphics[trim={2cm 0cm 2cm 2cm},clip, width=1.0\columnwidth]{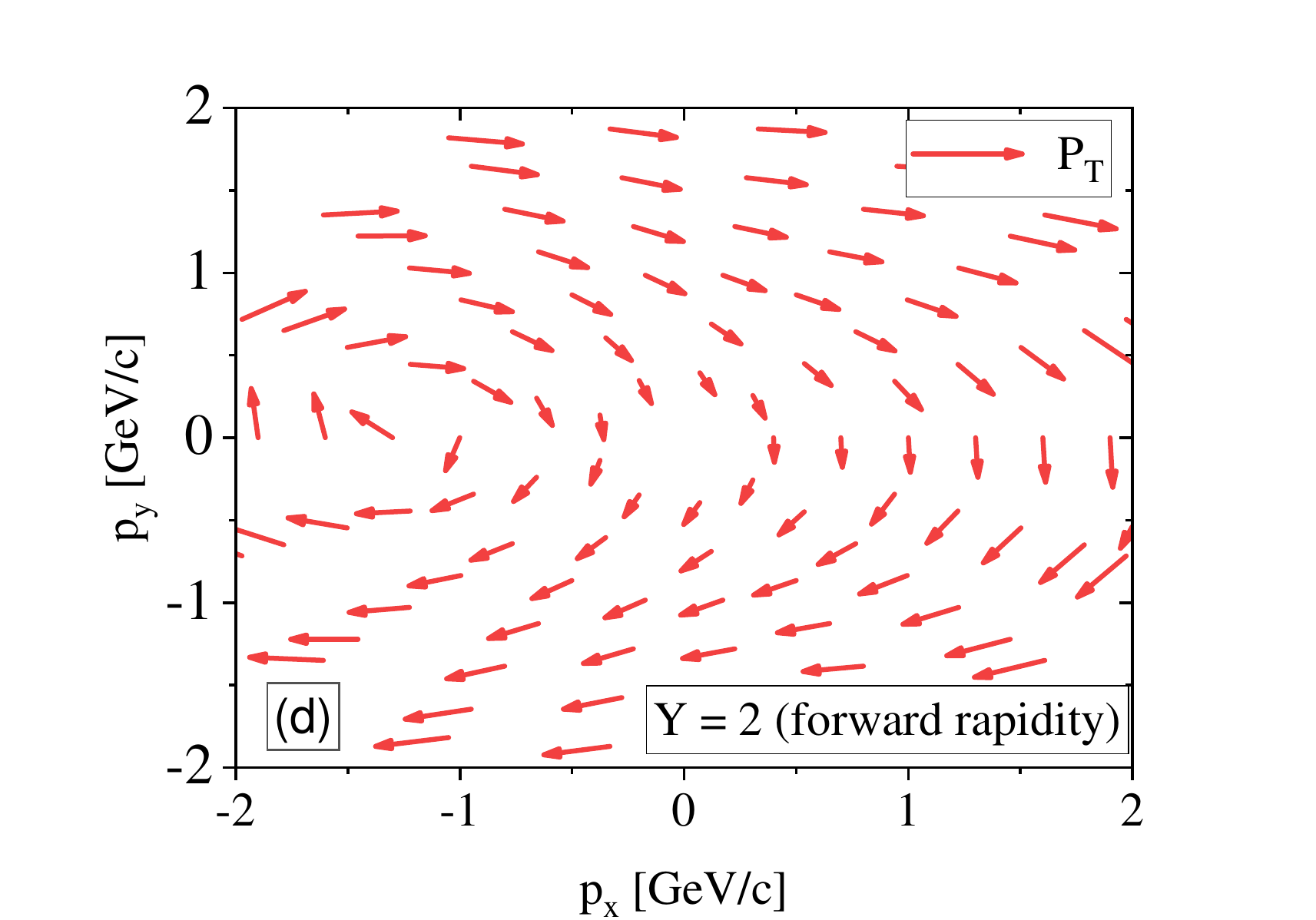}\vspace{4pt}
  \end{minipage}%
  }%
  \subfigure{\label{fig:circle_c}}%
  \subfigure{\label{fig:circle_d}}%
  \caption{
   (a) Distribution of $P_y$ on $p_x$-$Y$ plane in momentum space.
   (b-d) Transverse spin polarization ${\bm P}_\perp$ at mid-rapidity ($Y=0$),  at backward rapidity ($Y=-2$), and at forward rapidity ($Y=2$) in 19.6 A GeV Au + Au collisions at $20$-$50\%$ centrality, calculated from AMPT+MUSIC. }\label{fig:circle}
\end{figure*}


\begin{figure*}[htbp]
    \centering
\subfigure{
\begin{minipage}[b]{0.4\linewidth}
\centering
\center{\hspace*{-0.2cm}\includegraphics[trim={1cm 0cm 2cm 2cm},clip, width=1.05\columnwidth]{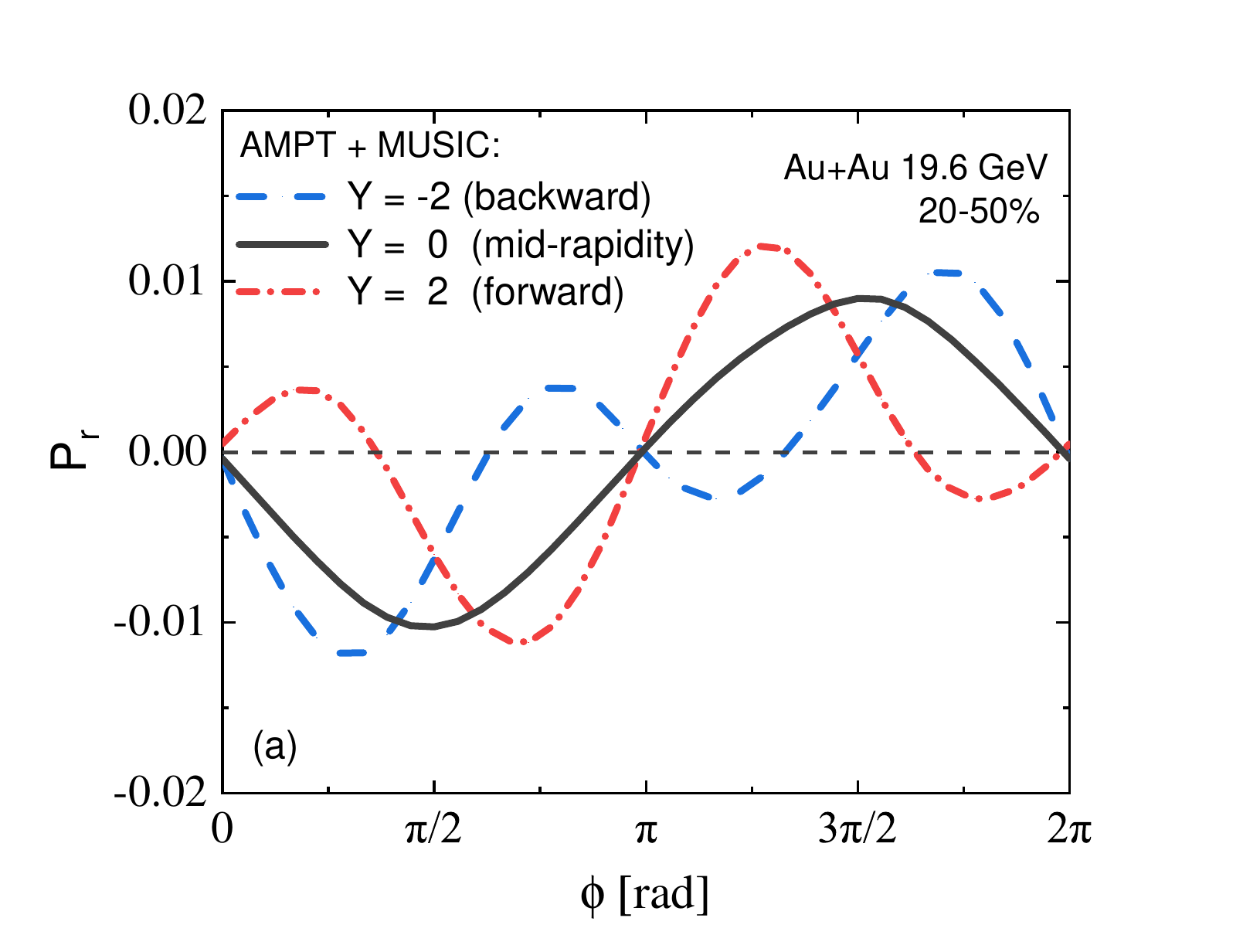}}
\end{minipage}%
\hspace*{1cm}
}%
\subfigure{
\begin{minipage}[b]{0.4\linewidth}
\center{\hspace*{-1.3cm} \includegraphics[trim={1cm 0cm 2cm 2cm},clip, width=1.05\columnwidth]{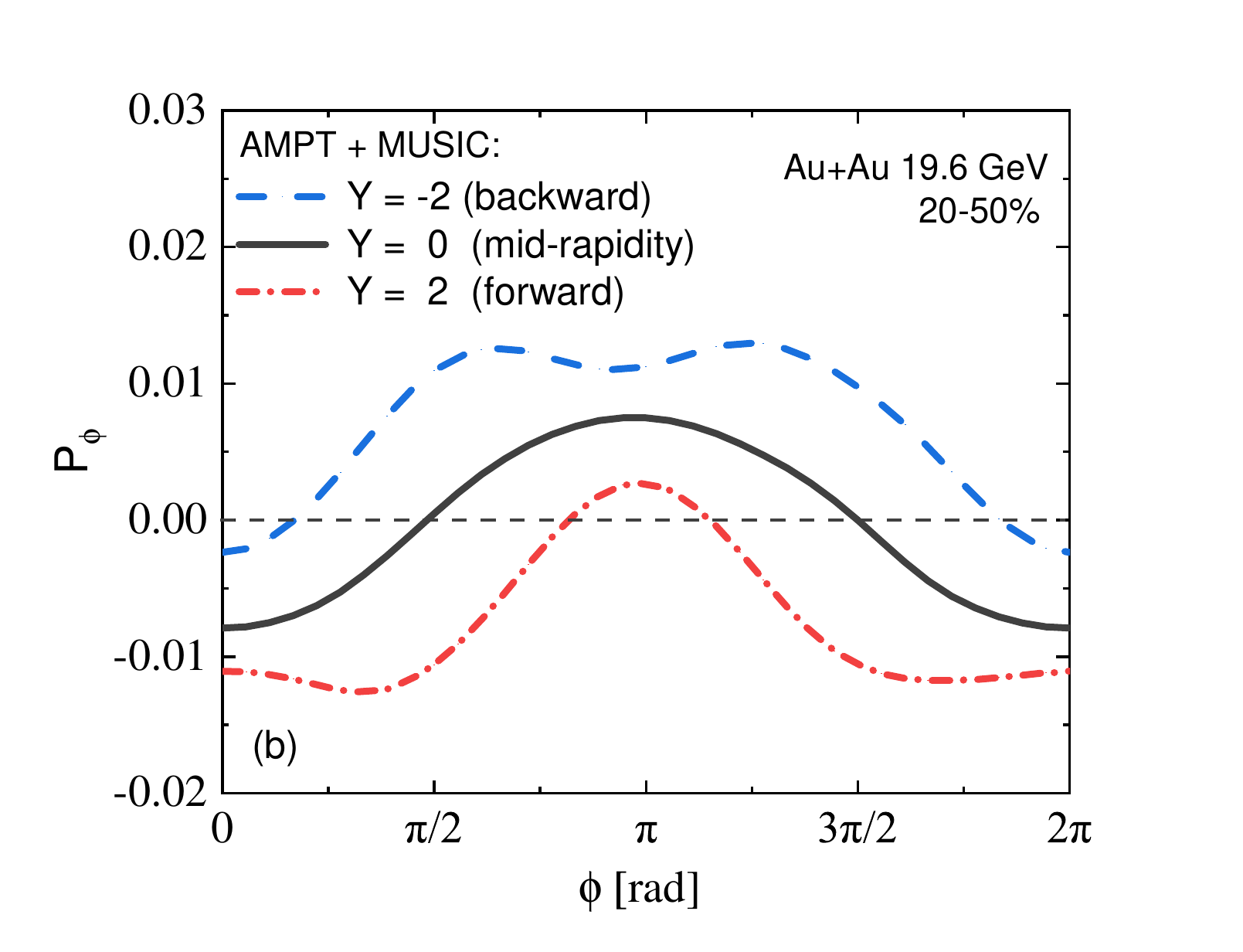}}
\end{minipage}}
    \centering
    \caption{\label{fig:phi} 
               The azimuthal-angle $\phi$ dependence of the (a) radial ($r$) and (b) azimuthal ($\phi$) components of the transverse spin polarization ${\bm P}_\perp$ in momentum space at forward, backward, and mid rapidity, calculated from AMPT+MUSIC. }
\end{figure*}

\begin{figure*}[htbp]  
    \centering
    \subfigure{\label{fig:azimuthal_a}
    {
        \begin{minipage}[t]{0.45\linewidth}
            \centering
            \hspace*{-0.5cm}
            \includegraphics[trim={1cm 0 0 1cm},clip, width=1.05\columnwidth]{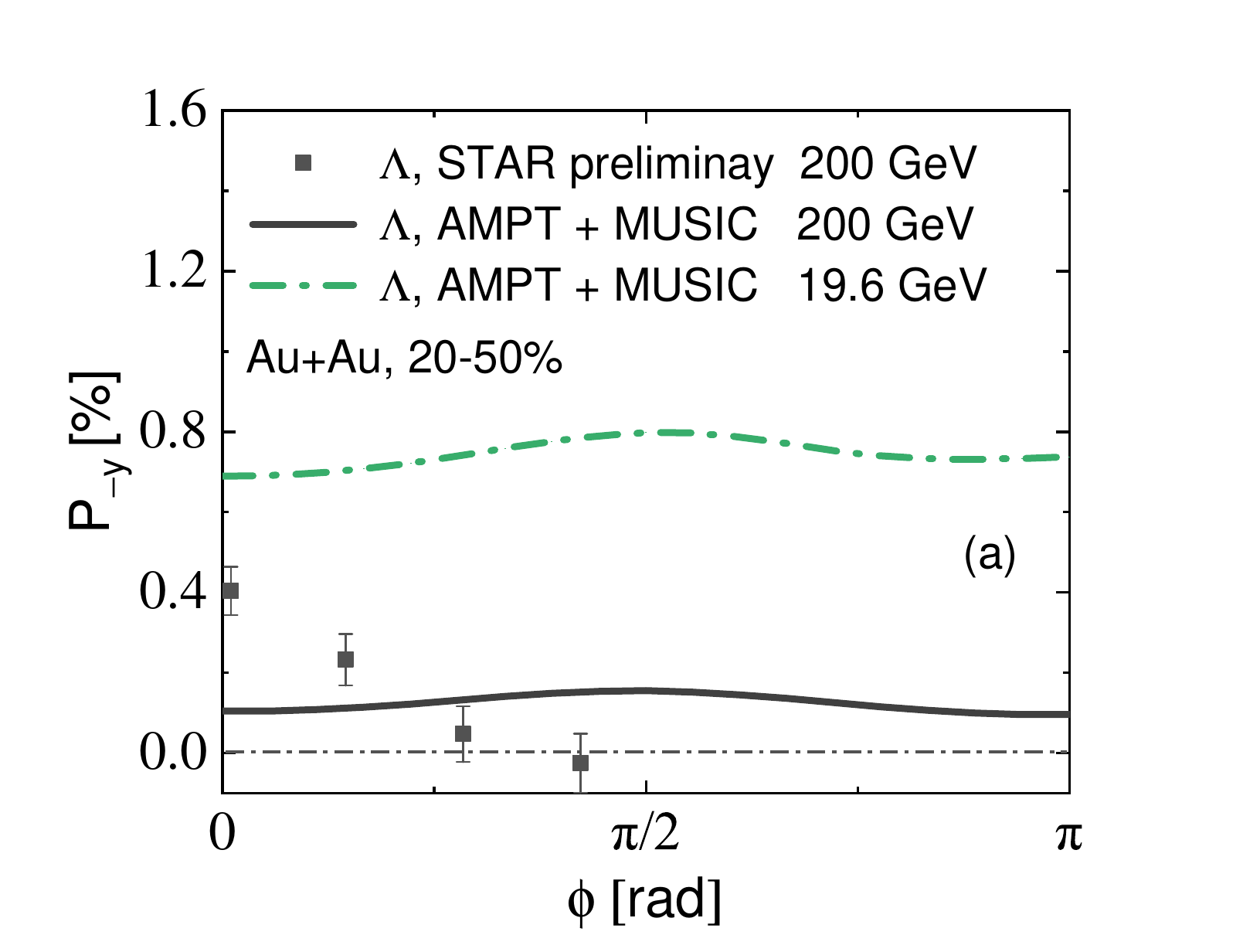}
        \end{minipage}%
    }
    }%
    \subfigure{\label{fig:azimuthal_b}
    {
        \begin{minipage}[t]{0.45\linewidth}
            \centering
            \hspace*{-1cm}
            \includegraphics[trim={0.5cm 0 0 0cm},clip, width=1.05\columnwidth]{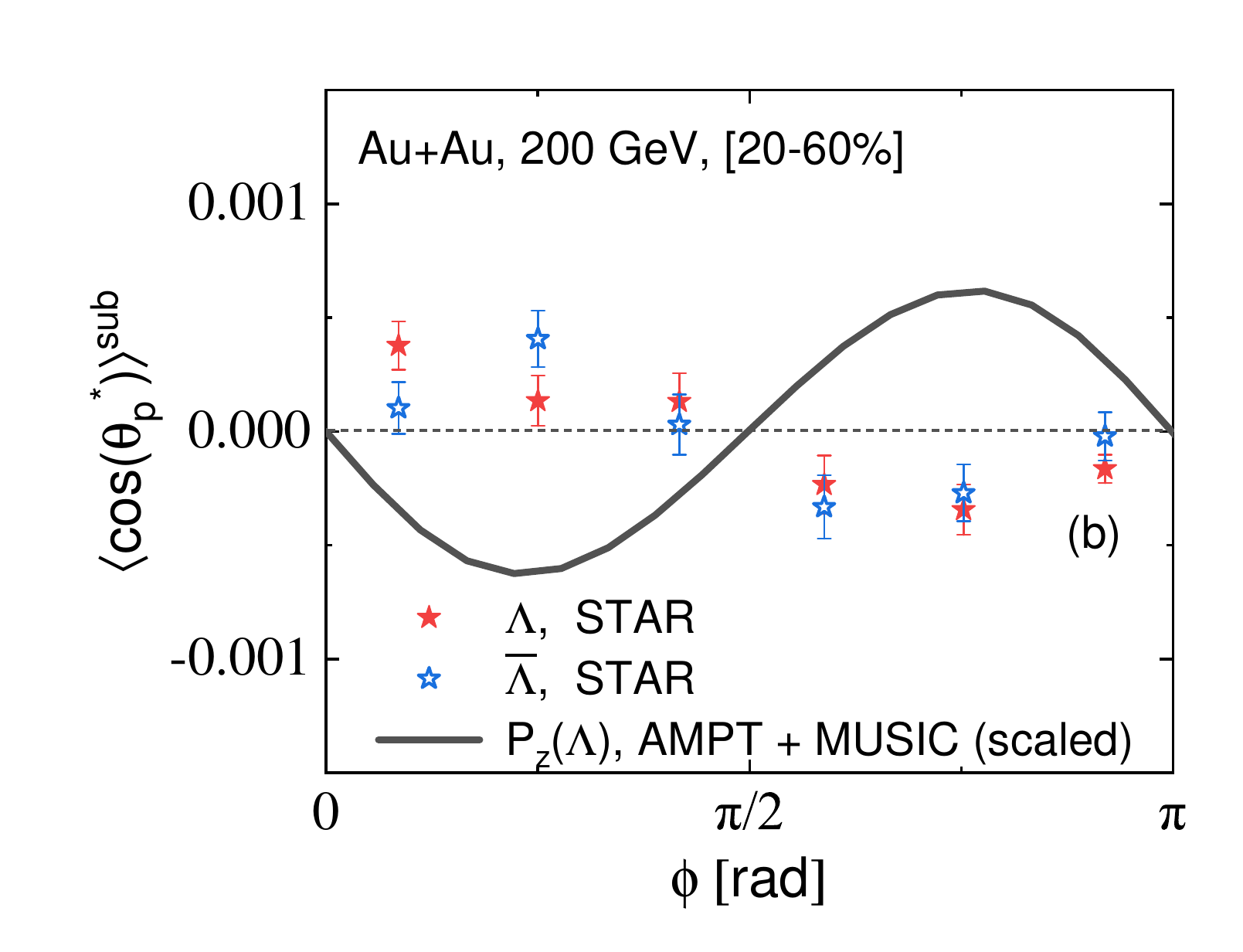}
        \end{minipage}%
    }
    }%
    \centering
    \caption{\label{fig:azimuthal}
               Azimuthal distribution of (a) $P_{-y}$  and (b) $\langle \cos(\theta_p^*) \rangle^{\rm sub}$  in 200 A GeV and 19.6 A GeV Au + Au collisions at $20$-$50\%$ centrality. In sub-figure (b), the calculated $P_z$ is scaled by a factor 0.33 to make the comparison more explicit. The STAR data are taken from~\cite{Niida:2018hfw, STAR:2019srw}.}
\end{figure*}

\subsection{\label{sec:Results}Local spin polarization}

After studying the global spin polarization,  we now focus on the local spin polarization, namely, the differential properties of the spin polarization. Figure \ref{fig:pt_and_y} shows the transverse momentum and pseudo-rapidity dependent $\Lambda$ polarization. AMPT+MUSIC calculations show no significant dependence on $p_T$ or $\eta$,
which qualitatively describe the experimental data in  200 A GeV and 27 A GeV Au+Au collisions~\cite{STAR:2018ivw,Adams:qm}.

Besides the mid-rapidity results, forward and backward spin polarization provide more nuanced information for the local vortical structure. Fig.~\ref{fig:circle_a} shows the local $\Lambda$ polarization in $y$ direction, $P_y$, distributed in the $p_x$-$Y$ plane. At finite rapidity region, $P_y$ shows opposite signs on two sides of the longitudinal axis and form a quadrupole structure in the $p_x$-$Y$ plane. As Refs. \cite{Ivanov:2018eej, Xia:2018tes, Wei:2018zfb} pointed out,
this quadrupole pattern is a projection of a ring structure of the transverse spin polarization ${\bm P}_\perp=(P_x, P_y)$ onto $p_x$-$Y$ plane at forward or backward region. Fig. \ref{fig:circle_c} and Fig. \ref{fig:circle_d} show such a ring structure of ${\bm P}_\perp$ at forward and backward rapidity $Y = \pm2$, where the direction of the ring depends on the sign of rapidity. In the mid-rapidity region, the global spin polarization dominates and the ring structure degenerates to direct at $-y$ direction evenly, as shown in Fig. \ref{fig:circle_b}. Physically, the ring structure of ${\bm P}_\perp$ at finite rapidity is understood by the ring structure of thermal vorticity at finite spacetime rapidity due to the anisotropic and inhomogenous expansion of the system. For  more discussions, please refer to Refs.~\cite{Xia:2018tes, Wei:2018zfb}.

Figure \ref{fig:circle} also demonstrates that the transverse spin polarization has striking angular distribution at different rapidity. To visualize such angular distribution more clearly, we decompose ${\bm P}_\perp$ into radial ($r$) and azimuthal ($\phi$) components and show their azimuthal-angle dependence in Fig. \ref{fig:phi}. The special modulation feature in $\phi$ is clearly seen, which could provide an specific way to detect the vortex-ring structure by measuring such $\phi$-modulation behavior of radial, $P_r$, and azimuthal, $P_\phi$, components of ${\bm P}_\perp$ at finite rapidity.

\begin{figure}[htb] 
\center{\includegraphics[trim={1.5cm 0cm 1cm 0cm},clip, width=0.95\columnwidth]  {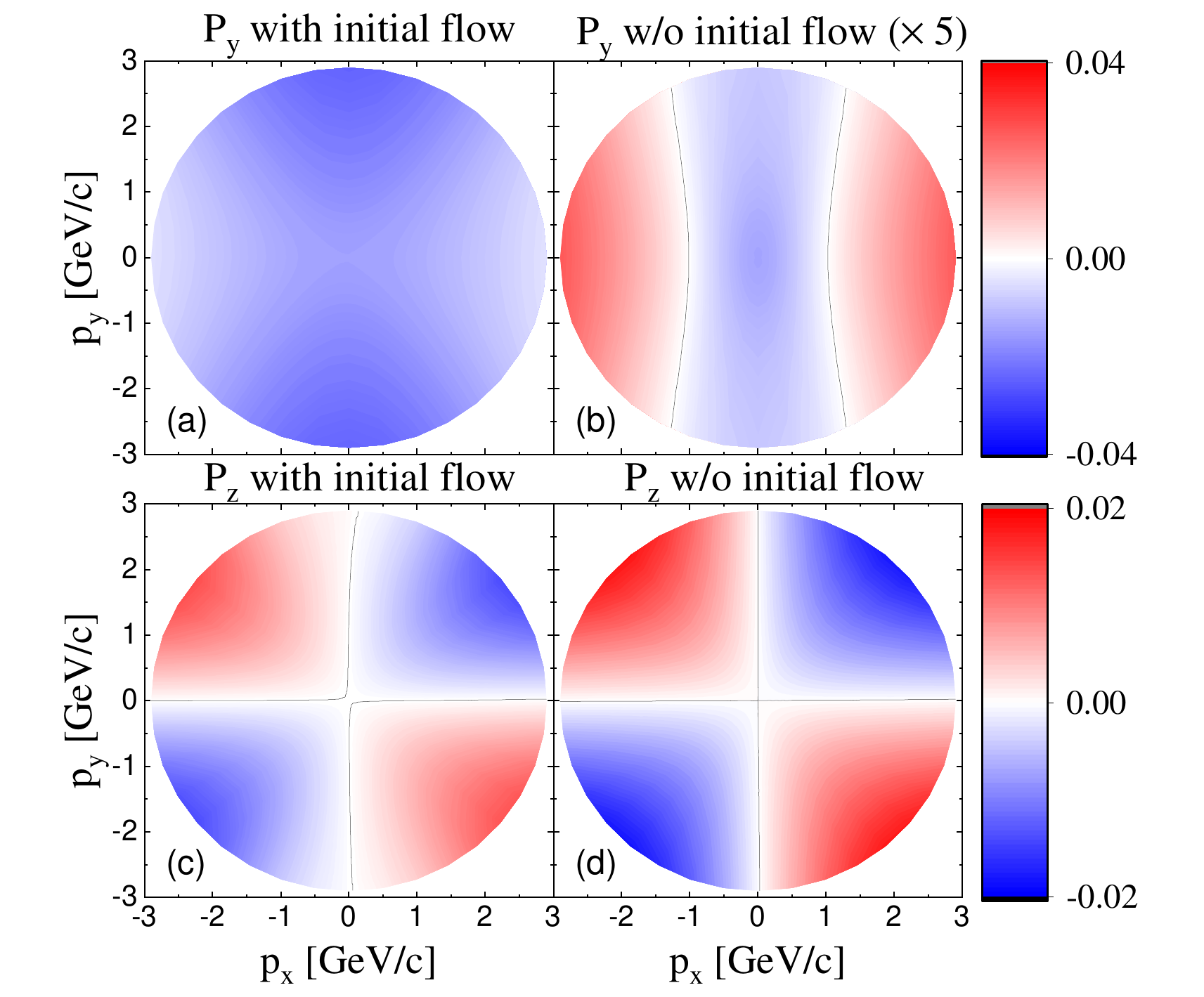}}
\subfigure{\label{fig:pz_initial_flow_a}}%
\subfigure{\label{fig:pz_initial_flow_b}}%
\caption{\label{fig:pz_initial_flow}
The distribution of $P_y$ and $P_z$ at mid-rapidity in transverse $p_x$ -$p_y$ plane in 19.6 A GeV Au + Au collisions, calculated from AMPT+MUSIC with and without initial flow.
 }
\end{figure}

Fig. \ref{fig:azimuthal_a} shows the azimuthal dependence of $\Lambda$ polarization $P_{-y}$ measured in experiment and calculated by model. Our calculated $P_{-y}$ slightly increases with the azimuthal angle $\phi$ from $0$ to $\pi/2$,  while, the measured $P_{-y}$ shows an opposite trend and decreases with $\phi$, which is strong at the in-plane direction but almost vanished at the out-of-plane direction. Fig. \ref{fig:azimuthal_b} shows $P_z$ calculated from our model and compared with the $\langle \cos(\theta_p^*) \rangle^{\rm sub}$  measured in experiment. Note that $\langle \cos(\theta_p^*) \rangle^{\rm sub}$ is related to the longitudinal polarization by $ P_z = \frac{\langle \cos \theta_p^* \rangle}{\alpha_H \langle \cos^2 \theta_p^* \rangle}$, where $\alpha_H$ is the decay factor and $\langle\cdots \rangle^{\rm sub}$ denotes the subtraction of the acceptance effect in experiment. As demonstrated by the lower panels of Fig. \ref{fig:pz_initial_flow}, the distribution of longitudinal spin polarization $P_z$ in the $p_x$-$p_y$  transverse momentum plane from our model calculations shows an obvious quadrupole structure. While Fig. \ref{fig:azimuthal_b} illustrates that although both model and data present a quadrupole pattern, the sign is opposite. Such situations are similar to many hydrodynamics or transport model calculations based on the spin Cooper-Frye formula, which could successfully describe the magnitude and energy dependence of the global $\Lambda$ polarization \cite{Becattini:2013vja,Becattini:2015ska,Becattini:2016gvu,Karpenko:2016jyx,Xie:2016fjj,Xie:2017upb,Li:2017slc,Ivanov:2019ern,Shi:2017wpk,Wei:2018zfb}, but fail to reproduce the local spin polarization, namely, the azimuthal dependence of $P_{-y}$ and $P_z$. This is the challenging spin ``sign problem'' which has attracted a lot of attention but has not been solved till now (For recent development, please refer to Refs.~\cite{Becattini:2017gcx, Florkowski:2019voj,Wu:2019eyi, Xie:2019jun, Xia:2019fjf,Becattini:2019ntv,Liu:2019krs}.

\begin{figure*}[ht] 
\center{\includegraphics[trim={10cm 0 0cm 0},clip,width=2.05\columnwidth]  {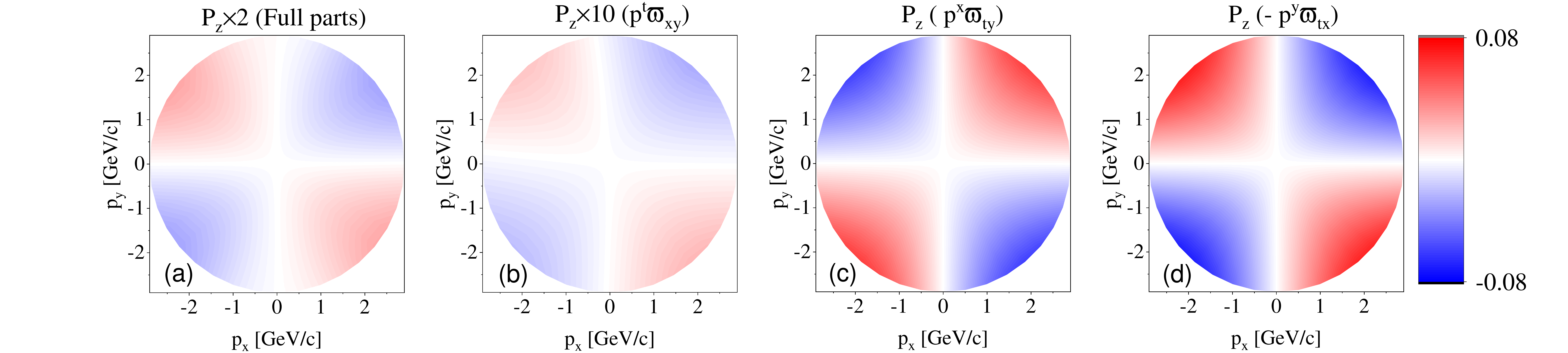}}
\caption{\label{fig:pz_decomposation} Contributions of $P_z$ from AMPT+MUSIC calculations: (a) total $P_z$, (b) the non-relativistic contribution, (c) and (d) Thomas-precession contributions in Eq.~(\ref{thomas}).
}
\end{figure*}

\begin{figure*}[htbp]
\centering
\subfigure{
\begin{minipage}[b]{0.4\linewidth}
\centering
\center{\hspace*{-1.5cm}\includegraphics[trim={0cm 0cm 2cm 2cm},clip, width=1.05\columnwidth]{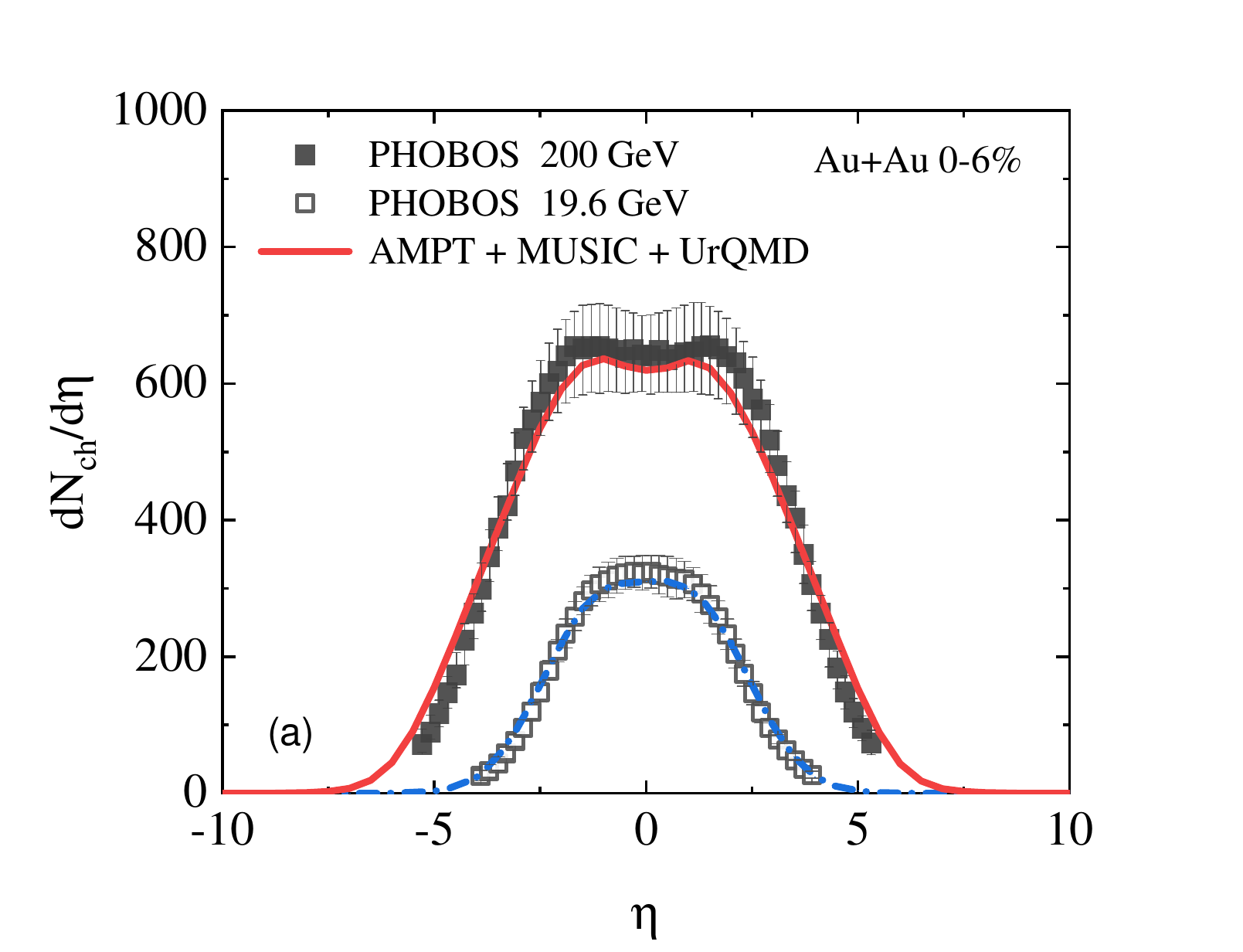}}
\end{minipage}%
}%
\subfigure{
\begin{minipage}[b]{0.4\linewidth}
\center{\hspace*{-1cm}\includegraphics[trim={0cm 0cm 2cm 2cm},clip, width=1.05\columnwidth]{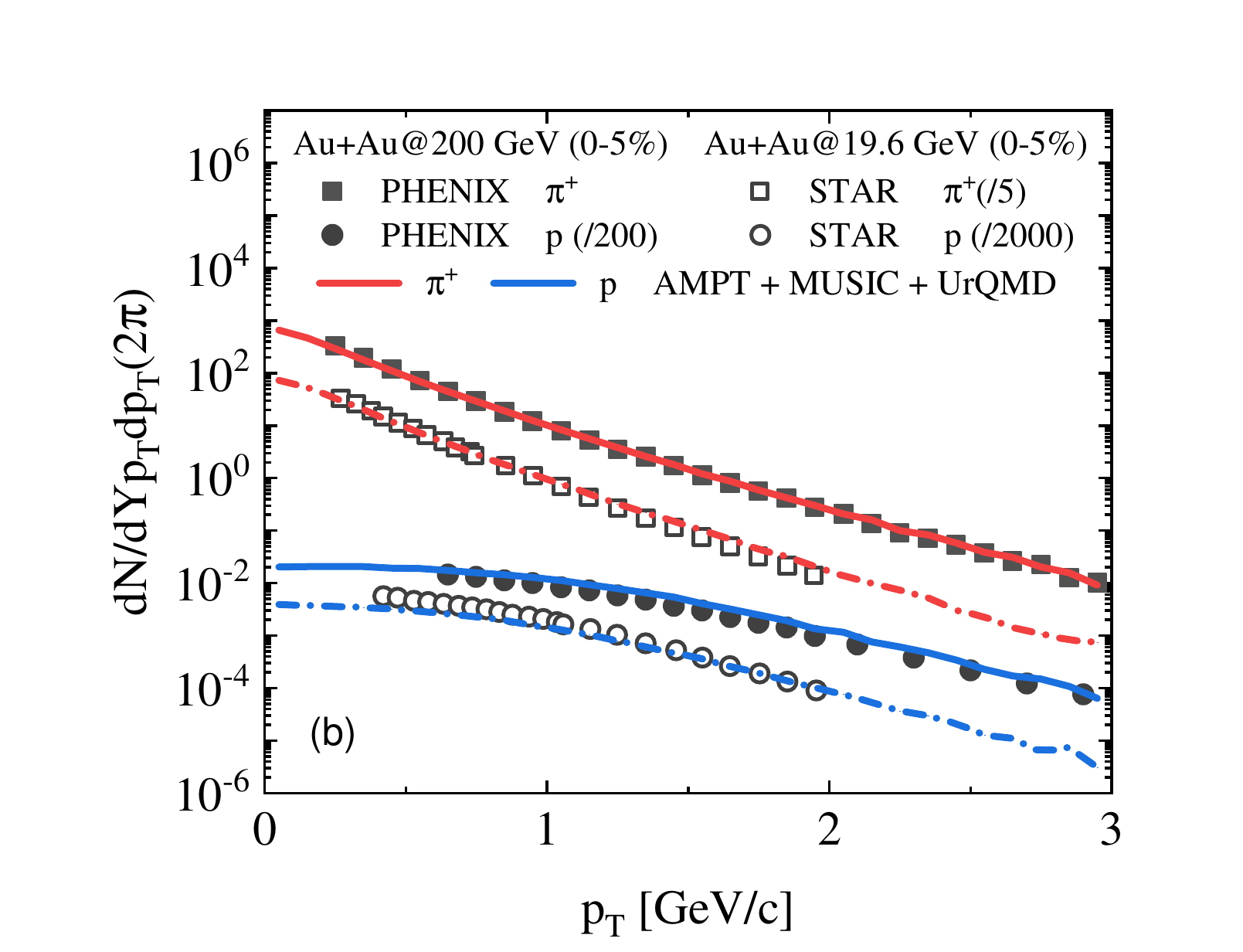}}
\end{minipage}}
  \caption{ (a) Pseudo-rapidity distribution of all charged hadron and (b) transverse momentum spectra of pions and protons in 200 and 19.6 A GeV Au+Au collisions. The experimental data are from the PHOBOS, PHENIX, and STAR Collaborations \cite{PHOBOS:2002wb, STAR_BES:2017iwn, PHENIX:2003cb}. }\label{fig:bulk}
\end{figure*}
\begin{figure*}[htbp]
\centering
\subfigure{
\begin{minipage}[b]{0.49\linewidth}
\centering
\center{\hspace*{0.5cm}\includegraphics[trim={0cm 0cm 2cm 2cm},clip, width=0.9\columnwidth]{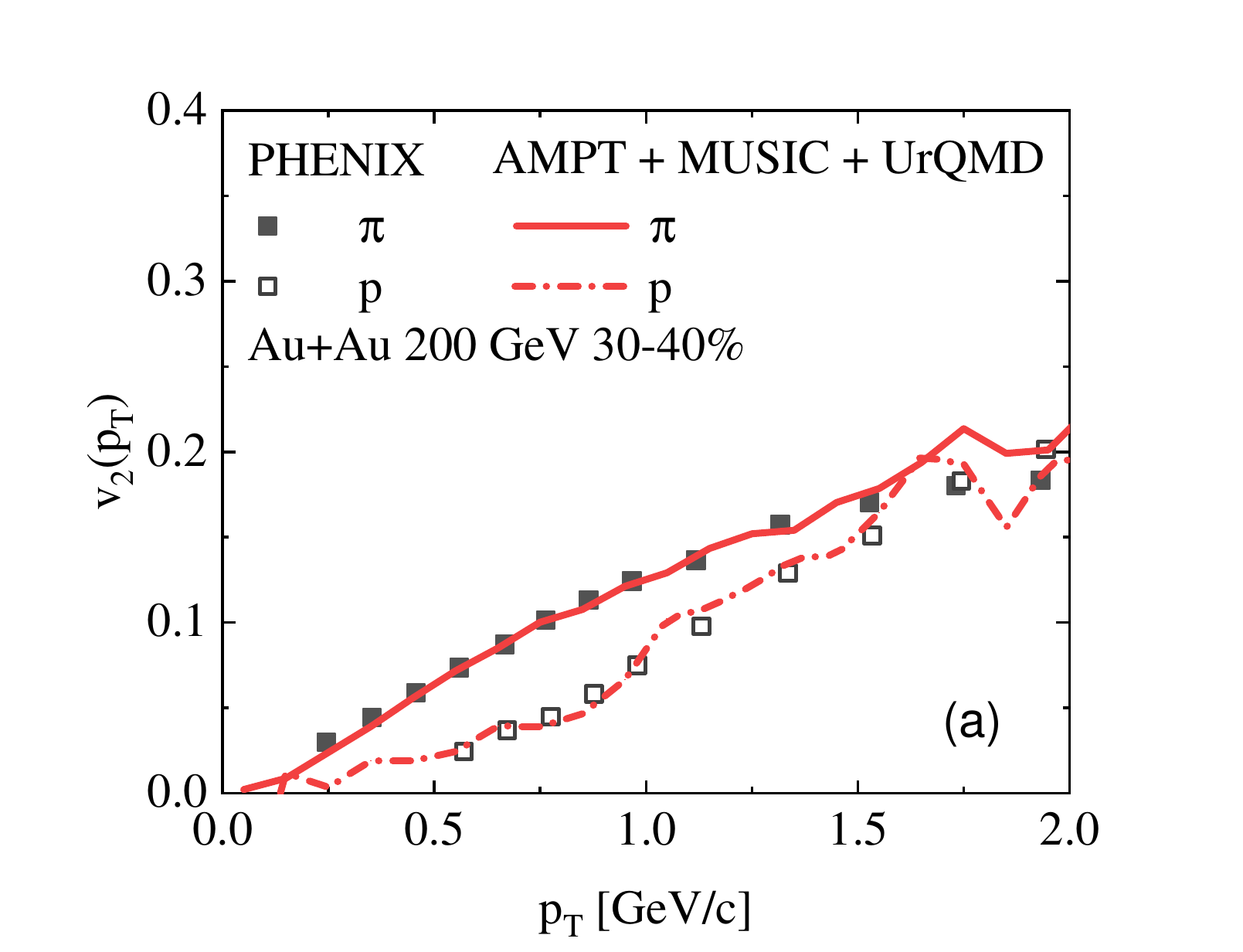}}
\end{minipage}%
}%
\subfigure{
\begin{minipage}[b]{0.49\linewidth}
\center{\hspace*{-2.5cm}\includegraphics[trim={0cm 0cm 2cm 2cm},clip, width=0.9\columnwidth]{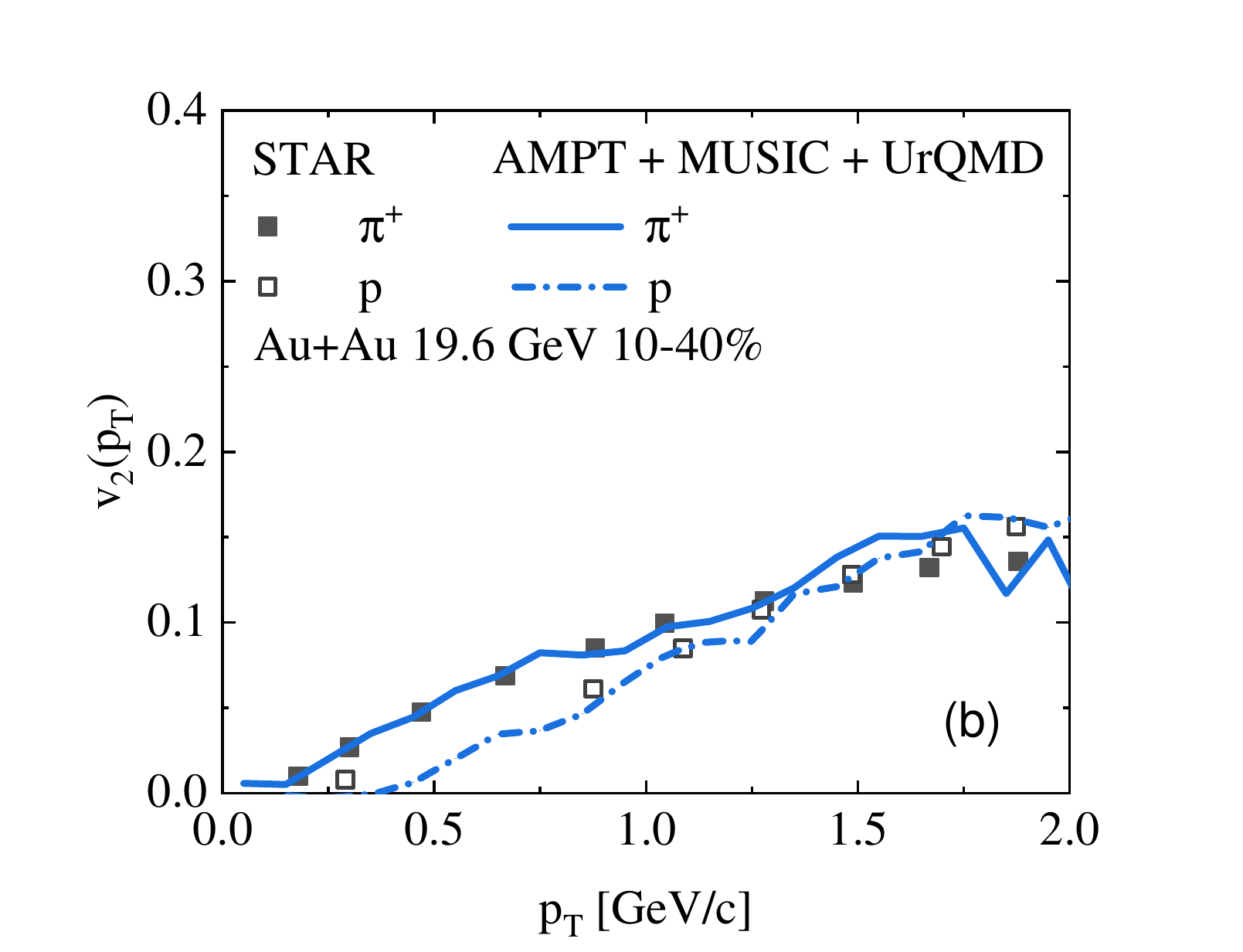}}
\end{minipage}}
  \caption{Differential elliptic flow $v_2(p_T)$ of pions and protons in 200 A GeV and 19.6 A GeV Au+Au collisions. The experimental data are from PHENIX and STAR Collaborations \cite{PHENIX:2014bga, STAR:2015fum}. }\label{fig:v2}
\end{figure*}

\begin{figure*}[htbp] 
\centering
\subfigure{
\label{fig:global_compare_a}
\begin{minipage}[b]{0.49\linewidth}
\centering
\center{\hspace*{1cm}\includegraphics[trim={2cm 0.5cm 2cm 2cm},clip, width=0.85\columnwidth]  {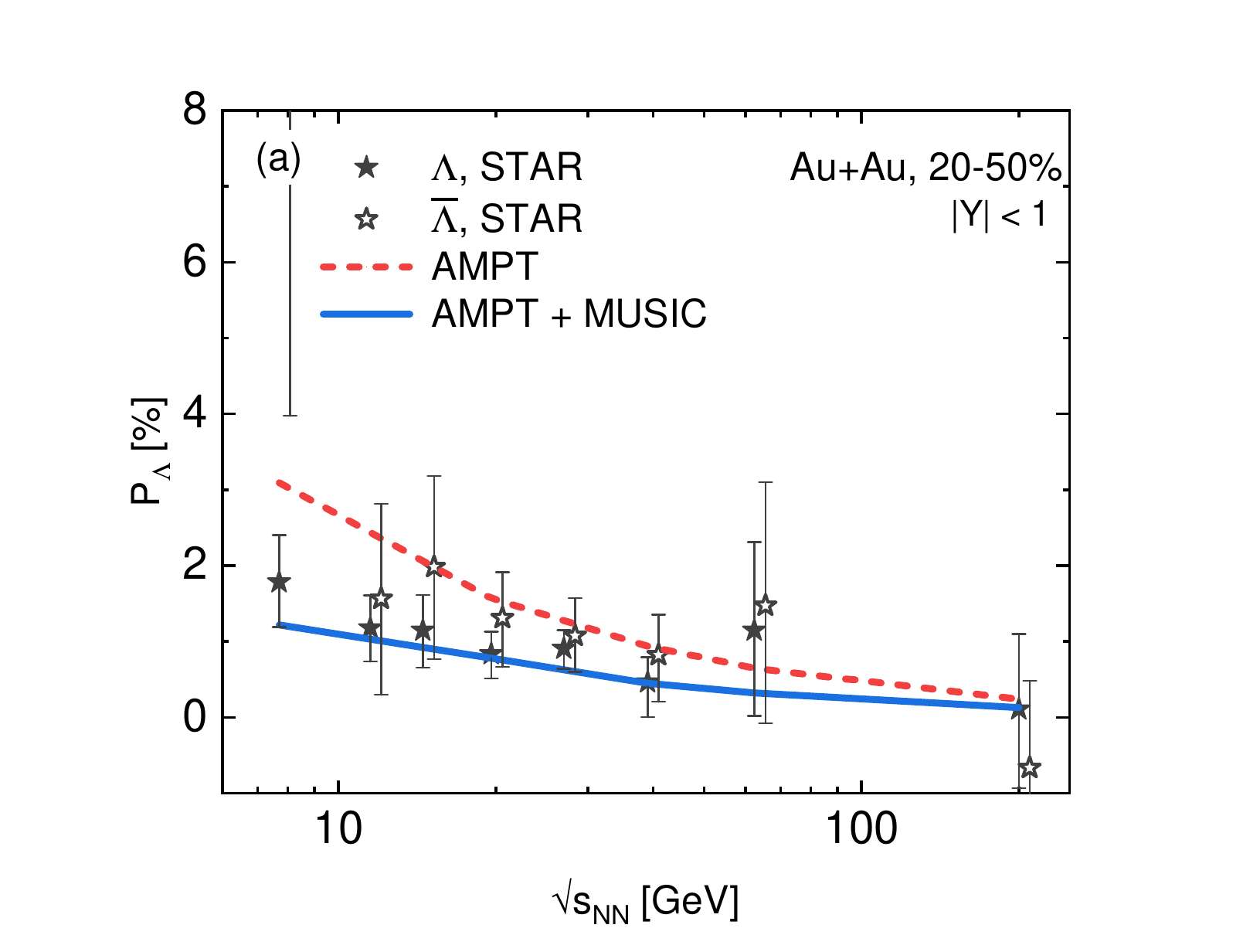} }
\end{minipage}%
}%
\subfigure{
\label{fig:global_compare_b}
\begin{minipage}[b]{0.49\linewidth}
\center{\hspace*{-2cm}\includegraphics[trim={2cm 0.5cm 2cm 2cm},clip, width=0.85\columnwidth]  {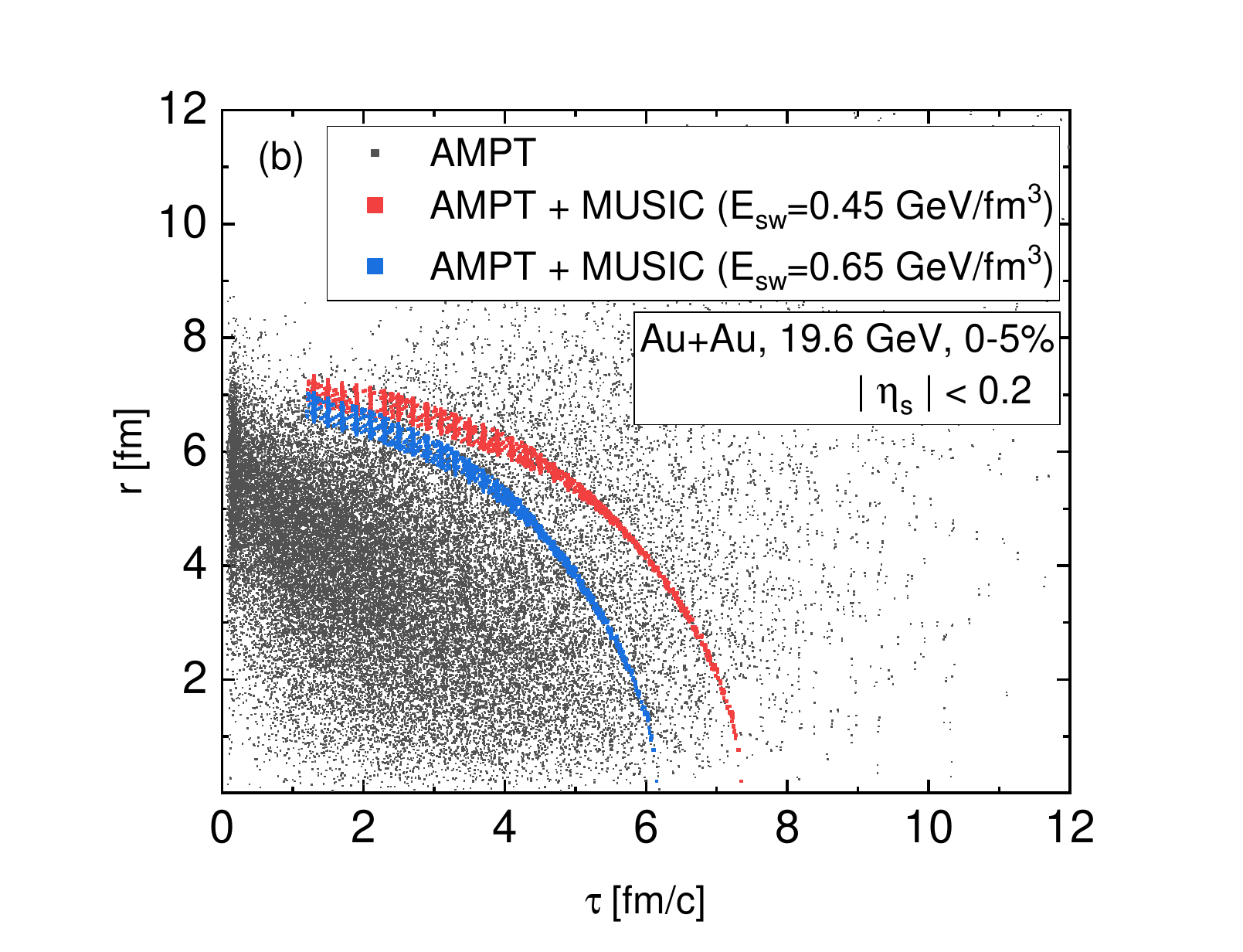} }
\end{minipage}}
  \caption{  (a) Energy dependent global $\Lambda$ polarization, calculated from AMPT \cite{Wei:2018zfb},
and from AMPT + MUSIC. (b) Hadronization distribution from AMPT and the freeze-out hypersurface from AMPT+MUSIC hydrodynamic model.}\label{fig:global_compare}
\end{figure*}

The spin Cooper-Frye formula (\ref{eq:cooper}) used in this paper and early calculations can be regarded as a mapping between thermal vorticity in coordinate space to spin polarization in momentum space, where the thermal vorticity is mainly contributed from the initial condition and the hydrodynamic evolution. Since the initial condition imprints the total angular momentum of the system, it also largely influences the global spin polarization in the final state  due to the angular momentum conservation. In contrast, both the initial condition and hydrodynamic evolution could largely affect the local spin polarization, which reflect the vorticity structure at late stage evolution.

In Fig.~\ref{fig:pz_initial_flow}, we plot the transverse distribution of spin polarization $P_y$ and $P_z$ in 20-50\% Au + Au collision at mid-rapidity, calculated from AMPT+MUSIC with and without initial flow. The full AMPT initial condition with initial flow has been described in Sec. \ref{subsec:AMPT}. For the AMPT initial condition without initial flow, we directly set  $u^x(\tau_0)=u^y(\tau_0)=u^{\eta_s}(\tau_0)=0$ and obtain the initial energy density as described in~\cite{Xu:2016hmp}.  The initial flow directly influences the initial angular momentum of the created QGP fireball.  When it is turned off, the vorticity of the initial fluid reduces to almost zero value. Correspondingly, the final global polarization $P_y$ almost vanishes, as demonstrated in Fig. \ref{fig:pz_initial_flow_b} (Note that we re-scaled the result of $P_y$ without initial flow by a factor of 5 to make it visible.).

It is generally believed that the longitudinal polarization $P_z$ is directly associated with the anisotropic transverse expansion of the systems but insensitive to the initial angular momentum~\cite{Becattini:2017gcx}. This is confirmed by our AMPT+MUSIC calculations with and without
initial flow, which demonstrate that $P_z$ has similar structure in these two comparison runs, as shown by the lower panels of Fig. \ref{fig:pz_initial_flow}. This also means that
the longitudinal vorticity $P_z$ mainly probes the vortical structure developed during the hydrodynamic evolution.

Note that, according to Eqs. (\ref{eq:defP})-(\ref{eq:p_integration}), the longitudinal component $P_z$ is contributed by 3 parts :
\begin{eqnarray}
\label{thomas}
P^z (p) \sim p^t \varpi_{xy} + p^x \varpi_{ty} - p^y \varpi_{tx}.
\end{eqnarray}
The first term is related to the non-relativistic vorticity $\omega_z\sim (\bm\nabla\times\bm v)_z$, which arises when the system expands anisotropically. The second and third terms are relativistic effect and can be considered as Thomas precession $\omega^z_{\rm Tho}\sim (\bm p\times\bm a)^z/m$ (taking into account the fact that for nearly ideal fluid $\bm a\approx-T^{-1} \bm\nabla T$) with $\bm a$ the acceleration of the fluid. Figure \ref{fig:pz_decomposation} shows that the contributions from the last two terms are much bigger than the first term, which indicates that the spin ``sign problem" is possibly a relativistic effect. Such analysis works for  all the hydrodynamic or transport calculations of $P_z$ based on the spin Cooper-Frye formula with thermal vorticity as the spin chemical potential. Similar analysis were also discussed in Ref. \cite{Karpenko:2018erl}. The discrepancy between the theoretical calculations based on the spin Cooper-Frye formula and the experimental data is still not fully understood. Some attempts to resolve such problem can be found in Refs. \cite{Florkowski:2019voj,Wu:2019eyi, Xie:2019jun,Xia:2019fjf,Becattini:2019ntv,Liu:2019krs}.

\section{\label{sec:Summary}Summary}

In this paper, we used MUSIC viscous hydrodynamics with the AMPT pre-equilibrium dynamics to study the hyperon spin polarization in relativistic heavy-ion collisions. With a UrQMD hadron cascade afterburner, this hybrid model can nicely describe various soft hadron observables at RHIC-BES energies, such as the rapidity distribution of all charged hadrons, transverse momentum spectra, and differential elliptic flows of identified hadrons; see Appendix \ref{app:soft} for the details. In order to study the spin polarization of hyperons, we implemented the spin Cooper-Frye formula that associates the momentum-space distribution of the hyperon polarization with the position-space vorticity of the fluid, and obtained a nice description of the collision-energy  dependence, centrality dependence, and the $p_T$ and $\eta$ dependence of the $\Lambda$ polarization. We also studied and predicted the global spin polarization of $\Xi^-$ and $\Omega^-$ as a function of collision energy, which provides a baseline for the studies of the magnetic moment, spin, and mass dependence of the spin polarization.

For the local spin polarization, we calculated the radial and azimuthal components of the transverse $\Lambda$ polarization and found specific modulating behavior that could reflect the circular vortical structure.
However, for the azimuthal dependence of the transverse and the longitudinal spin polarization, our approach, like most of the previous theoretical calculations, gives an opposite trend compared with the experimental data. These results suggest that the spin Cooper-Frye formula, which was derived under the assumption of thermal equilibrium of spin degree of freedom, needs to be improved. A promising direction for such a purpose is the spin hydrodynamics which has been intensively discussed recently~\cite{Florkowski:2017ruc,Hattori:2019lfp,Montenegro:2018bcf,
Florkowski:2018fap,Bhadury:2020puc,Bhadury:2020cop,Shi:2020htn,Fukushima:2020ucl}. With the future numerical implementation, such framework might give important insight to the puzzle of local spin polarization and may also be used to study the global and local spin alignment of vector mesons~\cite{Liang:2004xn,Sheng:2019kmk,Sheng:2020ghv,Xia:2020tyd}.

\appendix
\section{\label{app:soft}Soft hadronic observables}

In this appendix, we check the soft hadronic observables from AMPT + MUSIC simulations
which serve as baseline calculations before study the spin-polarization in relativistic heavy ion collisions.
Figures~\ref{fig:bulk} and \ref{fig:v2} present the rapidity distribution of all charged hadrons, transverse momentum spectra and differential elliptic flow of pions and protons in 200 A GeV and 19.6 A GeV Au+Au collisions. With a UrQMD hadronic afterburner and properly tuned parameters listed in Table \ref{tab:parameter}, our AMPT + MUSIC simulations nicely fit these soft observables measured in experiment. These agreements indicate that our hybrid model give a nice description for the bulk evolution of the QGP fireball.

\section{Global spin polarization --- a comparison between AMPT and AMPT+MUSIC}
In this appendix, we compare the global spin polarization calculated from our {AMPT+MUSIC} hybrid model and from the pure transport approach { AMPT}. As shown in Fig.~\ref{fig:global_compare_a}, the global polarization from AMPT is larger than that from MUSIC, although the pre-equilibrium dynamics at the early stage are both provided by AMPT.

To understand the difference between these two results, in Fig.~\ref{fig:global_compare_b}, we compare the hadronization distribution in AMPT and
the freeze-out hypersurface from AMPT+MUSIC hydrodynamic model. For hydrodynamic simulations, the particlization happens on the freeze-out hypersurface with constant energy density. For AMPT simulation, the hadronization process is realized through parton recombination with the Monte-Carlo sampling~\cite{Lin:2004en}. As a result, the hadronization in AMPT happens all the time during the system evolution. In contrast, the hydrodynamic particlization happens at a specific freeze-out hypersurface, which, on average, happens later than the AMPT  hadronization. Noting that the thermal vorticity gradually decreases with the evolution, AMPT thus gives a larger global polarization than the hydrodynamic simulations.

\begin{acknowledgments}
We thank T. Nidda, Z.-W. Lin, D.-X. Wei, and X.-L. Xia for helpful discussions. The work of B. F, K. X, and H. S. is supported by are supported by the NSFC under grant Nos.~11675004 and No.~12075007.  The work of X. G. H is supported by NSFC through Grants No.~11535012 and No.~11675041. B. F, K. X, and H. S. also gratefully acknowledge the extensive computing resources provided by the Super-computing Center of Chinese Academy of Science (SCCAS), Tianhe-1A from the National Supercomputing Center in Tianjin, China and the High-performance Computing Platform of Peking University.
\end{acknowledgments}

\bibliography{hydro_spin}

\providecommand{\noopsort}[1]{}\providecommand{\singleletter}[1]{#1}%
\begin{thebibliography}{97}%
\makeatletter
\providecommand \@ifxundefined [1]{%
 \@ifx{#1\undefined}
}%
\providecommand \@ifnum [1]{%
 \ifnum #1\expandafter \@firstoftwo
 \else \expandafter \@secondoftwo
 \fi
}%
\providecommand \@ifx [1]{%
 \ifx #1\expandafter \@firstoftwo
 \else \expandafter \@secondoftwo
 \fi
}%
\providecommand \natexlab [1]{#1}%
\providecommand \enquote  [1]{``#1''}%
\providecommand \bibnamefont  [1]{#1}%
\providecommand \bibfnamefont [1]{#1}%
\providecommand \citenamefont [1]{#1}%
\providecommand \href@noop [0]{\@secondoftwo}%
\providecommand \href [0]{\begingroup \@sanitize@url \@href}%
\providecommand \@href[1]{\@@startlink{#1}\@@href}%
\providecommand \@@href[1]{\endgroup#1\@@endlink}%
\providecommand \@sanitize@url [0]{\catcode `\\12\catcode `\$12\catcode
  `\&12\catcode `\#12\catcode `\^12\catcode `\_12\catcode `\%12\relax}%
\providecommand \@@startlink[1]{}%
\providecommand \@@endlink[0]{}%
\providecommand \url  [0]{\begingroup\@sanitize@url \@url }%
\providecommand \@url [1]{\endgroup\@href {#1}{\urlprefix }}%
\providecommand \urlprefix  [0]{URL }%
\providecommand \Eprint [0]{\href }%
\providecommand \doibase [0]{http://dx.doi.org/}%
\providecommand \selectlanguage [0]{\@gobble}%
\providecommand \bibinfo  [0]{\@secondoftwo}%
\providecommand \bibfield  [0]{\@secondoftwo}%
\providecommand \translation [1]{[#1]}%
\providecommand \BibitemOpen [0]{}%
\providecommand \bibitemStop [0]{}%
\providecommand \bibitemNoStop [0]{.\EOS\space}%
\providecommand \EOS [0]{\spacefactor3000\relax}%
\providecommand \BibitemShut  [1]{\csname bibitem#1\endcsname}%
\let\auto@bib@innerbib\@empty
\bibitem [{\citenamefont {Liang}\ and\ \citenamefont
  {Wang}(2005{\natexlab{a}})}]{Liang:2004ph}%
  \BibitemOpen
  \bibfield  {author} {\bibinfo {author} {\bibfnamefont {Z.-T.}\ \bibnamefont
  {Liang}}\ and\ \bibinfo {author} {\bibfnamefont {X.-N.}\ \bibnamefont
  {Wang}},\ }\href {\doibase 10.1103/PhysRevLett.94.102301} {\bibfield
  {journal} {\bibinfo  {journal} {Phys. Rev. Lett.}\ }\textbf {\bibinfo
  {volume} {94}},\ \bibinfo {pages} {102301} (\bibinfo {year}
  {2005}{\natexlab{a}})},\ \bibinfo {note} {[Erratum: Phys.Rev.Lett. 96, 039901
  (2006)]},\ \Eprint {http://arxiv.org/abs/nucl-th/0410079}
  {arXiv:nucl-th/0410079} \BibitemShut {NoStop}%
\bibitem [{\citenamefont {Voloshin}(2004)}]{Voloshin:2004ha}%
  \BibitemOpen
  \bibfield  {author} {\bibinfo {author} {\bibfnamefont {S.~A.}\ \bibnamefont
  {Voloshin}},\ }\href@noop {} {\  (\bibinfo {year} {2004})},\ \Eprint
  {http://arxiv.org/abs/nucl-th/0410089} {arXiv:nucl-th/0410089} \BibitemShut
  {NoStop}%
\bibitem [{\citenamefont {Abelev}\ \emph {et~al.}(2007)\citenamefont {Abelev}
  \emph {et~al.}}]{STAR:2007}%
  \BibitemOpen
  \bibfield  {author} {\bibinfo {author} {\bibfnamefont {B.}~\bibnamefont
  {Abelev}} \emph {et~al.} (\bibinfo {collaboration} {STAR}),\ }\href {\doibase
  10.1103/PhysRevC.76.024915} {\bibfield  {journal} {\bibinfo  {journal} {Phys.
  Rev. C}\ }\textbf {\bibinfo {volume} {76}},\ \bibinfo {pages} {024915}
  (\bibinfo {year} {2007})},\ \bibinfo {note} {[Erratum: Phys.Rev.C 95, 039906
  (2017)]},\ \Eprint {http://arxiv.org/abs/0705.1691} {arXiv:0705.1691
  [nucl-ex]} \BibitemShut {NoStop}%
\bibitem [{\citenamefont {Adamczyk}\ \emph
  {et~al.}(2017{\natexlab{a}})\citenamefont {Adamczyk} \emph
  {et~al.}}]{STAR:2017ckg}%
  \BibitemOpen
  \bibfield  {author} {\bibinfo {author} {\bibfnamefont {L.}~\bibnamefont
  {Adamczyk}} \emph {et~al.} (\bibinfo {collaboration} {STAR}),\ }\href
  {\doibase 10.1038/nature23004} {\bibfield  {journal} {\bibinfo  {journal}
  {Nature}\ }\textbf {\bibinfo {volume} {548}},\ \bibinfo {pages} {62}
  (\bibinfo {year} {2017}{\natexlab{a}})},\ \Eprint
  {http://arxiv.org/abs/1701.06657} {arXiv:1701.06657 [nucl-ex]} \BibitemShut
  {NoStop}%
\bibitem [{\citenamefont {Adam}\ \emph {et~al.}(2018)\citenamefont {Adam} \emph
  {et~al.}}]{STAR:2018ivw}%
  \BibitemOpen
  \bibfield  {author} {\bibinfo {author} {\bibfnamefont {J.}~\bibnamefont
  {Adam}} \emph {et~al.} (\bibinfo {collaboration} {STAR}),\ }\href {\doibase
  10.1103/PhysRevC.98.014910} {\bibfield  {journal} {\bibinfo  {journal} {Phys.
  Rev. C}\ }\textbf {\bibinfo {volume} {98}},\ \bibinfo {pages} {014910}
  (\bibinfo {year} {2018})},\ \Eprint {http://arxiv.org/abs/1805.04400}
  {arXiv:1805.04400 [nucl-ex]} \BibitemShut {NoStop}%
\bibitem [{\citenamefont {Niida}(2019)}]{Niida:2018hfw}%
  \BibitemOpen
  \bibfield  {author} {\bibinfo {author} {\bibfnamefont {T.}~\bibnamefont
  {Niida}} (\bibinfo {collaboration} {STAR}),\ }\href {\doibase
  10.1016/j.nuclphysa.2018.08.034} {\bibfield  {journal} {\bibinfo  {journal}
  {Nucl. Phys. A}\ }\textbf {\bibinfo {volume} {982}},\ \bibinfo {pages} {511}
  (\bibinfo {year} {2019})},\ \Eprint {http://arxiv.org/abs/1808.10482}
  {arXiv:1808.10482 [nucl-ex]} \BibitemShut {NoStop}%
\bibitem [{\citenamefont {Acharya}\ \emph {et~al.}(2020)\citenamefont {Acharya}
  \emph {et~al.}}]{ALICE:2019}%
  \BibitemOpen
  \bibfield  {author} {\bibinfo {author} {\bibfnamefont {S.}~\bibnamefont
  {Acharya}} \emph {et~al.} (\bibinfo {collaboration} {ALICE}),\ }\href
  {\doibase 10.1103/PhysRevC.101.044611} {\bibfield  {journal} {\bibinfo
  {journal} {Phys. Rev. C}\ }\textbf {\bibinfo {volume} {101}},\ \bibinfo
  {pages} {044611} (\bibinfo {year} {2020})},\ \Eprint
  {http://arxiv.org/abs/1909.01281} {arXiv:1909.01281 [nucl-ex]} \BibitemShut
  {NoStop}%
\bibitem [{\citenamefont {Adam}\ \emph {et~al.}(2019)\citenamefont {Adam} \emph
  {et~al.}}]{STAR:2019srw}%
  \BibitemOpen
  \bibfield  {author} {\bibinfo {author} {\bibfnamefont {J.}~\bibnamefont
  {Adam}} \emph {et~al.} (\bibinfo {collaboration} {STAR}),\ }\href {\doibase
  10.1103/PhysRevLett.123.132301} {\bibfield  {journal} {\bibinfo  {journal}
  {Phys. Rev. Lett.}\ }\textbf {\bibinfo {volume} {123}},\ \bibinfo {pages}
  {132301} (\bibinfo {year} {2019})},\ \Eprint
  {http://arxiv.org/abs/1905.11917} {arXiv:1905.11917 [nucl-ex]} \BibitemShut
  {NoStop}%
\bibitem [{\citenamefont {Gao}\ \emph {et~al.}(2008)\citenamefont {Gao},
  \citenamefont {Chen}, \citenamefont {Deng}, \citenamefont {Liang},
  \citenamefont {Wang},\ and\ \citenamefont {Wang}}]{Gao:2007bc}%
  \BibitemOpen
  \bibfield  {author} {\bibinfo {author} {\bibfnamefont {J.-H.}\ \bibnamefont
  {Gao}}, \bibinfo {author} {\bibfnamefont {S.-W.}\ \bibnamefont {Chen}},
  \bibinfo {author} {\bibfnamefont {W.-T.}\ \bibnamefont {Deng}}, \bibinfo
  {author} {\bibfnamefont {Z.-T.}\ \bibnamefont {Liang}}, \bibinfo {author}
  {\bibfnamefont {Q.}~\bibnamefont {Wang}}, \ and\ \bibinfo {author}
  {\bibfnamefont {X.-N.}\ \bibnamefont {Wang}},\ }\href {\doibase
  10.1103/PhysRevC.77.044902} {\bibfield  {journal} {\bibinfo  {journal} {Phys.
  Rev. C}\ }\textbf {\bibinfo {volume} {77}},\ \bibinfo {pages} {044902}
  (\bibinfo {year} {2008})},\ \Eprint {http://arxiv.org/abs/0710.2943}
  {arXiv:0710.2943 [nucl-th]} \BibitemShut {NoStop}%
\bibitem [{\citenamefont {Betz}\ \emph {et~al.}(2007)\citenamefont {Betz},
  \citenamefont {Gyulassy},\ and\ \citenamefont {Torrieri}}]{Betz:2007kg}%
  \BibitemOpen
  \bibfield  {author} {\bibinfo {author} {\bibfnamefont {B.}~\bibnamefont
  {Betz}}, \bibinfo {author} {\bibfnamefont {M.}~\bibnamefont {Gyulassy}}, \
  and\ \bibinfo {author} {\bibfnamefont {G.}~\bibnamefont {Torrieri}},\ }\href
  {\doibase 10.1103/PhysRevC.76.044901} {\bibfield  {journal} {\bibinfo
  {journal} {Phys. Rev. C}\ }\textbf {\bibinfo {volume} {76}},\ \bibinfo
  {pages} {044901} (\bibinfo {year} {2007})},\ \Eprint
  {http://arxiv.org/abs/0708.0035} {arXiv:0708.0035 [nucl-th]} \BibitemShut
  {NoStop}%
\bibitem [{\citenamefont {Huang}\ \emph {et~al.}(2011)\citenamefont {Huang},
  \citenamefont {Huovinen},\ and\ \citenamefont {Wang}}]{Huang:2011ru}%
  \BibitemOpen
  \bibfield  {author} {\bibinfo {author} {\bibfnamefont {X.-G.}\ \bibnamefont
  {Huang}}, \bibinfo {author} {\bibfnamefont {P.}~\bibnamefont {Huovinen}}, \
  and\ \bibinfo {author} {\bibfnamefont {X.-N.}\ \bibnamefont {Wang}},\ }\href
  {\doibase 10.1103/PhysRevC.84.054910} {\bibfield  {journal} {\bibinfo
  {journal} {Phys. Rev. C}\ }\textbf {\bibinfo {volume} {84}},\ \bibinfo
  {pages} {054910} (\bibinfo {year} {2011})},\ \Eprint
  {http://arxiv.org/abs/1108.5649} {arXiv:1108.5649 [nucl-th]} \BibitemShut
  {NoStop}%
\bibitem [{\citenamefont {Becattini}\ \emph
  {et~al.}(2013{\natexlab{a}})\citenamefont {Becattini}, \citenamefont
  {Chandra}, \citenamefont {Del~Zanna},\ and\ \citenamefont
  {Grossi}}]{Becattini:2013fla}%
  \BibitemOpen
  \bibfield  {author} {\bibinfo {author} {\bibfnamefont {F.}~\bibnamefont
  {Becattini}}, \bibinfo {author} {\bibfnamefont {V.}~\bibnamefont {Chandra}},
  \bibinfo {author} {\bibfnamefont {L.}~\bibnamefont {Del~Zanna}}, \ and\
  \bibinfo {author} {\bibfnamefont {E.}~\bibnamefont {Grossi}},\ }\href
  {\doibase 10.1016/j.aop.2013.07.004} {\bibfield  {journal} {\bibinfo
  {journal} {Annals Phys.}\ }\textbf {\bibinfo {volume} {338}},\ \bibinfo
  {pages} {32} (\bibinfo {year} {2013}{\natexlab{a}})},\ \Eprint
  {http://arxiv.org/abs/1303.3431} {arXiv:1303.3431 [nucl-th]} \BibitemShut
  {NoStop}%
\bibitem [{\citenamefont {Fang}\ \emph {et~al.}(2016)\citenamefont {Fang},
  \citenamefont {Pang}, \citenamefont {Wang},\ and\ \citenamefont
  {Wang}}]{Fang:2016vpj}%
  \BibitemOpen
  \bibfield  {author} {\bibinfo {author} {\bibfnamefont {R.-h.}\ \bibnamefont
  {Fang}}, \bibinfo {author} {\bibfnamefont {L.-g.}\ \bibnamefont {Pang}},
  \bibinfo {author} {\bibfnamefont {Q.}~\bibnamefont {Wang}}, \ and\ \bibinfo
  {author} {\bibfnamefont {X.-n.}\ \bibnamefont {Wang}},\ }\href {\doibase
  10.1103/PhysRevC.94.024904} {\bibfield  {journal} {\bibinfo  {journal} {Phys.
  Rev. C}\ }\textbf {\bibinfo {volume} {94}},\ \bibinfo {pages} {024904}
  (\bibinfo {year} {2016})},\ \Eprint {http://arxiv.org/abs/1604.04036}
  {arXiv:1604.04036 [nucl-th]} \BibitemShut {NoStop}%
\bibitem [{\citenamefont {Liu}\ \emph {et~al.}(2020{\natexlab{a}})\citenamefont
  {Liu}, \citenamefont {Mameda},\ and\ \citenamefont {Huang}}]{Liu:2020flb}%
  \BibitemOpen
  \bibfield  {author} {\bibinfo {author} {\bibfnamefont {Y.-C.}\ \bibnamefont
  {Liu}}, \bibinfo {author} {\bibfnamefont {K.}~\bibnamefont {Mameda}}, \ and\
  \bibinfo {author} {\bibfnamefont {X.-G.}\ \bibnamefont {Huang}},\ }\href
  {\doibase 10.1088/1674-1137/44/9/094101} {\bibfield  {journal} {\bibinfo
  {journal} {Chin. Phys. C}\ }\textbf {\bibinfo {volume} {44}},\ \bibinfo
  {pages} {094101} (\bibinfo {year} {2020}{\natexlab{a}})},\ \Eprint
  {http://arxiv.org/abs/2002.03753} {arXiv:2002.03753 [hep-ph]} \BibitemShut
  {NoStop}%
\bibitem [{\citenamefont {Becattini}\ \emph
  {et~al.}(2013{\natexlab{b}})\citenamefont {Becattini}, \citenamefont
  {Csernai},\ and\ \citenamefont {Wang}}]{Becattini:2013vja}%
  \BibitemOpen
  \bibfield  {author} {\bibinfo {author} {\bibfnamefont {F.}~\bibnamefont
  {Becattini}}, \bibinfo {author} {\bibfnamefont {L.}~\bibnamefont {Csernai}},
  \ and\ \bibinfo {author} {\bibfnamefont {D.}~\bibnamefont {Wang}},\ }\href
  {\doibase 10.1103/PhysRevC.88.034905} {\bibfield  {journal} {\bibinfo
  {journal} {Phys. Rev. C}\ }\textbf {\bibinfo {volume} {88}},\ \bibinfo
  {pages} {034905} (\bibinfo {year} {2013}{\natexlab{b}})},\ \bibinfo {note}
  {[Erratum: Phys.Rev.C 93, 069901 (2016)]},\ \Eprint
  {http://arxiv.org/abs/1304.4427} {arXiv:1304.4427 [nucl-th]} \BibitemShut
  {NoStop}%
\bibitem [{\citenamefont {Becattini}\ \emph {et~al.}(2015)\citenamefont
  {Becattini}, \citenamefont {Inghirami}, \citenamefont {Rolando},
  \citenamefont {Beraudo}, \citenamefont {Del~Zanna}, \citenamefont {De~Pace},
  \citenamefont {Nardi}, \citenamefont {Pagliara},\ and\ \citenamefont
  {Chandra}}]{Becattini:2015ska}%
  \BibitemOpen
  \bibfield  {author} {\bibinfo {author} {\bibfnamefont {F.}~\bibnamefont
  {Becattini}}, \bibinfo {author} {\bibfnamefont {G.}~\bibnamefont
  {Inghirami}}, \bibinfo {author} {\bibfnamefont {V.}~\bibnamefont {Rolando}},
  \bibinfo {author} {\bibfnamefont {A.}~\bibnamefont {Beraudo}}, \bibinfo
  {author} {\bibfnamefont {L.}~\bibnamefont {Del~Zanna}}, \bibinfo {author}
  {\bibfnamefont {A.}~\bibnamefont {De~Pace}}, \bibinfo {author} {\bibfnamefont
  {M.}~\bibnamefont {Nardi}}, \bibinfo {author} {\bibfnamefont
  {G.}~\bibnamefont {Pagliara}}, \ and\ \bibinfo {author} {\bibfnamefont
  {V.}~\bibnamefont {Chandra}},\ }\href {\doibase
  10.1140/epjc/s10052-015-3624-1} {\bibfield  {journal} {\bibinfo  {journal}
  {Eur. Phys. J. C}\ }\textbf {\bibinfo {volume} {75}},\ \bibinfo {pages} {406}
  (\bibinfo {year} {2015})},\ \bibinfo {note} {[Erratum: Eur.Phys.J.C 78, 354
  (2018)]},\ \Eprint {http://arxiv.org/abs/1501.04468} {arXiv:1501.04468
  [nucl-th]} \BibitemShut {NoStop}%
\bibitem [{\citenamefont {Becattini}\ \emph {et~al.}(2017)\citenamefont
  {Becattini}, \citenamefont {Karpenko}, \citenamefont {Lisa}, \citenamefont
  {Upsal},\ and\ \citenamefont {Voloshin}}]{Becattini:2016gvu}%
  \BibitemOpen
  \bibfield  {author} {\bibinfo {author} {\bibfnamefont {F.}~\bibnamefont
  {Becattini}}, \bibinfo {author} {\bibfnamefont {I.}~\bibnamefont {Karpenko}},
  \bibinfo {author} {\bibfnamefont {M.}~\bibnamefont {Lisa}}, \bibinfo {author}
  {\bibfnamefont {I.}~\bibnamefont {Upsal}}, \ and\ \bibinfo {author}
  {\bibfnamefont {S.}~\bibnamefont {Voloshin}},\ }\href {\doibase
  10.1103/PhysRevC.95.054902} {\bibfield  {journal} {\bibinfo  {journal} {Phys.
  Rev. C}\ }\textbf {\bibinfo {volume} {95}},\ \bibinfo {pages} {054902}
  (\bibinfo {year} {2017})},\ \Eprint {http://arxiv.org/abs/1610.02506}
  {arXiv:1610.02506 [nucl-th]} \BibitemShut {NoStop}%
\bibitem [{\citenamefont {Karpenko}\ and\ \citenamefont
  {Becattini}(2017)}]{Karpenko:2016jyx}%
  \BibitemOpen
  \bibfield  {author} {\bibinfo {author} {\bibfnamefont {I.}~\bibnamefont
  {Karpenko}}\ and\ \bibinfo {author} {\bibfnamefont {F.}~\bibnamefont
  {Becattini}},\ }\href {\doibase 10.1140/epjc/s10052-017-4765-1} {\bibfield
  {journal} {\bibinfo  {journal} {Eur. Phys. J. C}\ }\textbf {\bibinfo {volume}
  {77}},\ \bibinfo {pages} {213} (\bibinfo {year} {2017})},\ \Eprint
  {http://arxiv.org/abs/1610.04717} {arXiv:1610.04717 [nucl-th]} \BibitemShut
  {NoStop}%
\bibitem [{\citenamefont {Xie}\ \emph {et~al.}(2016)\citenamefont {Xie},
  \citenamefont {Bleicher}, \citenamefont {St\"ocker}, \citenamefont {Wang},\
  and\ \citenamefont {Csernai}}]{Xie:2016fjj}%
  \BibitemOpen
  \bibfield  {author} {\bibinfo {author} {\bibfnamefont {Y.}~\bibnamefont
  {Xie}}, \bibinfo {author} {\bibfnamefont {M.}~\bibnamefont {Bleicher}},
  \bibinfo {author} {\bibfnamefont {H.}~\bibnamefont {St\"ocker}}, \bibinfo
  {author} {\bibfnamefont {D.}~\bibnamefont {Wang}}, \ and\ \bibinfo {author}
  {\bibfnamefont {L.}~\bibnamefont {Csernai}},\ }\href {\doibase
  10.1103/PhysRevC.94.054907} {\bibfield  {journal} {\bibinfo  {journal} {Phys.
  Rev. C}\ }\textbf {\bibinfo {volume} {94}},\ \bibinfo {pages} {054907}
  (\bibinfo {year} {2016})},\ \Eprint {http://arxiv.org/abs/1610.08678}
  {arXiv:1610.08678 [nucl-th]} \BibitemShut {NoStop}%
\bibitem [{\citenamefont {Xie}\ \emph {et~al.}(2017)\citenamefont {Xie},
  \citenamefont {Wang},\ and\ \citenamefont {Csernai}}]{Xie:2017upb}%
  \BibitemOpen
  \bibfield  {author} {\bibinfo {author} {\bibfnamefont {Y.}~\bibnamefont
  {Xie}}, \bibinfo {author} {\bibfnamefont {D.}~\bibnamefont {Wang}}, \ and\
  \bibinfo {author} {\bibfnamefont {L.~P.}\ \bibnamefont {Csernai}},\ }\href
  {\doibase 10.1103/PhysRevC.95.031901} {\bibfield  {journal} {\bibinfo
  {journal} {Phys. Rev. C}\ }\textbf {\bibinfo {volume} {95}},\ \bibinfo
  {pages} {031901} (\bibinfo {year} {2017})},\ \Eprint
  {http://arxiv.org/abs/1703.03770} {arXiv:1703.03770 [nucl-th]} \BibitemShut
  {NoStop}%
\bibitem [{\citenamefont {Li}\ \emph {et~al.}(2017)\citenamefont {Li},
  \citenamefont {Pang}, \citenamefont {Wang},\ and\ \citenamefont
  {Xia}}]{Li:2017slc}%
  \BibitemOpen
  \bibfield  {author} {\bibinfo {author} {\bibfnamefont {H.}~\bibnamefont
  {Li}}, \bibinfo {author} {\bibfnamefont {L.-G.}\ \bibnamefont {Pang}},
  \bibinfo {author} {\bibfnamefont {Q.}~\bibnamefont {Wang}}, \ and\ \bibinfo
  {author} {\bibfnamefont {X.-L.}\ \bibnamefont {Xia}},\ }\href {\doibase
  10.1103/PhysRevC.96.054908} {\bibfield  {journal} {\bibinfo  {journal} {Phys.
  Rev. C}\ }\textbf {\bibinfo {volume} {96}},\ \bibinfo {pages} {054908}
  (\bibinfo {year} {2017})},\ \Eprint {http://arxiv.org/abs/1704.01507}
  {arXiv:1704.01507 [nucl-th]} \BibitemShut {NoStop}%
\bibitem [{\citenamefont {Ivanov}\ \emph {et~al.}(2019)\citenamefont {Ivanov},
  \citenamefont {Toneev},\ and\ \citenamefont {Soldatov}}]{Ivanov:2019ern}%
  \BibitemOpen
  \bibfield  {author} {\bibinfo {author} {\bibfnamefont {Y.~B.}\ \bibnamefont
  {Ivanov}}, \bibinfo {author} {\bibfnamefont {V.}~\bibnamefont {Toneev}}, \
  and\ \bibinfo {author} {\bibfnamefont {A.}~\bibnamefont {Soldatov}},\ }\href
  {\doibase 10.1103/PhysRevC.100.014908} {\bibfield  {journal} {\bibinfo
  {journal} {Phys. Rev. C}\ }\textbf {\bibinfo {volume} {100}},\ \bibinfo
  {pages} {014908} (\bibinfo {year} {2019})},\ \Eprint
  {http://arxiv.org/abs/1903.05455} {arXiv:1903.05455 [nucl-th]} \BibitemShut
  {NoStop}%
\bibitem [{\citenamefont {Shi}\ \emph {et~al.}(2019)\citenamefont {Shi},
  \citenamefont {Li},\ and\ \citenamefont {Liao}}]{Shi:2017wpk}%
  \BibitemOpen
  \bibfield  {author} {\bibinfo {author} {\bibfnamefont {S.}~\bibnamefont
  {Shi}}, \bibinfo {author} {\bibfnamefont {K.}~\bibnamefont {Li}}, \ and\
  \bibinfo {author} {\bibfnamefont {J.}~\bibnamefont {Liao}},\ }\href {\doibase
  10.1016/j.physletb.2018.09.066} {\bibfield  {journal} {\bibinfo  {journal}
  {Phys. Lett. B}\ }\textbf {\bibinfo {volume} {788}},\ \bibinfo {pages} {409}
  (\bibinfo {year} {2019})},\ \Eprint {http://arxiv.org/abs/1712.00878}
  {arXiv:1712.00878 [nucl-th]} \BibitemShut {NoStop}%
\bibitem [{\citenamefont {Wei}\ \emph {et~al.}(2019)\citenamefont {Wei},
  \citenamefont {Deng},\ and\ \citenamefont {Huang}}]{Wei:2018zfb}%
  \BibitemOpen
  \bibfield  {author} {\bibinfo {author} {\bibfnamefont {D.-X.}\ \bibnamefont
  {Wei}}, \bibinfo {author} {\bibfnamefont {W.-T.}\ \bibnamefont {Deng}}, \
  and\ \bibinfo {author} {\bibfnamefont {X.-G.}\ \bibnamefont {Huang}},\ }\href
  {\doibase 10.1103/PhysRevC.99.014905} {\bibfield  {journal} {\bibinfo
  {journal} {Phys. Rev. C}\ }\textbf {\bibinfo {volume} {99}},\ \bibinfo
  {pages} {014905} (\bibinfo {year} {2019})},\ \Eprint
  {http://arxiv.org/abs/1810.00151} {arXiv:1810.00151 [nucl-th]} \BibitemShut
  {NoStop}%
\bibitem [{\citenamefont {Ivanov}\ and\ \citenamefont
  {Soldatov}(2020)}]{Ivanov:2020wak}%
  \BibitemOpen
  \bibfield  {author} {\bibinfo {author} {\bibfnamefont {Y.}~\bibnamefont
  {Ivanov}}\ and\ \bibinfo {author} {\bibfnamefont {A.}~\bibnamefont
  {Soldatov}},\ }\href {\doibase 10.1103/PhysRevC.102.024916} {\bibfield
  {journal} {\bibinfo  {journal} {Phys. Rev. C}\ }\textbf {\bibinfo {volume}
  {102}},\ \bibinfo {pages} {024916} (\bibinfo {year} {2020})},\ \Eprint
  {http://arxiv.org/abs/2004.05166} {arXiv:2004.05166 [nucl-th]} \BibitemShut
  {NoStop}%
\bibitem [{\citenamefont {Becattini}\ and\ \citenamefont
  {Karpenko}(2018)}]{Becattini:2017gcx}%
  \BibitemOpen
  \bibfield  {author} {\bibinfo {author} {\bibfnamefont {F.}~\bibnamefont
  {Becattini}}\ and\ \bibinfo {author} {\bibfnamefont {I.}~\bibnamefont
  {Karpenko}},\ }\href {\doibase 10.1103/PhysRevLett.120.012302} {\bibfield
  {journal} {\bibinfo  {journal} {Phys. Rev. Lett.}\ }\textbf {\bibinfo
  {volume} {120}},\ \bibinfo {pages} {012302} (\bibinfo {year} {2018})},\
  \Eprint {http://arxiv.org/abs/1707.07984} {arXiv:1707.07984 [nucl-th]}
  \BibitemShut {NoStop}%
\bibitem [{\citenamefont {Florkowski}\ \emph
  {et~al.}(2019{\natexlab{a}})\citenamefont {Florkowski}, \citenamefont
  {Kumar}, \citenamefont {Ryblewski},\ and\ \citenamefont
  {Mazeliauskas}}]{Florkowski:2019voj}%
  \BibitemOpen
  \bibfield  {author} {\bibinfo {author} {\bibfnamefont {W.}~\bibnamefont
  {Florkowski}}, \bibinfo {author} {\bibfnamefont {A.}~\bibnamefont {Kumar}},
  \bibinfo {author} {\bibfnamefont {R.}~\bibnamefont {Ryblewski}}, \ and\
  \bibinfo {author} {\bibfnamefont {A.}~\bibnamefont {Mazeliauskas}},\ }\href
  {\doibase 10.1103/PhysRevC.100.054907} {\bibfield  {journal} {\bibinfo
  {journal} {Phys. Rev. C}\ }\textbf {\bibinfo {volume} {100}},\ \bibinfo
  {pages} {054907} (\bibinfo {year} {2019}{\natexlab{a}})},\ \Eprint
  {http://arxiv.org/abs/1904.00002} {arXiv:1904.00002 [nucl-th]} \BibitemShut
  {NoStop}%
\bibitem [{\citenamefont {Wu}\ \emph {et~al.}(2019)\citenamefont {Wu},
  \citenamefont {Pang}, \citenamefont {Huang},\ and\ \citenamefont
  {Wang}}]{Wu:2019eyi}%
  \BibitemOpen
  \bibfield  {author} {\bibinfo {author} {\bibfnamefont {H.-Z.}\ \bibnamefont
  {Wu}}, \bibinfo {author} {\bibfnamefont {L.-G.}\ \bibnamefont {Pang}},
  \bibinfo {author} {\bibfnamefont {X.-G.}\ \bibnamefont {Huang}}, \ and\
  \bibinfo {author} {\bibfnamefont {Q.}~\bibnamefont {Wang}},\ }\href {\doibase
  10.1103/PhysRevResearch.1.033058} {\bibfield  {journal} {\bibinfo  {journal}
  {Phys. Rev. Research.}\ }\textbf {\bibinfo {volume} {1}},\ \bibinfo {pages}
  {033058} (\bibinfo {year} {2019})},\ \Eprint
  {http://arxiv.org/abs/1906.09385} {arXiv:1906.09385 [nucl-th]} \BibitemShut
  {NoStop}%
\bibitem [{\citenamefont {Xie}\ \emph {et~al.}(2020)\citenamefont {Xie},
  \citenamefont {Wang},\ and\ \citenamefont {Csernai}}]{Xie:2019jun}%
  \BibitemOpen
  \bibfield  {author} {\bibinfo {author} {\bibfnamefont {Y.}~\bibnamefont
  {Xie}}, \bibinfo {author} {\bibfnamefont {D.}~\bibnamefont {Wang}}, \ and\
  \bibinfo {author} {\bibfnamefont {L.~P.}\ \bibnamefont {Csernai}},\ }\href
  {\doibase 10.1140/epjc/s10052-019-7576-8} {\bibfield  {journal} {\bibinfo
  {journal} {Eur. Phys. J. C}\ }\textbf {\bibinfo {volume} {80}},\ \bibinfo
  {pages} {39} (\bibinfo {year} {2020})},\ \Eprint
  {http://arxiv.org/abs/1907.00773} {arXiv:1907.00773 [hep-ph]} \BibitemShut
  {NoStop}%
\bibitem [{\citenamefont {Xia}\ \emph {et~al.}(2019)\citenamefont {Xia},
  \citenamefont {Li}, \citenamefont {Huang},\ and\ \citenamefont
  {Huang}}]{Xia:2019fjf}%
  \BibitemOpen
  \bibfield  {author} {\bibinfo {author} {\bibfnamefont {X.-L.}\ \bibnamefont
  {Xia}}, \bibinfo {author} {\bibfnamefont {H.}~\bibnamefont {Li}}, \bibinfo
  {author} {\bibfnamefont {X.-G.}\ \bibnamefont {Huang}}, \ and\ \bibinfo
  {author} {\bibfnamefont {H.~Z.}\ \bibnamefont {Huang}},\ }\href {\doibase
  10.1103/PhysRevC.100.014913} {\bibfield  {journal} {\bibinfo  {journal}
  {Phys. Rev. C}\ }\textbf {\bibinfo {volume} {100}},\ \bibinfo {pages}
  {014913} (\bibinfo {year} {2019})},\ \Eprint
  {http://arxiv.org/abs/1905.03120} {arXiv:1905.03120 [nucl-th]} \BibitemShut
  {NoStop}%
\bibitem [{\citenamefont {Becattini}\ \emph {et~al.}(2019)\citenamefont
  {Becattini}, \citenamefont {Cao},\ and\ \citenamefont
  {Speranza}}]{Becattini:2019ntv}%
  \BibitemOpen
  \bibfield  {author} {\bibinfo {author} {\bibfnamefont {F.}~\bibnamefont
  {Becattini}}, \bibinfo {author} {\bibfnamefont {G.}~\bibnamefont {Cao}}, \
  and\ \bibinfo {author} {\bibfnamefont {E.}~\bibnamefont {Speranza}},\ }\href
  {\doibase 10.1140/epjc/s10052-019-7213-6} {\bibfield  {journal} {\bibinfo
  {journal} {Eur. Phys. J. C}\ }\textbf {\bibinfo {volume} {79}},\ \bibinfo
  {pages} {741} (\bibinfo {year} {2019})},\ \Eprint
  {http://arxiv.org/abs/1905.03123} {arXiv:1905.03123 [nucl-th]} \BibitemShut
  {NoStop}%
\bibitem [{\citenamefont {Liu}\ \emph {et~al.}(2020{\natexlab{b}})\citenamefont
  {Liu}, \citenamefont {Sun},\ and\ \citenamefont {Ko}}]{Liu:2019krs}%
  \BibitemOpen
  \bibfield  {author} {\bibinfo {author} {\bibfnamefont {S.~Y.}\ \bibnamefont
  {Liu}}, \bibinfo {author} {\bibfnamefont {Y.}~\bibnamefont {Sun}}, \ and\
  \bibinfo {author} {\bibfnamefont {C.~M.}\ \bibnamefont {Ko}},\ }\href
  {\doibase 10.1103/PhysRevLett.125.062301} {\bibfield  {journal} {\bibinfo
  {journal} {Phys. Rev. Lett.}\ }\textbf {\bibinfo {volume} {125}},\ \bibinfo
  {pages} {062301} (\bibinfo {year} {2020}{\natexlab{b}})},\ \Eprint
  {http://arxiv.org/abs/1910.06774} {arXiv:1910.06774 [nucl-th]} \BibitemShut
  {NoStop}%
\bibitem [{\citenamefont {Huang}(2020)}]{Huang:2020xyr}%
  \BibitemOpen
  \bibfield  {author} {\bibinfo {author} {\bibfnamefont {X.-G.}\ \bibnamefont
  {Huang}},\ }\href@noop {} {\  (\bibinfo {year} {2020})},\ \Eprint
  {http://arxiv.org/abs/2002.07549} {arXiv:2002.07549 [nucl-th]} \BibitemShut
  {NoStop}%
\bibitem [{\citenamefont {Liu}\ and\ \citenamefont
  {Huang}(2020)}]{Liu:2020ymh}%
  \BibitemOpen
  \bibfield  {author} {\bibinfo {author} {\bibfnamefont {Y.-C.}\ \bibnamefont
  {Liu}}\ and\ \bibinfo {author} {\bibfnamefont {X.-G.}\ \bibnamefont
  {Huang}},\ }\href {\doibase 10.1007/s41365-020-00764-z} {\bibfield  {journal}
  {\bibinfo  {journal} {Nucl. Sci. Tech.}\ }\textbf {\bibinfo {volume} {31}},\
  \bibinfo {pages} {56} (\bibinfo {year} {2020})},\ \Eprint
  {http://arxiv.org/abs/2003.12482} {arXiv:2003.12482 [nucl-th]} \BibitemShut
  {NoStop}%
\bibitem [{\citenamefont {Becattini}\ and\ \citenamefont
  {Lisa}(2020)}]{Becattini:2020ngo}%
  \BibitemOpen
  \bibfield  {author} {\bibinfo {author} {\bibfnamefont {F.}~\bibnamefont
  {Becattini}}\ and\ \bibinfo {author} {\bibfnamefont {M.~A.}\ \bibnamefont
  {Lisa}},\ }\href {\doibase 10.1146/annurev-nucl-021920-095245} {\  (\bibinfo
  {year} {2020}),\ 10.1146/annurev-nucl-021920-095245},\ \Eprint
  {http://arxiv.org/abs/2003.03640} {arXiv:2003.03640 [nucl-ex]} \BibitemShut
  {NoStop}%
\bibitem [{\citenamefont {Gao}\ \emph {et~al.}(2020{\natexlab{a}})\citenamefont
  {Gao}, \citenamefont {Ma}, \citenamefont {Pu},\ and\ \citenamefont
  {Wang}}]{Gao:2020vbh}%
  \BibitemOpen
  \bibfield  {author} {\bibinfo {author} {\bibfnamefont {J.-H.}\ \bibnamefont
  {Gao}}, \bibinfo {author} {\bibfnamefont {G.-L.}\ \bibnamefont {Ma}},
  \bibinfo {author} {\bibfnamefont {S.}~\bibnamefont {Pu}}, \ and\ \bibinfo
  {author} {\bibfnamefont {Q.}~\bibnamefont {Wang}},\ }\href {\doibase
  10.1007/s41365-020-00801-x} {\bibfield  {journal} {\bibinfo  {journal} {Nucl.
  Sci. Tech.}\ }\textbf {\bibinfo {volume} {31}},\ \bibinfo {pages} {90}
  (\bibinfo {year} {2020}{\natexlab{a}})},\ \Eprint
  {http://arxiv.org/abs/2005.10432} {arXiv:2005.10432 [hep-ph]} \BibitemShut
  {NoStop}%
\bibitem [{\citenamefont {Gao}\ \emph {et~al.}(2020{\natexlab{b}})\citenamefont
  {Gao}, \citenamefont {Liang}, \citenamefont {Wang},\ and\ \citenamefont
  {Wang}}]{Gao:2020lxh}%
  \BibitemOpen
  \bibfield  {author} {\bibinfo {author} {\bibfnamefont {J.-H.}\ \bibnamefont
  {Gao}}, \bibinfo {author} {\bibfnamefont {Z.-T.}\ \bibnamefont {Liang}},
  \bibinfo {author} {\bibfnamefont {Q.}~\bibnamefont {Wang}}, \ and\ \bibinfo
  {author} {\bibfnamefont {X.-N.}\ \bibnamefont {Wang}},\ }\href@noop {} {\
  (\bibinfo {year} {2020}{\natexlab{b}})},\ \Eprint
  {http://arxiv.org/abs/2009.04803} {arXiv:2009.04803 [nucl-th]} \BibitemShut
  {NoStop}%
\bibitem [{\citenamefont {Huang}\ \emph {et~al.}(2020)\citenamefont {Huang},
  \citenamefont {Liao}, \citenamefont {Wang},\ and\ \citenamefont
  {Xia}}]{Huang:2020dtn}%
  \BibitemOpen
  \bibfield  {author} {\bibinfo {author} {\bibfnamefont {X.-G.}\ \bibnamefont
  {Huang}}, \bibinfo {author} {\bibfnamefont {J.}~\bibnamefont {Liao}},
  \bibinfo {author} {\bibfnamefont {Q.}~\bibnamefont {Wang}}, \ and\ \bibinfo
  {author} {\bibfnamefont {X.-L.}\ \bibnamefont {Xia}},\ }\href@noop {} {\
  (\bibinfo {year} {2020})},\ \Eprint {http://arxiv.org/abs/2010.08937}
  {arXiv:2010.08937 [nucl-th]} \BibitemShut {NoStop}%
\bibitem [{\citenamefont {Liang}\ and\ \citenamefont
  {Wang}(2005{\natexlab{b}})}]{Liang:2004xn}%
  \BibitemOpen
  \bibfield  {author} {\bibinfo {author} {\bibfnamefont {Z.-T.}\ \bibnamefont
  {Liang}}\ and\ \bibinfo {author} {\bibfnamefont {X.-N.}\ \bibnamefont
  {Wang}},\ }\href {\doibase 10.1016/j.physletb.2005.09.060} {\bibfield
  {journal} {\bibinfo  {journal} {Phys. Lett. B}\ }\textbf {\bibinfo {volume}
  {629}},\ \bibinfo {pages} {20} (\bibinfo {year} {2005}{\natexlab{b}})},\
  \Eprint {http://arxiv.org/abs/nucl-th/0411101} {arXiv:nucl-th/0411101}
  \BibitemShut {NoStop}%
\bibitem [{\citenamefont {Sheng}\ \emph
  {et~al.}(2020{\natexlab{a}})\citenamefont {Sheng}, \citenamefont {Oliva},\
  and\ \citenamefont {Wang}}]{Sheng:2019kmk}%
  \BibitemOpen
  \bibfield  {author} {\bibinfo {author} {\bibfnamefont {X.-L.}\ \bibnamefont
  {Sheng}}, \bibinfo {author} {\bibfnamefont {L.}~\bibnamefont {Oliva}}, \ and\
  \bibinfo {author} {\bibfnamefont {Q.}~\bibnamefont {Wang}},\ }\href {\doibase
  10.1103/PhysRevD.101.096005} {\bibfield  {journal} {\bibinfo  {journal}
  {Phys. Rev. D}\ }\textbf {\bibinfo {volume} {101}},\ \bibinfo {pages}
  {096005} (\bibinfo {year} {2020}{\natexlab{a}})},\ \Eprint
  {http://arxiv.org/abs/1910.13684} {arXiv:1910.13684 [nucl-th]} \BibitemShut
  {NoStop}%
\bibitem [{\citenamefont {Sheng}\ \emph
  {et~al.}(2020{\natexlab{b}})\citenamefont {Sheng}, \citenamefont {Wang},\
  and\ \citenamefont {Wang}}]{Sheng:2020ghv}%
  \BibitemOpen
  \bibfield  {author} {\bibinfo {author} {\bibfnamefont {X.-L.}\ \bibnamefont
  {Sheng}}, \bibinfo {author} {\bibfnamefont {Q.}~\bibnamefont {Wang}}, \ and\
  \bibinfo {author} {\bibfnamefont {X.-N.}\ \bibnamefont {Wang}},\ }\href
  {\doibase 10.1103/PhysRevD.102.056013} {\bibfield  {journal} {\bibinfo
  {journal} {Phys. Rev. D}\ }\textbf {\bibinfo {volume} {102}},\ \bibinfo
  {pages} {056013} (\bibinfo {year} {2020}{\natexlab{b}})},\ \Eprint
  {http://arxiv.org/abs/2007.05106} {arXiv:2007.05106 [nucl-th]} \BibitemShut
  {NoStop}%
\bibitem [{\citenamefont {Xia}\ \emph {et~al.}(2020)\citenamefont {Xia},
  \citenamefont {Li}, \citenamefont {Huang},\ and\ \citenamefont
  {Huang}}]{Xia:2020tyd}%
  \BibitemOpen
  \bibfield  {author} {\bibinfo {author} {\bibfnamefont {X.-L.}\ \bibnamefont
  {Xia}}, \bibinfo {author} {\bibfnamefont {H.}~\bibnamefont {Li}}, \bibinfo
  {author} {\bibfnamefont {X.-G.}\ \bibnamefont {Huang}}, \ and\ \bibinfo
  {author} {\bibfnamefont {H.~Z.}\ \bibnamefont {Huang}},\ }\href@noop {} {\
  (\bibinfo {year} {2020})},\ \Eprint {http://arxiv.org/abs/2010.01474}
  {arXiv:2010.01474 [nucl-th]} \BibitemShut {NoStop}%
\bibitem [{\citenamefont {Taya}\ \emph {et~al.}(2020)\citenamefont {Taya} \emph
  {et~al.}}]{Taya:2020sej}%
  \BibitemOpen
  \bibfield  {author} {\bibinfo {author} {\bibfnamefont {H.}~\bibnamefont
  {Taya}} \emph {et~al.} (\bibinfo {collaboration} {ExHIC-P}),\ }\href
  {\doibase 10.1103/PhysRevC.102.021901} {\bibfield  {journal} {\bibinfo
  {journal} {Phys. Rev. C}\ }\textbf {\bibinfo {volume} {102}},\ \bibinfo
  {pages} {021901} (\bibinfo {year} {2020})},\ \Eprint
  {http://arxiv.org/abs/2002.10082} {arXiv:2002.10082 [nucl-th]} \BibitemShut
  {NoStop}%
\bibitem [{\citenamefont {Liu}\ and\ \citenamefont {Yin}(2020)}]{Liu:2020dxg}%
  \BibitemOpen
  \bibfield  {author} {\bibinfo {author} {\bibfnamefont {S.~Y.}\ \bibnamefont
  {Liu}}\ and\ \bibinfo {author} {\bibfnamefont {Y.}~\bibnamefont {Yin}},\
  }\href@noop {} {\  (\bibinfo {year} {2020})},\ \Eprint
  {http://arxiv.org/abs/2006.12421} {arXiv:2006.12421 [nucl-th]} \BibitemShut
  {NoStop}%
\bibitem [{\citenamefont {Schenke}\ \emph {et~al.}(2011)\citenamefont
  {Schenke}, \citenamefont {Jeon},\ and\ \citenamefont
  {Gale}}]{Schenke:2010rr}%
  \BibitemOpen
  \bibfield  {author} {\bibinfo {author} {\bibfnamefont {B.}~\bibnamefont
  {Schenke}}, \bibinfo {author} {\bibfnamefont {S.}~\bibnamefont {Jeon}}, \
  and\ \bibinfo {author} {\bibfnamefont {C.}~\bibnamefont {Gale}},\ }\href
  {\doibase 10.1103/PhysRevLett.106.042301} {\bibfield  {journal} {\bibinfo
  {journal} {Phys. Rev. Lett.}\ }\textbf {\bibinfo {volume} {106}},\ \bibinfo
  {pages} {042301} (\bibinfo {year} {2011})},\ \Eprint
  {http://arxiv.org/abs/1009.3244} {arXiv:1009.3244 [hep-ph]} \BibitemShut
  {NoStop}%
\bibitem [{\citenamefont {Schenke}\ \emph {et~al.}(2012)\citenamefont
  {Schenke}, \citenamefont {Jeon},\ and\ \citenamefont
  {Gale}}]{Schenke:2011bn}%
  \BibitemOpen
  \bibfield  {author} {\bibinfo {author} {\bibfnamefont {B.}~\bibnamefont
  {Schenke}}, \bibinfo {author} {\bibfnamefont {S.}~\bibnamefont {Jeon}}, \
  and\ \bibinfo {author} {\bibfnamefont {C.}~\bibnamefont {Gale}},\ }\href
  {\doibase 10.1103/PhysRevC.85.024901} {\bibfield  {journal} {\bibinfo
  {journal} {Phys. Rev. C}\ }\textbf {\bibinfo {volume} {85}},\ \bibinfo
  {pages} {024901} (\bibinfo {year} {2012})},\ \Eprint
  {http://arxiv.org/abs/1109.6289} {arXiv:1109.6289 [hep-ph]} \BibitemShut
  {NoStop}%
\bibitem [{\citenamefont {Gale}\ \emph {et~al.}(2013)\citenamefont {Gale},
  \citenamefont {Jeon}, \citenamefont {Schenke}, \citenamefont {Tribedy},\ and\
  \citenamefont {Venugopalan}}]{Gale:2012rq}%
  \BibitemOpen
  \bibfield  {author} {\bibinfo {author} {\bibfnamefont {C.}~\bibnamefont
  {Gale}}, \bibinfo {author} {\bibfnamefont {S.}~\bibnamefont {Jeon}}, \bibinfo
  {author} {\bibfnamefont {B.}~\bibnamefont {Schenke}}, \bibinfo {author}
  {\bibfnamefont {P.}~\bibnamefont {Tribedy}}, \ and\ \bibinfo {author}
  {\bibfnamefont {R.}~\bibnamefont {Venugopalan}},\ }\href {\doibase
  10.1103/PhysRevLett.110.012302} {\bibfield  {journal} {\bibinfo  {journal}
  {Phys. Rev. Lett.}\ }\textbf {\bibinfo {volume} {110}},\ \bibinfo {pages}
  {012302} (\bibinfo {year} {2013})},\ \Eprint {http://arxiv.org/abs/1209.6330}
  {arXiv:1209.6330 [nucl-th]} \BibitemShut {NoStop}%
\bibitem [{\citenamefont {Schenke}\ and\ \citenamefont
  {Venugopalan}(2014)}]{Schenke:2014zha}%
  \BibitemOpen
  \bibfield  {author} {\bibinfo {author} {\bibfnamefont {B.}~\bibnamefont
  {Schenke}}\ and\ \bibinfo {author} {\bibfnamefont {R.}~\bibnamefont
  {Venugopalan}},\ }\href {\doibase 10.1103/PhysRevLett.113.102301} {\bibfield
  {journal} {\bibinfo  {journal} {Phys. Rev. Lett.}\ }\textbf {\bibinfo
  {volume} {113}},\ \bibinfo {pages} {102301} (\bibinfo {year} {2014})},\
  \Eprint {http://arxiv.org/abs/1405.3605} {arXiv:1405.3605 [nucl-th]}
  \BibitemShut {NoStop}%
\bibitem [{\citenamefont {Karpenko}\ \emph {et~al.}(2015)\citenamefont
  {Karpenko}, \citenamefont {Huovinen}, \citenamefont {Petersen},\ and\
  \citenamefont {Bleicher}}]{Karpenko:2015xea}%
  \BibitemOpen
  \bibfield  {author} {\bibinfo {author} {\bibfnamefont {I.}~\bibnamefont
  {Karpenko}}, \bibinfo {author} {\bibfnamefont {P.}~\bibnamefont {Huovinen}},
  \bibinfo {author} {\bibfnamefont {H.}~\bibnamefont {Petersen}}, \ and\
  \bibinfo {author} {\bibfnamefont {M.}~\bibnamefont {Bleicher}},\ }\href
  {\doibase 10.1103/PhysRevC.91.064901} {\bibfield  {journal} {\bibinfo
  {journal} {Phys. Rev. C}\ }\textbf {\bibinfo {volume} {91}},\ \bibinfo
  {pages} {064901} (\bibinfo {year} {2015})},\ \Eprint
  {http://arxiv.org/abs/1502.01978} {arXiv:1502.01978 [nucl-th]} \BibitemShut
  {NoStop}%
\bibitem [{\citenamefont {Song}\ \emph {et~al.}(2017)\citenamefont {Song},
  \citenamefont {Zhou},\ and\ \citenamefont {Gajdosova}}]{Song:2017wtw}%
  \BibitemOpen
  \bibfield  {author} {\bibinfo {author} {\bibfnamefont {H.}~\bibnamefont
  {Song}}, \bibinfo {author} {\bibfnamefont {Y.}~\bibnamefont {Zhou}}, \ and\
  \bibinfo {author} {\bibfnamefont {K.}~\bibnamefont {Gajdosova}},\ }\href
  {\doibase 10.1007/s41365-017-0245-4} {\bibfield  {journal} {\bibinfo
  {journal} {Nucl. Sci. Tech.}\ }\textbf {\bibinfo {volume} {28}},\ \bibinfo
  {pages} {99} (\bibinfo {year} {2017})},\ \Eprint
  {http://arxiv.org/abs/1703.00670} {arXiv:1703.00670 [nucl-th]} \BibitemShut
  {NoStop}%
\bibitem [{\citenamefont {Zhao}\ \emph {et~al.}(2017)\citenamefont {Zhao},
  \citenamefont {Xu},\ and\ \citenamefont {Song}}]{Zhao:2017yhj}%
  \BibitemOpen
  \bibfield  {author} {\bibinfo {author} {\bibfnamefont {W.}~\bibnamefont
  {Zhao}}, \bibinfo {author} {\bibfnamefont {H.-j.}\ \bibnamefont {Xu}}, \ and\
  \bibinfo {author} {\bibfnamefont {H.}~\bibnamefont {Song}},\ }\href {\doibase
  10.1140/epjc/s10052-017-5186-x} {\bibfield  {journal} {\bibinfo  {journal}
  {Eur. Phys. J. C}\ }\textbf {\bibinfo {volume} {77}},\ \bibinfo {pages} {645}
  (\bibinfo {year} {2017})},\ \Eprint {http://arxiv.org/abs/1703.10792}
  {arXiv:1703.10792 [nucl-th]} \BibitemShut {NoStop}%
\bibitem [{\citenamefont {Zhu}\ \emph {et~al.}(2017)\citenamefont {Zhu},
  \citenamefont {Zhou}, \citenamefont {Xu},\ and\ \citenamefont
  {Song}}]{Zhu:2016puf}%
  \BibitemOpen
  \bibfield  {author} {\bibinfo {author} {\bibfnamefont {X.}~\bibnamefont
  {Zhu}}, \bibinfo {author} {\bibfnamefont {Y.}~\bibnamefont {Zhou}}, \bibinfo
  {author} {\bibfnamefont {H.}~\bibnamefont {Xu}}, \ and\ \bibinfo {author}
  {\bibfnamefont {H.}~\bibnamefont {Song}},\ }\href {\doibase
  10.1103/PhysRevC.95.044902} {\bibfield  {journal} {\bibinfo  {journal} {Phys.
  Rev. C}\ }\textbf {\bibinfo {volume} {95}},\ \bibinfo {pages} {044902}
  (\bibinfo {year} {2017})},\ \Eprint {http://arxiv.org/abs/1608.05305}
  {arXiv:1608.05305 [nucl-th]} \BibitemShut {NoStop}%
\bibitem [{\citenamefont {Zhao}\ \emph
  {et~al.}(2018{\natexlab{a}})\citenamefont {Zhao}, \citenamefont {Zhu},
  \citenamefont {Zheng}, \citenamefont {Ko},\ and\ \citenamefont
  {Song}}]{Zhao:2018lyf}%
  \BibitemOpen
  \bibfield  {author} {\bibinfo {author} {\bibfnamefont {W.}~\bibnamefont
  {Zhao}}, \bibinfo {author} {\bibfnamefont {L.}~\bibnamefont {Zhu}}, \bibinfo
  {author} {\bibfnamefont {H.}~\bibnamefont {Zheng}}, \bibinfo {author}
  {\bibfnamefont {C.~M.}\ \bibnamefont {Ko}}, \ and\ \bibinfo {author}
  {\bibfnamefont {H.}~\bibnamefont {Song}},\ }\href {\doibase
  10.1103/PhysRevC.98.054905} {\bibfield  {journal} {\bibinfo  {journal} {Phys.
  Rev. C}\ }\textbf {\bibinfo {volume} {98}},\ \bibinfo {pages} {054905}
  (\bibinfo {year} {2018}{\natexlab{a}})},\ \Eprint
  {http://arxiv.org/abs/1807.02813} {arXiv:1807.02813 [nucl-th]} \BibitemShut
  {NoStop}%
\bibitem [{\citenamefont {Denicol}\ \emph {et~al.}(2018)\citenamefont
  {Denicol}, \citenamefont {Gale}, \citenamefont {Jeon}, \citenamefont
  {Monnai}, \citenamefont {Schenke},\ and\ \citenamefont
  {Shen}}]{Denicol:2018wdp}%
  \BibitemOpen
  \bibfield  {author} {\bibinfo {author} {\bibfnamefont {G.~S.}\ \bibnamefont
  {Denicol}}, \bibinfo {author} {\bibfnamefont {C.}~\bibnamefont {Gale}},
  \bibinfo {author} {\bibfnamefont {S.}~\bibnamefont {Jeon}}, \bibinfo {author}
  {\bibfnamefont {A.}~\bibnamefont {Monnai}}, \bibinfo {author} {\bibfnamefont
  {B.}~\bibnamefont {Schenke}}, \ and\ \bibinfo {author} {\bibfnamefont
  {C.}~\bibnamefont {Shen}},\ }\href {\doibase 10.1103/PhysRevC.98.034916}
  {\bibfield  {journal} {\bibinfo  {journal} {Phys. Rev. C}\ }\textbf {\bibinfo
  {volume} {98}},\ \bibinfo {pages} {034916} (\bibinfo {year} {2018})},\
  \Eprint {http://arxiv.org/abs/1804.10557} {arXiv:1804.10557 [nucl-th]}
  \BibitemShut {NoStop}%
\bibitem [{\citenamefont {Schenke}\ \emph
  {et~al.}(2020{\natexlab{a}})\citenamefont {Schenke}, \citenamefont {Shen},\
  and\ \citenamefont {Tribedy}}]{Schenke:2020mbo}%
  \BibitemOpen
  \bibfield  {author} {\bibinfo {author} {\bibfnamefont {B.}~\bibnamefont
  {Schenke}}, \bibinfo {author} {\bibfnamefont {C.}~\bibnamefont {Shen}}, \
  and\ \bibinfo {author} {\bibfnamefont {P.}~\bibnamefont {Tribedy}},\
  }\href@noop {} {\  (\bibinfo {year} {2020}{\natexlab{a}})},\ \Eprint
  {http://arxiv.org/abs/2005.14682} {arXiv:2005.14682 [nucl-th]} \BibitemShut
  {NoStop}%
\bibitem [{\citenamefont {Schenke}\ \emph
  {et~al.}(2020{\natexlab{b}})\citenamefont {Schenke}, \citenamefont {Shen},\
  and\ \citenamefont {Tribedy}}]{Schenke:2019pmk}%
  \BibitemOpen
  \bibfield  {author} {\bibinfo {author} {\bibfnamefont {B.}~\bibnamefont
  {Schenke}}, \bibinfo {author} {\bibfnamefont {C.}~\bibnamefont {Shen}}, \
  and\ \bibinfo {author} {\bibfnamefont {P.}~\bibnamefont {Tribedy}},\ }\href
  {\doibase 10.1016/j.physletb.2020.135322} {\bibfield  {journal} {\bibinfo
  {journal} {Phys. Lett. B}\ }\textbf {\bibinfo {volume} {803}},\ \bibinfo
  {pages} {135322} (\bibinfo {year} {2020}{\natexlab{b}})},\ \Eprint
  {http://arxiv.org/abs/1908.06212} {arXiv:1908.06212 [nucl-th]} \BibitemShut
  {NoStop}%
\bibitem [{\citenamefont {Lin}\ \emph {et~al.}(2005)\citenamefont {Lin},
  \citenamefont {Ko}, \citenamefont {Li}, \citenamefont {Zhang},\ and\
  \citenamefont {Pal}}]{Lin:2004en}%
  \BibitemOpen
  \bibfield  {author} {\bibinfo {author} {\bibfnamefont {Z.-W.}\ \bibnamefont
  {Lin}}, \bibinfo {author} {\bibfnamefont {C.~M.}\ \bibnamefont {Ko}},
  \bibinfo {author} {\bibfnamefont {B.-A.}\ \bibnamefont {Li}}, \bibinfo
  {author} {\bibfnamefont {B.}~\bibnamefont {Zhang}}, \ and\ \bibinfo {author}
  {\bibfnamefont {S.}~\bibnamefont {Pal}},\ }\href {\doibase
  10.1103/PhysRevC.72.064901} {\bibfield  {journal} {\bibinfo  {journal} {Phys.
  Rev. C}\ }\textbf {\bibinfo {volume} {72}},\ \bibinfo {pages} {064901}
  (\bibinfo {year} {2005})},\ \Eprint {http://arxiv.org/abs/nucl-th/0411110}
  {arXiv:nucl-th/0411110} \BibitemShut {NoStop}%
\bibitem [{\citenamefont {Schenke}\ \emph {et~al.}(2010)\citenamefont
  {Schenke}, \citenamefont {Jeon},\ and\ \citenamefont
  {Gale}}]{Schenke:2010nt}%
  \BibitemOpen
  \bibfield  {author} {\bibinfo {author} {\bibfnamefont {B.}~\bibnamefont
  {Schenke}}, \bibinfo {author} {\bibfnamefont {S.}~\bibnamefont {Jeon}}, \
  and\ \bibinfo {author} {\bibfnamefont {C.}~\bibnamefont {Gale}},\ }\href
  {\doibase 10.1103/PhysRevC.82.014903} {\bibfield  {journal} {\bibinfo
  {journal} {Phys. Rev. C}\ }\textbf {\bibinfo {volume} {82}},\ \bibinfo
  {pages} {014903} (\bibinfo {year} {2010})},\ \Eprint
  {http://arxiv.org/abs/1004.1408} {arXiv:1004.1408 [hep-ph]} \BibitemShut
  {NoStop}%
\bibitem [{\citenamefont {Bass}\ \emph {et~al.}(1998)\citenamefont {Bass} \emph
  {et~al.}}]{Bass:1998ca}%
  \BibitemOpen
  \bibfield  {author} {\bibinfo {author} {\bibfnamefont {S.}~\bibnamefont
  {Bass}} \emph {et~al.},\ }\href {\doibase 10.1016/S0146-6410(98)00058-1}
  {\bibfield  {journal} {\bibinfo  {journal} {Prog. Part. Nucl. Phys.}\
  }\textbf {\bibinfo {volume} {41}},\ \bibinfo {pages} {255} (\bibinfo {year}
  {1998})},\ \Eprint {http://arxiv.org/abs/nucl-th/9803035}
  {arXiv:nucl-th/9803035} \BibitemShut {NoStop}%
\bibitem [{\citenamefont {Bleicher}\ \emph {et~al.}(1999)\citenamefont
  {Bleicher} \emph {et~al.}}]{Bleicher:1999xi}%
  \BibitemOpen
  \bibfield  {author} {\bibinfo {author} {\bibfnamefont {M.}~\bibnamefont
  {Bleicher}} \emph {et~al.},\ }\href {\doibase 10.1088/0954-3899/25/9/308}
  {\bibfield  {journal} {\bibinfo  {journal} {J. Phys. G}\ }\textbf {\bibinfo
  {volume} {25}},\ \bibinfo {pages} {1859} (\bibinfo {year} {1999})},\ \Eprint
  {http://arxiv.org/abs/hep-ph/9909407} {arXiv:hep-ph/9909407} \BibitemShut
  {NoStop}%
\bibitem [{\citenamefont {Xu}\ \emph {et~al.}(2016)\citenamefont {Xu},
  \citenamefont {Li},\ and\ \citenamefont {Song}}]{Xu:2016hmp}%
  \BibitemOpen
  \bibfield  {author} {\bibinfo {author} {\bibfnamefont {H.-j.}\ \bibnamefont
  {Xu}}, \bibinfo {author} {\bibfnamefont {Z.}~\bibnamefont {Li}}, \ and\
  \bibinfo {author} {\bibfnamefont {H.}~\bibnamefont {Song}},\ }\href {\doibase
  10.1103/PhysRevC.93.064905} {\bibfield  {journal} {\bibinfo  {journal} {Phys.
  Rev. C}\ }\textbf {\bibinfo {volume} {93}},\ \bibinfo {pages} {064905}
  (\bibinfo {year} {2016})},\ \Eprint {http://arxiv.org/abs/1602.02029}
  {arXiv:1602.02029 [nucl-th]} \BibitemShut {NoStop}%
\bibitem [{\citenamefont {Zhao}\ \emph
  {et~al.}(2018{\natexlab{b}})\citenamefont {Zhao}, \citenamefont {Zhou},
  \citenamefont {Xu}, \citenamefont {Deng},\ and\ \citenamefont
  {Song}}]{Zhao:2017rgg}%
  \BibitemOpen
  \bibfield  {author} {\bibinfo {author} {\bibfnamefont {W.}~\bibnamefont
  {Zhao}}, \bibinfo {author} {\bibfnamefont {Y.}~\bibnamefont {Zhou}}, \bibinfo
  {author} {\bibfnamefont {H.}~\bibnamefont {Xu}}, \bibinfo {author}
  {\bibfnamefont {W.}~\bibnamefont {Deng}}, \ and\ \bibinfo {author}
  {\bibfnamefont {H.}~\bibnamefont {Song}},\ }\href {\doibase
  10.1016/j.physletb.2018.03.022} {\bibfield  {journal} {\bibinfo  {journal}
  {Phys. Lett. B}\ }\textbf {\bibinfo {volume} {780}},\ \bibinfo {pages} {495}
  (\bibinfo {year} {2018}{\natexlab{b}})},\ \Eprint
  {http://arxiv.org/abs/1801.00271} {arXiv:1801.00271 [nucl-th]} \BibitemShut
  {NoStop}%
\bibitem [{\citenamefont {Zhao}\ \emph {et~al.}(2020)\citenamefont {Zhao},
  \citenamefont {Zhou}, \citenamefont {Murase},\ and\ \citenamefont
  {Song}}]{Zhao:2020pty}%
  \BibitemOpen
  \bibfield  {author} {\bibinfo {author} {\bibfnamefont {W.}~\bibnamefont
  {Zhao}}, \bibinfo {author} {\bibfnamefont {Y.}~\bibnamefont {Zhou}}, \bibinfo
  {author} {\bibfnamefont {K.}~\bibnamefont {Murase}}, \ and\ \bibinfo {author}
  {\bibfnamefont {H.}~\bibnamefont {Song}},\ }\href {\doibase
  10.1140/epjc/s10052-020-8376-x} {\bibfield  {journal} {\bibinfo  {journal}
  {Eur. Phys. J. C}\ }\textbf {\bibinfo {volume} {80}},\ \bibinfo {pages} {846}
  (\bibinfo {year} {2020})},\ \Eprint {http://arxiv.org/abs/2001.06742}
  {arXiv:2001.06742 [nucl-th]} \BibitemShut {NoStop}%
\bibitem [{\citenamefont {Wang}\ and\ \citenamefont
  {Gyulassy}(1991)}]{Wang:1991hta}%
  \BibitemOpen
  \bibfield  {author} {\bibinfo {author} {\bibfnamefont {X.-N.}\ \bibnamefont
  {Wang}}\ and\ \bibinfo {author} {\bibfnamefont {M.}~\bibnamefont
  {Gyulassy}},\ }\href {\doibase 10.1103/PhysRevD.44.3501} {\bibfield
  {journal} {\bibinfo  {journal} {Phys. Rev. D}\ }\textbf {\bibinfo {volume}
  {44}},\ \bibinfo {pages} {3501} (\bibinfo {year} {1991})}\BibitemShut
  {NoStop}%
\bibitem [{\citenamefont {Gyulassy}\ and\ \citenamefont
  {Wang}(1994)}]{Gyulassy:1994ew}%
  \BibitemOpen
  \bibfield  {author} {\bibinfo {author} {\bibfnamefont {M.}~\bibnamefont
  {Gyulassy}}\ and\ \bibinfo {author} {\bibfnamefont {X.-N.}\ \bibnamefont
  {Wang}},\ }\href {\doibase 10.1016/0010-4655(94)90057-4} {\bibfield
  {journal} {\bibinfo  {journal} {Comput. Phys. Commun.}\ }\textbf {\bibinfo
  {volume} {83}},\ \bibinfo {pages} {307} (\bibinfo {year} {1994})},\ \Eprint
  {http://arxiv.org/abs/nucl-th/9502021} {arXiv:nucl-th/9502021} \BibitemShut
  {NoStop}%
\bibitem [{\citenamefont {Pang}\ \emph {et~al.}(2012)\citenamefont {Pang},
  \citenamefont {Wang},\ and\ \citenamefont {Wang}}]{Pang:2012he}%
  \BibitemOpen
  \bibfield  {author} {\bibinfo {author} {\bibfnamefont {L.}~\bibnamefont
  {Pang}}, \bibinfo {author} {\bibfnamefont {Q.}~\bibnamefont {Wang}}, \ and\
  \bibinfo {author} {\bibfnamefont {X.-N.}\ \bibnamefont {Wang}},\ }\href
  {\doibase 10.1103/PhysRevC.86.024911} {\bibfield  {journal} {\bibinfo
  {journal} {Phys. Rev. C}\ }\textbf {\bibinfo {volume} {86}},\ \bibinfo
  {pages} {024911} (\bibinfo {year} {2012})},\ \Eprint
  {http://arxiv.org/abs/1205.5019} {arXiv:1205.5019 [nucl-th]} \BibitemShut
  {NoStop}%
\bibitem [{\citenamefont {Pang}\ \emph {et~al.}(2016)\citenamefont {Pang},
  \citenamefont {Petersen}, \citenamefont {Wang},\ and\ \citenamefont
  {Wang}}]{Pang:2016igs}%
  \BibitemOpen
  \bibfield  {author} {\bibinfo {author} {\bibfnamefont {L.-G.}\ \bibnamefont
  {Pang}}, \bibinfo {author} {\bibfnamefont {H.}~\bibnamefont {Petersen}},
  \bibinfo {author} {\bibfnamefont {Q.}~\bibnamefont {Wang}}, \ and\ \bibinfo
  {author} {\bibfnamefont {X.-N.}\ \bibnamefont {Wang}},\ }\href {\doibase
  10.1103/PhysRevLett.117.192301} {\bibfield  {journal} {\bibinfo  {journal}
  {Phys. Rev. Lett.}\ }\textbf {\bibinfo {volume} {117}},\ \bibinfo {pages}
  {192301} (\bibinfo {year} {2016})},\ \Eprint
  {http://arxiv.org/abs/1605.04024} {arXiv:1605.04024 [hep-ph]} \BibitemShut
  {NoStop}%
\bibitem [{\citenamefont {Denicol}\ \emph {et~al.}(2012)\citenamefont
  {Denicol}, \citenamefont {Niemi}, \citenamefont {Molnar},\ and\ \citenamefont
  {Rischke}}]{Denicol:2012cn}%
  \BibitemOpen
  \bibfield  {author} {\bibinfo {author} {\bibfnamefont {G.}~\bibnamefont
  {Denicol}}, \bibinfo {author} {\bibfnamefont {H.}~\bibnamefont {Niemi}},
  \bibinfo {author} {\bibfnamefont {E.}~\bibnamefont {Molnar}}, \ and\ \bibinfo
  {author} {\bibfnamefont {D.}~\bibnamefont {Rischke}},\ }\href {\doibase
  10.1103/PhysRevD.85.114047} {\bibfield  {journal} {\bibinfo  {journal} {Phys.
  Rev. D}\ }\textbf {\bibinfo {volume} {85}},\ \bibinfo {pages} {114047}
  (\bibinfo {year} {2012})},\ \bibinfo {note} {[Erratum: Phys.Rev.D 91, 039902
  (2015)]},\ \Eprint {http://arxiv.org/abs/1202.4551} {arXiv:1202.4551
  [nucl-th]} \BibitemShut {NoStop}%
\bibitem [{\citenamefont {Denicol}\ \emph {et~al.}(2014)\citenamefont
  {Denicol}, \citenamefont {Jeon},\ and\ \citenamefont
  {Gale}}]{Denicol:2014vaa}%
  \BibitemOpen
  \bibfield  {author} {\bibinfo {author} {\bibfnamefont {G.}~\bibnamefont
  {Denicol}}, \bibinfo {author} {\bibfnamefont {S.}~\bibnamefont {Jeon}}, \
  and\ \bibinfo {author} {\bibfnamefont {C.}~\bibnamefont {Gale}},\ }\href
  {\doibase 10.1103/PhysRevC.90.024912} {\bibfield  {journal} {\bibinfo
  {journal} {Phys. Rev. C}\ }\textbf {\bibinfo {volume} {90}},\ \bibinfo
  {pages} {024912} (\bibinfo {year} {2014})},\ \Eprint
  {http://arxiv.org/abs/1403.0962} {arXiv:1403.0962 [nucl-th]} \BibitemShut
  {NoStop}%
\bibitem [{\citenamefont {Ryu}\ \emph {et~al.}(2015)\citenamefont {Ryu},
  \citenamefont {Paquet}, \citenamefont {Shen}, \citenamefont {Denicol},
  \citenamefont {Schenke}, \citenamefont {Jeon},\ and\ \citenamefont
  {Gale}}]{Ryu:2015vwa}%
  \BibitemOpen
  \bibfield  {author} {\bibinfo {author} {\bibfnamefont {S.}~\bibnamefont
  {Ryu}}, \bibinfo {author} {\bibfnamefont {J.~F.}\ \bibnamefont {Paquet}},
  \bibinfo {author} {\bibfnamefont {C.}~\bibnamefont {Shen}}, \bibinfo {author}
  {\bibfnamefont {G.}~\bibnamefont {Denicol}}, \bibinfo {author} {\bibfnamefont
  {B.}~\bibnamefont {Schenke}}, \bibinfo {author} {\bibfnamefont
  {S.}~\bibnamefont {Jeon}}, \ and\ \bibinfo {author} {\bibfnamefont
  {C.}~\bibnamefont {Gale}},\ }\href {\doibase 10.1103/PhysRevLett.115.132301}
  {\bibfield  {journal} {\bibinfo  {journal} {Phys. Rev. Lett.}\ }\textbf
  {\bibinfo {volume} {115}},\ \bibinfo {pages} {132301} (\bibinfo {year}
  {2015})},\ \Eprint {http://arxiv.org/abs/1502.01675} {arXiv:1502.01675
  [nucl-th]} \BibitemShut {NoStop}%
\bibitem [{\citenamefont {Shen}\ \emph {et~al.}(2016)\citenamefont {Shen},
  \citenamefont {Qiu}, \citenamefont {Song}, \citenamefont {Bernhard},
  \citenamefont {Bass},\ and\ \citenamefont {Heinz}}]{Shen:2014vra}%
  \BibitemOpen
  \bibfield  {author} {\bibinfo {author} {\bibfnamefont {C.}~\bibnamefont
  {Shen}}, \bibinfo {author} {\bibfnamefont {Z.}~\bibnamefont {Qiu}}, \bibinfo
  {author} {\bibfnamefont {H.}~\bibnamefont {Song}}, \bibinfo {author}
  {\bibfnamefont {J.}~\bibnamefont {Bernhard}}, \bibinfo {author}
  {\bibfnamefont {S.}~\bibnamefont {Bass}}, \ and\ \bibinfo {author}
  {\bibfnamefont {U.}~\bibnamefont {Heinz}},\ }\href {\doibase
  10.1016/j.cpc.2015.08.039} {\bibfield  {journal} {\bibinfo  {journal}
  {Comput. Phys. Commun.}\ }\textbf {\bibinfo {volume} {199}},\ \bibinfo
  {pages} {61} (\bibinfo {year} {2016})},\ \Eprint
  {http://arxiv.org/abs/1409.8164} {arXiv:1409.8164 [nucl-th]} \BibitemShut
  {NoStop}%
\bibitem [{\citenamefont {Cooper}\ and\ \citenamefont
  {Frye}(1974)}]{Cooper:1974mv}%
  \BibitemOpen
  \bibfield  {author} {\bibinfo {author} {\bibfnamefont {F.}~\bibnamefont
  {Cooper}}\ and\ \bibinfo {author} {\bibfnamefont {G.}~\bibnamefont {Frye}},\
  }\href {\doibase 10.1103/PhysRevD.10.186} {\bibfield  {journal} {\bibinfo
  {journal} {Phys. Rev. D}\ }\textbf {\bibinfo {volume} {10}},\ \bibinfo
  {pages} {186} (\bibinfo {year} {1974})}\BibitemShut {NoStop}%
\bibitem [{\citenamefont {Grad}(1949)}]{doi:10.1002}%
  \BibitemOpen
  \bibfield  {author} {\bibinfo {author} {\bibfnamefont {H.}~\bibnamefont
  {Grad}},\ }\href {\doibase 10.1002/cpa.3160020403} {\bibfield  {journal}
  {\bibinfo  {journal} {Communications on Pure and Applied Mathematics}\
  }\textbf {\bibinfo {volume} {2}},\ \bibinfo {pages} {331} (\bibinfo {year}
  {1949})},\ \Eprint
  {http://arxiv.org/abs/https://onlinelibrary.wiley.com/doi/pdf/10.1002/cpa.3160020403}
  {https://onlinelibrary.wiley.com/doi/pdf/10.1002/cpa.3160020403} \BibitemShut
  {NoStop}%
\bibitem [{\citenamefont {Adams}(2020)}]{Adams:AUM}%
  \BibitemOpen
  \bibfield  {author} {\bibinfo {author} {\bibfnamefont {J.}~\bibnamefont
  {Adams}} (\bibinfo {collaboration} {STAR}),\ }\href@noop {} {\enquote
  {\bibinfo {title} {{\it Global polarization of hyperons from STAR
  experiment}},}\ }\bibinfo {howpublished} {Talk given at RHIC and AGS Annual
  Users Meeting 2020} (\bibinfo {year} {2020})\BibitemShut {NoStop}%
\bibitem [{\citenamefont {Deng}\ and\ \citenamefont
  {Huang}(2016)}]{Deng:2016gyh}%
  \BibitemOpen
  \bibfield  {author} {\bibinfo {author} {\bibfnamefont {W.-T.}\ \bibnamefont
  {Deng}}\ and\ \bibinfo {author} {\bibfnamefont {X.-G.}\ \bibnamefont
  {Huang}},\ }\href {\doibase 10.1103/PhysRevC.93.064907} {\bibfield  {journal}
  {\bibinfo  {journal} {Phys. Rev. C}\ }\textbf {\bibinfo {volume} {93}},\
  \bibinfo {pages} {064907} (\bibinfo {year} {2016})},\ \Eprint
  {http://arxiv.org/abs/1603.06117} {arXiv:1603.06117 [nucl-th]} \BibitemShut
  {NoStop}%
\bibitem [{\citenamefont {Jiang}\ \emph {et~al.}(2016)\citenamefont {Jiang},
  \citenamefont {Lin},\ and\ \citenamefont {Liao}}]{Jiang:2016woz}%
  \BibitemOpen
  \bibfield  {author} {\bibinfo {author} {\bibfnamefont {Y.}~\bibnamefont
  {Jiang}}, \bibinfo {author} {\bibfnamefont {Z.-W.}\ \bibnamefont {Lin}}, \
  and\ \bibinfo {author} {\bibfnamefont {J.}~\bibnamefont {Liao}},\ }\href
  {\doibase 10.1103/PhysRevC.94.044910} {\bibfield  {journal} {\bibinfo
  {journal} {Phys. Rev. C}\ }\textbf {\bibinfo {volume} {94}},\ \bibinfo
  {pages} {044910} (\bibinfo {year} {2016})},\ \bibinfo {note} {[Erratum:
  Phys.Rev.C 95, 049904 (2017)]},\ \Eprint {http://arxiv.org/abs/1602.06580}
  {arXiv:1602.06580 [hep-ph]} \BibitemShut {NoStop}%
\bibitem [{\citenamefont {Deng}\ \emph {et~al.}(2020)\citenamefont {Deng},
  \citenamefont {Huang}, \citenamefont {Ma},\ and\ \citenamefont
  {Zhang}}]{Deng:2020ygd}%
  \BibitemOpen
  \bibfield  {author} {\bibinfo {author} {\bibfnamefont {X.-G.}\ \bibnamefont
  {Deng}}, \bibinfo {author} {\bibfnamefont {X.-G.}\ \bibnamefont {Huang}},
  \bibinfo {author} {\bibfnamefont {Y.-G.}\ \bibnamefont {Ma}}, \ and\ \bibinfo
  {author} {\bibfnamefont {S.}~\bibnamefont {Zhang}},\ }\href {\doibase
  10.1103/PhysRevC.101.064908} {\bibfield  {journal} {\bibinfo  {journal}
  {Phys. Rev. C}\ }\textbf {\bibinfo {volume} {101}},\ \bibinfo {pages}
  {064908} (\bibinfo {year} {2020})},\ \Eprint
  {http://arxiv.org/abs/2001.01371} {arXiv:2001.01371 [nucl-th]} \BibitemShut
  {NoStop}%
\bibitem [{\citenamefont {Fu}\ \emph {et~al.}()\citenamefont {Fu},
  \citenamefont {Xu}, \citenamefont {Huang},\ and\ \citenamefont
  {Song}}]{Note1}%
  \BibitemOpen
  \bibfield  {author} {\bibinfo {author} {\bibfnamefont {B.}~\bibnamefont
  {Fu}}, \bibinfo {author} {\bibfnamefont {K.}~\bibnamefont {Xu}}, \bibinfo
  {author} {\bibfnamefont {X.-G.}\ \bibnamefont {Huang}}, \ and\ \bibinfo
  {author} {\bibfnamefont {H.}~\bibnamefont {Song}},\ }\Eprint
  {http://arxiv.org/abs/privite notes} {privite notes} \BibitemShut {NoStop}%
\bibitem [{\citenamefont {Adams}(2019)}]{Adams:qm}%
  \BibitemOpen
  \bibfield  {author} {\bibinfo {author} {\bibfnamefont {J.}~\bibnamefont
  {Adams}} (\bibinfo {collaboration} {STAR}),\ }\href@noop {} {\enquote
  {\bibinfo {title} {{\it Differential measurements of Lambda polarization in
  Au + Au collisions and a search for the magnetic field by STAR}},}\ }\bibinfo
  {howpublished} {Talk given at Quark Matter Conference} (\bibinfo {year}
  {2019})\BibitemShut {NoStop}%
\bibitem [{\citenamefont {Guo}\ \emph {et~al.}(2019)\citenamefont {Guo},
  \citenamefont {Shi}, \citenamefont {Feng},\ and\ \citenamefont
  {Liao}}]{Guo:2019joy}%
  \BibitemOpen
  \bibfield  {author} {\bibinfo {author} {\bibfnamefont {Y.}~\bibnamefont
  {Guo}}, \bibinfo {author} {\bibfnamefont {S.}~\bibnamefont {Shi}}, \bibinfo
  {author} {\bibfnamefont {S.}~\bibnamefont {Feng}}, \ and\ \bibinfo {author}
  {\bibfnamefont {J.}~\bibnamefont {Liao}},\ }\href {\doibase
  10.1016/j.physletb.2019.134929} {\bibfield  {journal} {\bibinfo  {journal}
  {Phys. Lett. B}\ }\textbf {\bibinfo {volume} {798}},\ \bibinfo {pages}
  {134929} (\bibinfo {year} {2019})},\ \Eprint
  {http://arxiv.org/abs/1905.12613} {arXiv:1905.12613 [nucl-th]} \BibitemShut
  {NoStop}%
\bibitem [{\citenamefont {Ambrus}\ and\ \citenamefont
  {Chernodub}(2020)}]{Ambrus:2020oiw}%
  \BibitemOpen
  \bibfield  {author} {\bibinfo {author} {\bibfnamefont {V.~E.}\ \bibnamefont
  {Ambrus}}\ and\ \bibinfo {author} {\bibfnamefont {M.}~\bibnamefont
  {Chernodub}},\ }\href@noop {} {\  (\bibinfo {year} {2020})},\ \Eprint
  {http://arxiv.org/abs/2010.05831} {arXiv:2010.05831 [hep-ph]} \BibitemShut
  {NoStop}%
\bibitem [{\citenamefont {Ivanov}\ and\ \citenamefont
  {Soldatov}(2018)}]{Ivanov:2018eej}%
  \BibitemOpen
  \bibfield  {author} {\bibinfo {author} {\bibfnamefont {Y.~B.}\ \bibnamefont
  {Ivanov}}\ and\ \bibinfo {author} {\bibfnamefont {A.}~\bibnamefont
  {Soldatov}},\ }\href {\doibase 10.1103/PhysRevC.97.044915} {\bibfield
  {journal} {\bibinfo  {journal} {Phys. Rev. C}\ }\textbf {\bibinfo {volume}
  {97}},\ \bibinfo {pages} {044915} (\bibinfo {year} {2018})},\ \Eprint
  {http://arxiv.org/abs/1803.01525} {arXiv:1803.01525 [nucl-th]} \BibitemShut
  {NoStop}%
\bibitem [{\citenamefont {Xia}\ \emph {et~al.}(2018)\citenamefont {Xia},
  \citenamefont {Li}, \citenamefont {Tang},\ and\ \citenamefont
  {Wang}}]{Xia:2018tes}%
  \BibitemOpen
  \bibfield  {author} {\bibinfo {author} {\bibfnamefont {X.-L.}\ \bibnamefont
  {Xia}}, \bibinfo {author} {\bibfnamefont {H.}~\bibnamefont {Li}}, \bibinfo
  {author} {\bibfnamefont {Z.-B.}\ \bibnamefont {Tang}}, \ and\ \bibinfo
  {author} {\bibfnamefont {Q.}~\bibnamefont {Wang}},\ }\href {\doibase
  10.1103/PhysRevC.98.024905} {\bibfield  {journal} {\bibinfo  {journal} {Phys.
  Rev. C}\ }\textbf {\bibinfo {volume} {98}},\ \bibinfo {pages} {024905}
  (\bibinfo {year} {2018})},\ \Eprint {http://arxiv.org/abs/1803.00867}
  {arXiv:1803.00867 [nucl-th]} \BibitemShut {NoStop}%
\bibitem [{\citenamefont {Back}\ \emph {et~al.}(2003)\citenamefont {Back} \emph
  {et~al.}}]{PHOBOS:2002wb}%
  \BibitemOpen
  \bibfield  {author} {\bibinfo {author} {\bibfnamefont {B.}~\bibnamefont
  {Back}} \emph {et~al.},\ }\href {\doibase 10.1103/PhysRevLett.91.052303}
  {\bibfield  {journal} {\bibinfo  {journal} {Phys. Rev. Lett.}\ }\textbf
  {\bibinfo {volume} {91}},\ \bibinfo {pages} {052303} (\bibinfo {year}
  {2003})},\ \Eprint {http://arxiv.org/abs/nucl-ex/0210015}
  {arXiv:nucl-ex/0210015} \BibitemShut {NoStop}%
\bibitem [{\citenamefont {Adamczyk}\ \emph
  {et~al.}(2017{\natexlab{b}})\citenamefont {Adamczyk} \emph
  {et~al.}}]{STAR_BES:2017iwn}%
  \BibitemOpen
  \bibfield  {author} {\bibinfo {author} {\bibfnamefont {L.}~\bibnamefont
  {Adamczyk}} \emph {et~al.} (\bibinfo {collaboration} {STAR}),\ }\href
  {\doibase 10.1103/PhysRevC.96.044904} {\bibfield  {journal} {\bibinfo
  {journal} {Phys. Rev. C}\ }\textbf {\bibinfo {volume} {96}},\ \bibinfo
  {pages} {044904} (\bibinfo {year} {2017}{\natexlab{b}})},\ \Eprint
  {http://arxiv.org/abs/1701.07065} {arXiv:1701.07065 [nucl-ex]} \BibitemShut
  {NoStop}%
\bibitem [{\citenamefont {Adler}\ \emph {et~al.}(2004)\citenamefont {Adler}
  \emph {et~al.}}]{PHENIX:2003cb}%
  \BibitemOpen
  \bibfield  {author} {\bibinfo {author} {\bibfnamefont {S.}~\bibnamefont
  {Adler}} \emph {et~al.} (\bibinfo {collaboration} {PHENIX}),\ }\href
  {\doibase 10.1103/PhysRevC.69.034909} {\bibfield  {journal} {\bibinfo
  {journal} {Phys. Rev. C}\ }\textbf {\bibinfo {volume} {69}},\ \bibinfo
  {pages} {034909} (\bibinfo {year} {2004})},\ \Eprint
  {http://arxiv.org/abs/nucl-ex/0307022} {arXiv:nucl-ex/0307022} \BibitemShut
  {NoStop}%
\bibitem [{\citenamefont {Adare}\ \emph {et~al.}(2015)\citenamefont {Adare}
  \emph {et~al.}}]{PHENIX:2014bga}%
  \BibitemOpen
  \bibfield  {author} {\bibinfo {author} {\bibfnamefont {A.}~\bibnamefont
  {Adare}} \emph {et~al.} (\bibinfo {collaboration} {PHENIX}),\ }\href
  {\doibase 10.1103/PhysRevC.92.034913} {\bibfield  {journal} {\bibinfo
  {journal} {Phys. Rev. C}\ }\textbf {\bibinfo {volume} {92}},\ \bibinfo
  {pages} {034913} (\bibinfo {year} {2015})},\ \Eprint
  {http://arxiv.org/abs/1412.1043} {arXiv:1412.1043 [nucl-ex]} \BibitemShut
  {NoStop}%
\bibitem [{\citenamefont {Adamczyk}\ \emph {et~al.}(2016)\citenamefont
  {Adamczyk} \emph {et~al.}}]{STAR:2015fum}%
  \BibitemOpen
  \bibfield  {author} {\bibinfo {author} {\bibfnamefont {L.}~\bibnamefont
  {Adamczyk}} \emph {et~al.} (\bibinfo {collaboration} {STAR}),\ }\href
  {\doibase 10.1103/PhysRevC.93.014907} {\bibfield  {journal} {\bibinfo
  {journal} {Phys. Rev. C}\ }\textbf {\bibinfo {volume} {93}},\ \bibinfo
  {pages} {014907} (\bibinfo {year} {2016})},\ \Eprint
  {http://arxiv.org/abs/1509.08397} {arXiv:1509.08397 [nucl-ex]} \BibitemShut
  {NoStop}%
\bibitem [{\citenamefont {Karpenko}\ and\ \citenamefont
  {Becattini}(2019)}]{Karpenko:2018erl}%
  \BibitemOpen
  \bibfield  {author} {\bibinfo {author} {\bibfnamefont {I.}~\bibnamefont
  {Karpenko}}\ and\ \bibinfo {author} {\bibfnamefont {F.}~\bibnamefont
  {Becattini}},\ }\href {\doibase 10.1016/j.nuclphysa.2018.10.067} {\bibfield
  {journal} {\bibinfo  {journal} {Nucl. Phys. A}\ }\textbf {\bibinfo {volume}
  {982}},\ \bibinfo {pages} {519} (\bibinfo {year} {2019})},\ \Eprint
  {http://arxiv.org/abs/1811.00322} {arXiv:1811.00322 [nucl-th]} \BibitemShut
  {NoStop}%
\bibitem [{\citenamefont {Florkowski}\ \emph {et~al.}(2018)\citenamefont
  {Florkowski}, \citenamefont {Friman}, \citenamefont {Jaiswal},\ and\
  \citenamefont {Speranza}}]{Florkowski:2017ruc}%
  \BibitemOpen
  \bibfield  {author} {\bibinfo {author} {\bibfnamefont {W.}~\bibnamefont
  {Florkowski}}, \bibinfo {author} {\bibfnamefont {B.}~\bibnamefont {Friman}},
  \bibinfo {author} {\bibfnamefont {A.}~\bibnamefont {Jaiswal}}, \ and\
  \bibinfo {author} {\bibfnamefont {E.}~\bibnamefont {Speranza}},\ }\href
  {\doibase 10.1103/PhysRevC.97.041901} {\bibfield  {journal} {\bibinfo
  {journal} {Phys. Rev. C}\ }\textbf {\bibinfo {volume} {97}},\ \bibinfo
  {pages} {041901} (\bibinfo {year} {2018})},\ \Eprint
  {http://arxiv.org/abs/1705.00587} {arXiv:1705.00587 [nucl-th]} \BibitemShut
  {NoStop}%
\bibitem [{\citenamefont {Hattori}\ \emph {et~al.}(2019)\citenamefont
  {Hattori}, \citenamefont {Hongo}, \citenamefont {Huang}, \citenamefont
  {Matsuo},\ and\ \citenamefont {Taya}}]{Hattori:2019lfp}%
  \BibitemOpen
  \bibfield  {author} {\bibinfo {author} {\bibfnamefont {K.}~\bibnamefont
  {Hattori}}, \bibinfo {author} {\bibfnamefont {M.}~\bibnamefont {Hongo}},
  \bibinfo {author} {\bibfnamefont {X.-G.}\ \bibnamefont {Huang}}, \bibinfo
  {author} {\bibfnamefont {M.}~\bibnamefont {Matsuo}}, \ and\ \bibinfo {author}
  {\bibfnamefont {H.}~\bibnamefont {Taya}},\ }\href {\doibase
  10.1016/j.physletb.2019.05.040} {\bibfield  {journal} {\bibinfo  {journal}
  {Phys. Lett. B}\ }\textbf {\bibinfo {volume} {795}},\ \bibinfo {pages} {100}
  (\bibinfo {year} {2019})},\ \Eprint {http://arxiv.org/abs/1901.06615}
  {arXiv:1901.06615 [hep-th]} \BibitemShut {NoStop}%
\bibitem [{\citenamefont {Montenegro}\ and\ \citenamefont
  {Torrieri}(2019)}]{Montenegro:2018bcf}%
  \BibitemOpen
  \bibfield  {author} {\bibinfo {author} {\bibfnamefont {D.}~\bibnamefont
  {Montenegro}}\ and\ \bibinfo {author} {\bibfnamefont {G.}~\bibnamefont
  {Torrieri}},\ }\href {\doibase 10.1103/PhysRevD.100.056011} {\bibfield
  {journal} {\bibinfo  {journal} {Phys. Rev. D}\ }\textbf {\bibinfo {volume}
  {100}},\ \bibinfo {pages} {056011} (\bibinfo {year} {2019})},\ \Eprint
  {http://arxiv.org/abs/1807.02796} {arXiv:1807.02796 [hep-th]} \BibitemShut
  {NoStop}%
\bibitem [{\citenamefont {Florkowski}\ \emph
  {et~al.}(2019{\natexlab{b}})\citenamefont {Florkowski}, \citenamefont
  {Kumar},\ and\ \citenamefont {Ryblewski}}]{Florkowski:2018fap}%
  \BibitemOpen
  \bibfield  {author} {\bibinfo {author} {\bibfnamefont {W.}~\bibnamefont
  {Florkowski}}, \bibinfo {author} {\bibfnamefont {A.}~\bibnamefont {Kumar}}, \
  and\ \bibinfo {author} {\bibfnamefont {R.}~\bibnamefont {Ryblewski}},\ }\href
  {\doibase 10.1016/j.ppnp.2019.07.001} {\bibfield  {journal} {\bibinfo
  {journal} {Prog. Part. Nucl. Phys.}\ }\textbf {\bibinfo {volume} {108}},\
  \bibinfo {pages} {103709} (\bibinfo {year} {2019}{\natexlab{b}})},\ \Eprint
  {http://arxiv.org/abs/1811.04409} {arXiv:1811.04409 [nucl-th]} \BibitemShut
  {NoStop}%
\bibitem [{\citenamefont {Bhadury}\ \emph
  {et~al.}(2020{\natexlab{a}})\citenamefont {Bhadury}, \citenamefont
  {Florkowski}, \citenamefont {Jaiswal}, \citenamefont {Kumar},\ and\
  \citenamefont {Ryblewski}}]{Bhadury:2020puc}%
  \BibitemOpen
  \bibfield  {author} {\bibinfo {author} {\bibfnamefont {S.}~\bibnamefont
  {Bhadury}}, \bibinfo {author} {\bibfnamefont {W.}~\bibnamefont {Florkowski}},
  \bibinfo {author} {\bibfnamefont {A.}~\bibnamefont {Jaiswal}}, \bibinfo
  {author} {\bibfnamefont {A.}~\bibnamefont {Kumar}}, \ and\ \bibinfo {author}
  {\bibfnamefont {R.}~\bibnamefont {Ryblewski}},\ }\href@noop {} {\  (\bibinfo
  {year} {2020}{\natexlab{a}})},\ \Eprint {http://arxiv.org/abs/2002.03937}
  {arXiv:2002.03937 [hep-ph]} \BibitemShut {NoStop}%
\bibitem [{\citenamefont {Bhadury}\ \emph
  {et~al.}(2020{\natexlab{b}})\citenamefont {Bhadury}, \citenamefont
  {Florkowski}, \citenamefont {Jaiswal}, \citenamefont {Kumar},\ and\
  \citenamefont {Ryblewski}}]{Bhadury:2020cop}%
  \BibitemOpen
  \bibfield  {author} {\bibinfo {author} {\bibfnamefont {S.}~\bibnamefont
  {Bhadury}}, \bibinfo {author} {\bibfnamefont {W.}~\bibnamefont {Florkowski}},
  \bibinfo {author} {\bibfnamefont {A.}~\bibnamefont {Jaiswal}}, \bibinfo
  {author} {\bibfnamefont {A.}~\bibnamefont {Kumar}}, \ and\ \bibinfo {author}
  {\bibfnamefont {R.}~\bibnamefont {Ryblewski}},\ }\href@noop {} {\  (\bibinfo
  {year} {2020}{\natexlab{b}})},\ \Eprint {http://arxiv.org/abs/2008.10976}
  {arXiv:2008.10976 [nucl-th]} \BibitemShut {NoStop}%
\bibitem [{\citenamefont {Shi}\ \emph {et~al.}(2020)\citenamefont {Shi},
  \citenamefont {Gale},\ and\ \citenamefont {Jeon}}]{Shi:2020htn}%
  \BibitemOpen
  \bibfield  {author} {\bibinfo {author} {\bibfnamefont {S.}~\bibnamefont
  {Shi}}, \bibinfo {author} {\bibfnamefont {C.}~\bibnamefont {Gale}}, \ and\
  \bibinfo {author} {\bibfnamefont {S.}~\bibnamefont {Jeon}},\ }\href@noop {}
  {\  (\bibinfo {year} {2020})},\ \Eprint {http://arxiv.org/abs/2008.08618}
  {arXiv:2008.08618 [nucl-th]} \BibitemShut {NoStop}%
\bibitem [{\citenamefont {Fukushima}\ and\ \citenamefont
  {Pu}(2020)}]{Fukushima:2020ucl}%
  \BibitemOpen
  \bibfield  {author} {\bibinfo {author} {\bibfnamefont {K.}~\bibnamefont
  {Fukushima}}\ and\ \bibinfo {author} {\bibfnamefont {S.}~\bibnamefont {Pu}},\
  }\href@noop {} {\  (\bibinfo {year} {2020})},\ \Eprint
  {http://arxiv.org/abs/2010.01608} {arXiv:2010.01608 [hep-th]} \BibitemShut
  {NoStop}%
\end{thebibliography}%

\end{document}